\documentclass[twocolumn,twocolappendix]{aastex631}

\usepackage{amsmath}
\usepackage{bm}
\usepackage{rotating}
\usepackage{physics}
\usepackage{CJK}

\usepackage[dvipsnames]{xcolor}
\AtBeginDocument{\colorlet{defaultcolor}{.}}
\newcommand{\revcolor}{defaultcolor} % defaultcolor; BrickRed

\begin{document}

\begin{CJK*}{UTF8}{gkai}

\title{Stellar mergers and chemical element mixing: implications for metamorphic stellar evolution \\in AGN disks}

\correspondingauthor{Yanlong Shi}
\email{yanlong@cita.utoronto.ca}

\newcommand{\cita}{Canadian Institute for Theoretical Astrophysics, University of Toronto, Toronto, ON M5S 3H8, Canada}
\newcommand{\caltech}{TAPIR, MC 350-17, California Institute of Technology, Pasadena, CA 91125, USA}

\newcommand{\ucsc}{Department of Astronomy and Astrophysics, University of California, Santa Cruz, CA, 95064, USA}

\newcommand{\tsinghua}{Institute for Advanced Study, Tsinghua University, Beijing, 100084, China}

\newcommand{\westlake}{Department of Astronomy, Westlake University, Hangzhou, Zhejiang, 310030, China}

\author[0000-0002-0087-3237]{Yanlong Shi}
\affiliation{\cita}

\author[0000-0003-2868-489X]{Xiaoshan Huang (黄小珊)}
\affiliation{\caltech}

\author[0000-0001-5466-4628]{Douglas N. C. Lin (林潮)}
\affiliation{\ucsc}
\affiliation{\tsinghua}
\affiliation{\westlake}

\author[0000-0002-8659-3729]{Norman Murray}
\affiliation{\cita}

%\collaboration{20}{(AAS Journals Data Editors)}



\begin{abstract}

Chemical mixing during stellar mergers can significantly influence the subsequent evolution of the merger remnant. We perform a suite of three-dimensional hydrodynamical simulations of stellar mergers, each evolved for $\sim100$ stellar dynamical times until the remnant reaches a quasi-hydrostatic equilibrium. The simulations incorporate subgrid-scale diffusion models to capture the turbulent mixing of chemical elements. Starting with an idealized polytropic equation of state (EOS), we first identify the dominant mixing mechanisms and investigate how the merger outcome depends on the mass ratio, relative velocity, impact parameter, and stellar structure. We then extend our simulations to the context of stars embedded in active galactic nucleus (AGN) disks, using a realistic, composition-dependent EOS and AGN stellar models generated with the stellar evolution code MESA. We find that mergers with both younger metamorphic stars and H-rich accreting AGN stars can substantially rejuvenate old metamorphic stars through efficient core mixing after thermal relaxation. The merger remnants are nitrogen-enriched, with ${\rm N/O}\sim1$--3 and ${\rm N/C}\gtrsim5$, comparable to the abundances observed in the nebula surrounding SN~1987A. During subsequent stellar evolution, the remnants may converge onto the main sequence of isolated metamorphic AGN stars once they reach accretion--wind equilibrium. They may also deposit a significant amount of chemically enriched material into the AGN disk. This work provides a physical framework for connecting hydrodynamical stellar mergers with the long-term evolution of AGN stars and their observational and chemical signatures.

\end{abstract}



\section{Introduction} \label{sec:intro}

The merger of non-compact stars is a notable astrophysical process that can happen in both isolated binary evolution and dense stellar environments \citep{Schneider_2025arXiv250918421S}. For example, mass transfer from a supergiant to its binary companion can shrink the orbit, potentially driving overflow through the L$_2$ point and ultimately leading to a stellar merger \citep[][]{Pols_1994A&A...290..119P,WellsteinLangerBraun_2001A&A...369..939W,HennecoSchneiderLaplace_2024A&A...682A.169H}; in dense star clusters, repeated runaway stellar mergers may produce a central supermassive star, providing a promising channel for the formation of intermediate-mass black holes \citep[e.g.,][]{PortegiesZwartMcMillan_2002ApJ...576..899P,GurkanFreitagRasio_2004ApJ...604..632G,Mapelli_2016MNRAS.459.3432M,KremerSperaBecker_2020ApJ...903...45K,ShiGrudicHopkins_2021MNRAS.505.2753S,RantalaNaabLahen_2024MNRAS.531.3770R}. Stellar mergers are also speculated to be associated with blue stragglers \citep[][]{SillsAdamsDavies_2005MNRAS.358..716S}, SN 1987A \citep[][]{PodsiadlowskiJossRappaport_1990A&A...227L...9P}, and GW190521 \citep[][]{DiCarloMapelliPasquato_2021MNRAS.507.5132D}.

Another potential site for stellar mergers/collisions is the disks of active galactic nuclei (AGN). \textcolor{\revcolor}{In gravitationally unstable outliers of AGN disks, gas can fragment, collapse, and form stars \citep[e.g.,][]{Paczynski_1978AcA....28...91P,KolykhalovSyunyaev_1980SvAL....6..357K,Goodman_2003MNRAS.339..937G,CollinZahn_2008A&A...477..419C}. Mature stars in nuclear clusters may also be trapped by hydrodynamic drag as they pass through the AGN disks \citep[e.g.,][]{SyerClarkeRees_1991MNRAS.250..505S,artymowicz1993, davies2020}.} Similar to planets in protostellar disks \citep[e.g.,][]{linpap1986, ward1997, paardekooper2011, kleynelson2012, wu2024, ida2026}, these embedded stars migrate inward, outward, or stochastically. Stellar collisions may then occur frequently in these dense, dynamically evolving environments \citep[$10^{-3}$--1\,$\rm yr^{-1}$;][]{ChenLin_2024ApJ...967...88C}. The merger rate of embedded black holes in AGN disks is also significant \citep[e.g.,][]{BartosKocsisHaiman_2017ApJ...835..165B,TagawaHaimanKocsis_2020ApJ...898...25T,FordMcKernan_2022MNRAS.517.5827F}, making them a major host of gravitational-wave sources \citep[but also see][]{TomarHopkinsKremer_2026PhRvD.113f3036T}.

A key distinction of stellar evolution in AGN disks from that in the interstellar medium is that stars are continuously embedded in a dense gaseous background of $10^{-20}$--$10^{-10}\,\rm g\,cm^{-3}$, which enables sustained accretion \citep[][]{CantielloJermynLin_2021ApJ...910...94C,DittmannCantielloJermyn_2021ApJ...916...48D}, particularly of hydrogen that fuels nuclear burning during
the main-sequence evolution. \textcolor{\revcolor}{Meanwhile, these embedded stars eject radiation-driven winds, leading to chemical enrichment of the disk 
\citep[][]{CantielloJermynLin_2021ApJ...910...94C,DittmannCantielloJermyn_2021ApJ...916...48D}.} After an initial accretion-dominated mass-growing phase, these competing processes lead to an accretion--wind equilibrium with 
a mass-to-light ratio comparable to that of an Eddington limit
\citep{CantielloJermynLin_2021ApJ...910...94C}.  In this equilibrium, 
if the accretion of H-rich gas from the disk into its nuclear-burning 
core is faster than the H-exhaustion there, the star would indefinitely
remain on the main sequence and reach an ``immortal'' 
state \citep[]{JermynDittmannMcKernan_2022ApJ...929..133J,DittmannCantiello_2025ApJ...979..245D,FabjDittmannCantiello_2025ApJ...981...16F}. 

The ``immortal star'' scenario also predicts distinctive chemical feedback signatures. Since immortal stars remain indefinitely on the main sequence while continuously undergoing CNO burning, their stellar winds are expected to be enriched in He \citep[][]{CantielloJermynLin_2021ApJ...910...94C,JermynDittmannMcKernan_2022ApJ...929..133J}. However, observations of the broad-line regions (BLRs) of AGNs reveal only modest He enrichment; instead, the BLRs exhibit significant enhancement of $\alpha$-elements relative to the narrow-line regions (NLRs) \citep[][]{HuangLinShields_2023MNRAS.525.5702H}. If the chemical enrichment originates from stars embedded in the AGN disk, these abundance patterns imply that the stars must evolve beyond the main sequence and undergo He burning, in tension with the ``immortal star'' scenario.

\textcolor{\revcolor}{
\citet{CantielloJermynLin_2021ApJ...910...94C} noted that the ``immortal star'' scenario necessarily requires efficient internal mixing; without it, freshly accreted H-rich material cannot be efficiently transported from the stellar surface to the nuclear-burning core across the intervening radiative layer. In the absence of efficient mixing, the accumulated envelope eventually becomes feedback-dominated, leading to mass loss and, ultimately, the formation of a compact object. Stars evolving through this pathway were subsequently termed ``metamorphic stars'' \citep[][]{Ali-DibLin_2023MNRAS.526.5824A,Xu_2025RAA....25k5013X,XuChenLin_2026ApJ...997..206X}. Metamorphic stars can evolve off the main sequence, exhaust their He, and produce $\alpha$-elements, potentially explaining the chemical enrichment observed in AGN \citep[][]{HuangLinShields_2023MNRAS.525.5702H,XuChenLin_2026ApJ...997..206X}.
} 

Beyond internal mixing, an alternative pathway for delivering the H-rich gas in and outside the radiative envelope to the convective core is through episodic stellar mergers, particularly given the high collision rates expected in AGN disks \citep[][]{ChenLin_2024ApJ...967...88C}. In this scenario, an old metamorphic star with an H-poor and He-rich interior is impacted by a younger companion with an H-rich and He-poor core, allowing occasional resupply of H-rich gas to be efficiently transported into the old star's core, so its main-sequence lifespan would be significantly prolonged. 

Previous studies have shown that stellar mergers can redistribute chemical elements \citep[][]{LombardiWarrenRasio_2002ApJ...568..939L,GaburovLombardiPortegiesZwart_2008MNRAS.383L...5G,GlebbeekPolsHurley_2008A&A...488.1007G,GlebbeekPols_2008A&A...488.1017G,GlebbeekGaburovPortegiesZwart_2013MNRAS.434.3497G,SchneiderPodsiadlowskiLanger_2016MNRAS.457.2355S}, but a systematic exploration of the underlying chemical mixing processes is still lacking. Here, we investigate this problem using three-dimensional simulations that evolve the mergers for $\sim 100\times$ dynamical time, sufficient for them to reach quasi-hydrostatic equilibrium. 

Below, in \S~\ref{sec:method}, we introduce the merger simulation setup. {The modeling of stars relies on the equation of state (EOS) of matter, and we start with the polytropic EOS in \S~\ref{sec:res}, to show how the outputs of the merger simulations depend on the associated parameter space. Then in \S~\ref{sec:res_mesa}, we utilize tabulated EOS extracted from the stellar evolution code \texttt{MESA} \citep[][]{JermynBauerSchwab_2023ApJS..265...15J}, and construct 3D AGN star models based on 1D models in \citet{XuChenLin_2026ApJ...997..206X}. This allows us to obtain key objectives to be determined from simulations, particularly (1)~the distribution of all chemical elements involved with the nuclear network and (2)~the total mass retained after the merger event ($M_{\rm merger}$), that are crucial to the fate of the merger before thermal equilibrium. Based on these results, we discuss the mergers involved with metamorphic AGN stars in \S~\ref{sec:implications}. Finally, we summarize in \S~\ref{sec:summary}.
}

% In general, the more massive primaries have smaller He abundance ($Y_\star$) then the less massive companion.  If ``entropy sorting'' prevents mixing, the residual core would attain $Y_{\rm merger} \sim Y_\star$ of the low-mass companion.  In contrast, efficient mixing averages the He abundance in the core such that the merged star's main sequence lifespan is prolonged/shortened in comparison with that of the companion/primary and the occurrence rate of the ``metamorphic'' transition is reduced.

\section{Method}
\label{sec:method}

In this work, we use the hydrodynamical code \texttt{GIZMO} \citep[][]{Hopkins_2015MNRAS.450...53H} to simulate the merger between stars in the meshless-finite-mass (MFM) mode, which in principle does not allow inter-cell mass flows. This setup alone is sufficient to capture the large-scale splatter of elements, but misses small-scale turbulent mixing \citep[similar to SPH simulations, e.g.,][]{BenzHills_1987ApJ...323..614B,LaiRasioShapiro_1993ApJ...412..593L,RasioShapiro_1995ApJ...438..887R,SillsLombardiBailyn_1997ApJ...487..290S,YuLai_2025ApJ...993...88Y,RoseLombardiGonzalezPrieto_2026ApJ..1000..162R}. This is mitigated by sub-grid turbulent diffusion models \citep[as in][]{ColbrookMaHopkins_2017MNRAS.467.2421C,RennehanBabulHopkins_2019MNRAS.483.3810R,Rennehan_2021MNRAS.506.2836R} that allow the inter-cell flow of passive scalars, as reviewed below.

\subsection{Sub-grid turbulent diffusion}

We assume negligible new production of elements during the mergers (since the duration is very short relative to the nuclear-burning time), then the abundance of an arbitrary element ($Y$) can be treated as a passive scalar that preserves the total elemental mass. Ideally, the abundance $Y$ is transported by the continuity equation of $\rho Y$; but for actual simulations, turbulent diffusion can happen physically at unresolved scales \citep[][]{KlessenLin_2003PhRvE..67d6311K,ColbrookMaHopkins_2017MNRAS.467.2421C}. 

This issue is mitigated by sub-grid scale (SGS) models. Each explicit quantity $\phi$ in a finite-resolution simulation is effectively filtered: $\phi \to \bar \phi = \phi \star G$, where $G$ is a low-pass filter. We also define the volume-averaged quantity (Favre averaging) as $\widetilde{\phi} = \overline{\rho \phi}/\overline{\rho}$. For the Euler equation, the filtered equation is $\partial_t (\bar \rho \bar v_i) = \partial_j (\bar \rho \widetilde{v_i v_j} + \bar p \delta_{ij} )$, while the filtered quantity $\widetilde{v_i v_j}$ is not measurable from the simulation (unlike $\Tilde{v}_i$). \citet{Smagorinsky_1963MWRv...91...99S} first proposed the model that $\widetilde{v_i v_j} = \Tilde{v}_i \Tilde{v}_j + \tau_{ij}$ where $\tau_{ij}$ is the turbulent stress tensor, which is modeled as $\tau_{ij} = -2 \bar \rho \nu_{\rm sgs} S_{ij}$. Here $\nu_{\rm sgs}$ is the SGS turbulent viscosity and $S_{ij} = (\partial_i v_j + \partial_j v_i)/2 - (\partial_k v_k)\delta_{ij}/3$ is the shear tensor. Moreover, for scalars like $e$ (specific internal energy) and $Y$, the sub-grid turbulent terms are modeled with, e.g., $\widetilde{e v_j} - \tilde{e} \tilde{v}_j = -\nu_{\rm sgs} \partial_j \bar e$ ($\nu_{\rm sgs}$ is the SGS turbulent diffusivity). Then the governing equations are
\begin{align}
    \partial_t \rho & + \nabla \cdot ( \rho \bm{v})  = 0; \label{equ:conintuity} \\
    \partial_t (\rho \bm{v}) & + \nabla \cdot ( \rho \bm{v} \bm{v} + p \mathbf{I} - 2 \rho \nu_{\rm sgs} \mathbf{S} )  = \rho \bm{g}; \label{equ:momentum} \\
    \partial_t \left( \rho e+ \rho v^2/2\right) & + \nabla \cdot \left[\left(\rho e+\rho v^2/2 + p \right) \bm{v} - \rho \nu_{\rm sgs} \nabla e \right. \nonumber \\ & \left. - 2\rho \nu_{\rm sgs}\mathbf{S}\cdot \bm{v} \right] = \rho \bm{g}\cdot \bm{v}; \label{equ:energy} \\
    \partial_t (\rho Y) & + \nabla \cdot ( \rho Y \bm{v} - \rho \nu_{\rm sgs} \nabla Y )  = 0. \label{equ:element_transport}
\end{align}
Here $\bm{g}$ is the gravitational acceleration that will be computed by tree gravity. 

From Eqs.~\eqref{equ:conintuity}--\eqref{equ:element_transport}, sub-grid turbulence leads to diffusion (of $e$, $Y$) and dissipation (in $\bm{v}$, $e$). The SGS viscosity/diffusivity is prescribed as  $\nu_{\rm sgs} = C^2 h^2 \sqrt{2 \mathbf{S}:\mathbf{S}}$ \citep{Smagorinsky_1963MWRv...91...99S},
with $C$ being a constant ($\sim$0.1--0.2) and $h$ being the inter-grid/particle separation. The method has been applied to SPH or MFM simulations of stellar chemical feedback \citep[e.g.,][]{ShenWadsleyStinson_2010MNRAS.407.1581S,HopkinsWetzelKeres_2018MNRAS.480..800H,EscalaWetzelKirby_2018MNRAS.474.2194E, Rennehan_2021MNRAS.506.2836R,ShiDaiMurray_2026ApJ...997..309S}. But the classical Smagorinsky model can overestimate the diffusion of elements in some cases \citep[e.g., examples in][]{Rennehan_2021MNRAS.506.2836R}, so improved dynamical models \citep{GermanoPiomelliMoin_1991PhFlA...3.1760G} treat $C \to C_{\rm dyn}(\bm{x},t)$, and evaluate it with the filtered quantities at different scales. This study employs the dynamical Smagorinsky module implemented and tested in \citet{RennehanBabulHopkins_2019MNRAS.483.3810R} and \citet{Rennehan_2021MNRAS.506.2836R}. { We also test the effects of SGS terms in controlled simulations described in Appendix~\ref{app:convergence_tests}.}

\subsection{Equation of state}
\label{sec:method:mixing_of_gas}

{We adopt two kinds of EOS in this study, the simple polytropic EOS, and a tabulated EOS from \texttt{MESA}.}

\subsubsection{Polytropic EOS}

With a polytropic EOS, the pressure is analytically a function of $e$, i.e., $p = (\gamma-1)\rho e \propto \rho^\gamma$, where $\gamma$ is the polytropic index. For polytropes of $n$, $\gamma = 1+ 1/n$. Massive stars are radiation-dominated and are well approximated by $n = 3$ ($\gamma=4/3$) polytropes. On the other hand, $n=1.5$ polytropes are used to model less-massive, convective ($\gamma=5/3$) stars. 

Since the EOS also explicitly depends on $\gamma$, a natural question is how to deal with the mixing of gas with different $\gamma$, which is required for mergers between radiative ($\gamma=5/3$) and convective ($\gamma=4/3$) stars. Here, we assume that $\gamma$ evolves the same way as the chemical abundance (i.e., Eq.~\ref{equ:element_transport}). This argument can be validated with the definition of $\gamma$, as the ratio between constant-pressure and constant-volume specific heat capacities, $c_p/c_V$. For a mixture of gas with density $\rho$ and components $\rho_1,\cdots, \rho_i,\cdots$ (where $i$ denotes the species of gas components), the effective polytropic index is $\gamma_{\rm eff}  = (\sum_i c_{p,i} \rho_i) / (\sum_i c_{V,i} \rho_i)$. Note that $c_{p,i}/c_{V,i}$ for different gas components are similar (e.g., 4/3 versus 5/3 for $n=3$ and $n=1.5$ cases here), we find $\gamma_{\rm eff}  \approx \sum_i \rho_i \gamma_i / \rho$. This approximation implies that $\gamma$ can be mass-weighted just like $Y$. Practically, we treat $\gamma$ as an additional entry of the ``chemical abundance'' that follows Eq.~\eqref{equ:element_transport}.

To validate this treatment, we perform idealized one-dimensional radiation-hydrodynamical (RHD) simulations of gas mixing in Appendix~\ref{app:gas_mixing}. We find that the effective polytropic index of the radiation-gas mixture is in good agreement with another test that mixes polytropic gases of $\gamma=4/3$ and $\gamma=5/3$ (in the same way as Eq.~\ref{equ:element_transport}).

{

\subsubsection{Tabulated \texttt{MESA} EOS}
\label{sec:method:mesa_eos}

We also construct a more realistic EOS that depends not only on density and temperature (or internal energy), but also on the chemical composition, i.e.,
$p = p(X, Z, \rho, T)$,
where $X$ and $Z$ denote the hydrogen and metal mass fractions, respectively. This EOS is implemented by interpolating over a precomputed EOS table. To generate the table, we use the EOS interface of the \texttt{MESA} code \citep[r24.08.1;][]{JermynBauerSchwab_2023ApJS..265...15J} to calculate thermodynamic quantities, including $p_{\rm gas}$, $e$, and $\Gamma_1$, on a user-defined four-dimensional grid of $(X, Z, \rho, T)$. By default, \texttt{MESA} blends several EOSs \citep[see][]{JermynBauerSchwab_2023ApJS..265...15J}, including HELM \citep[][]{TimmesSwesty_2000ApJS..126..501T}, FreeEOS \citep[][]{Irwin_2012ascl.soft11002I}, OPAL \citep[][]{RogersNayfonov_2002ApJ...576.1064R}, SCVH \citep[][]{SaumonChabriervanHorn_1995ApJS...99..713S}, and Skye \citep[][]{JermynSchwabBauer_2021ApJ...913...72J}. We adopt the same smooth cubic-spline interpolation scheme as used in the OPAL EOS. We compare the interpolated thermodynamic quantities with the original \texttt{MESA} stellar model (will be introduced in \S~\ref{sec:res_mesa}) and find that the interpolated pressure agrees to a relative error of $\sim10^{-4}$ throughout the stellar interior. This level of accuracy enables high-fidelity mapping of one-dimensional \texttt{MESA} models into three-dimensional hydrodynamical initial conditions.

We also update the calculation of the sound speed, $c_{\rm s}=\sqrt{\Gamma_1 p/\rho}$, which is required by the Riemann solver. Temperature inversion is implemented by numerically solving $e(X, Z, \rho, T)=e_{\rm target}$ using Newton--Raphson iterations.

}

\begin{figure}
    \centering
    \includegraphics[width=\linewidth]{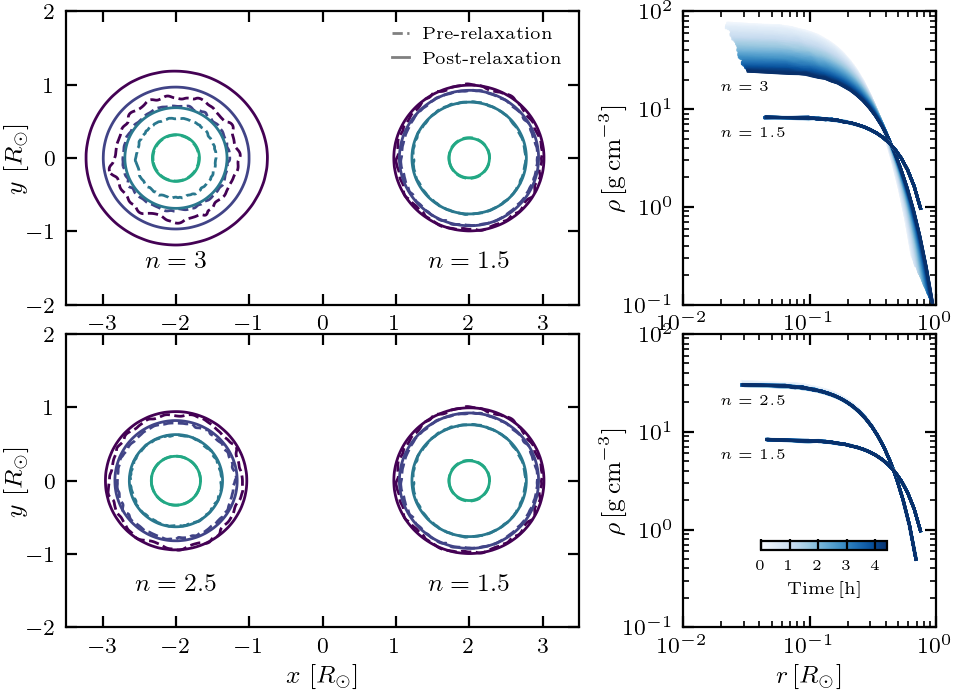}
    \caption{Density contours (\emph{left}) and density profiles (\emph{right}) of the binary stars before and after a $10\,t_{\rm dyn,\odot} \approx 4.4\,\rm h$ relaxation. \emph{Top}: A binary containing an $n=3$ star and an $n=1.5$ star (both of solar mass and radius), where the $n=3$ star fails to converge. \emph{Bottom}: Another binary of $n=2.5$ and $n=1.5$ stars which are both stable after relaxation.
    }
    \label{fig:relaxation}
\end{figure}

\subsection{Initial conditions}
\label{sec:method:ic}

To make a spherically symmetric 3D stellar model, we first construct 1D models that contain radial profiles like $\rho(r)$ and $e(r)$. These are natural for \texttt{MESA} stellar profiles; for polytropes, we assume a size-mass relation of $R/R_\odot \sim (M/M_\odot)^{0.6}$ \citep{ToutPolsEggleton_1996MNRAS.281..257T}. The 3D Lagrangian initial condition starts as a glassy, uniform particle distribution in a periodic box; then the desired radial coordinate of each particle is redistributed through $r(m)$ (where $m$ is the enclosed mass) to reproduce the target density profile $\rho(r)$. 

The binary configuration has a primary star (“star 1”) and a companion star (“star 2”), initially separated by $d_0 \equiv |\bm{x}_2 - \bm{x}_1|= 2(R_1 + R_2)$, where $\mathbf{x}_i$ and $R_i$ denote the center-of-mass position and radius of each star. The initial relative velocity $\bm{v}_{\rm rel,0}$ is in the $xy$ plane, which also defines the impact parameter $b \equiv |\hat{\bm{v}}_{\rm rel,0} \times (\bm{x}_2 - \bm{x}_1)|$.

We also relax the initial conditions for $10\,t_{\rm dyn,\odot} \approx 4.4\,\rm h$ (10 solar dynamical time) before assigning velocities \citep[like][but considering the mutual gravity of the stars]{OhlmannRopkePakmor_2017A&A...599A...5O}. These runs start with zero initial velocity for all gas cells, and each cell is decelerated by an additional damping force $\bm{a}_{\rm damp} = -\bm{v}/\tau_{\rm damp}$, where $\tau_{\rm damp} = 0.2\,t_{\rm dyn,\odot}$. Gas cells are labeled by their host star, and at each timestep, we subtract the center-of-mass velocity and acceleration of the corresponding star. This procedure relaxes the stars under both self-gravity and mutual gravity. For a solar-mass star, the post-relaxation residual velocity fluctuation is $\sim 5\,\rm km\,s^{-1}$.

An $n=3$ polytrope is marginally (un)stable, so instead of exactly $n=3$, we adopt slightly deviated, $n=2.5$ ($\gamma = 7/5$) polytropes to represent radiative stars. From our background tests (Fig.~\ref{fig:relaxation}), we find that $n=3$ stars are indeed unstable to a companion's gravity during the relaxation run, while the $n=2.5$ ones converge well.

\section{Mergers between polytropic stars}
\label{sec:res}

\subsection{Simulation setups}

\begin{deluxetable}{cccc}
\tablecaption{Parameter space of the polytropic merger simulations, including the mass ratio, impact parameter, and initial relative velocity. The direction of
initial motion is derived from the impact parameter $b$.
Here $v_{\rm circ,0} = \sqrt{G(M_1+M_2)/d_0}$, where $d_0=2(R_1+R_2)$ is the initial separation.
}
\label{tab:ics}
\tablehead{\colhead{$M_1:M_2$} & \colhead{$n_1, n_2$} & \colhead{$b\,[R_\odot]$} & \colhead{$v_{\rm rel,0}/v_{0}$ ($v_{\rm rel,0}\,[\rm km\,s^{-1}]$)}}
\startdata
1:1 & 2.5, 2.5 & 0, 1 & 0, $\sqrt{2}$ (437), 2 (618) \\
1:1 & 2.5, 1.5 & 0, 1 & 0, $\sqrt{2}$ (437), 2 (618) \\
10:1 & 2.5, 2.5 & 0, 1, 3 & 0, $\sqrt{2}$ (621), 2 (879) \\
10:1 & 2.5, 1.5 & 0, 1, 3 & 0, $\sqrt{2}$ (621), 2 (879) \\
\enddata
\end{deluxetable}

The parameter space explored in this study is summarized in Table~\ref{tab:ics}. We consider mergers with mass ratios of $1:1$ and $10:1$, polytropic indices $n=2.5$ (``radiative'') and $n=1.5$ (``convective''), impact parameters $b=0$--$3\,R_\odot$, and initial relative velocities ranging from bound ($v_{\rm rel,0}=0$) to hyperbolic encounters. The characteristic velocity is defined as $v_{\rm circ,0}=\sqrt{G(M_1+M_2)/d_0}$, where $d_0=2(R_1+R_2)$ is the initial separation. Consequently, $E_{\rm orbit}<0$, $=0$, and $>0$ correspond to $v_{\rm rel,0}/v_{\rm circ,0}=0$, $\sqrt{2}$, and $2$ or $4$, respectively. Nearly all encounters with $v_{\rm rel,0}/v_{\rm circ,0}=4$ ($\gtrsim1000\,\rm km\,s^{-1}$) disrupt rather than merge and are therefore excluded from the following analysis.

To visualize chemical mixing, we initialize each star with a uniform abundance tracer $Y$. The primary star is assigned $Y=0$, while the companion has $Y=1$, allowing the redistribution of stellar material to be directly tracked throughout the merger. Each simulation is labeled as \verb|n%d-n%d_b%g_v%g|; for example, \verb|n2.5-n1.5_b1_v1.4| denotes a merger between an $n=2.5$ primary and an $n=1.5$ companion with $b=R_\odot$ and $v_{\rm rel,0}/v_{\rm circ,0}=\sqrt{2}$.

The fiducial simulations have a gas mass resolution of $m_{\rm gas}=(1/64^3)\,M_\odot\approx4\times10^{-6}\,M_\odot$ (see Appendix~\ref{app:convergence_tests} for convergence tests) and are evolved for $100\,t_{\rm dyn,\odot}\approx44\,\rm h$, by which time the merger remnants have reached quasi-hydrostatic equilibrium.

\begin{figure*}
    \centering
    \gridline{
        \fig{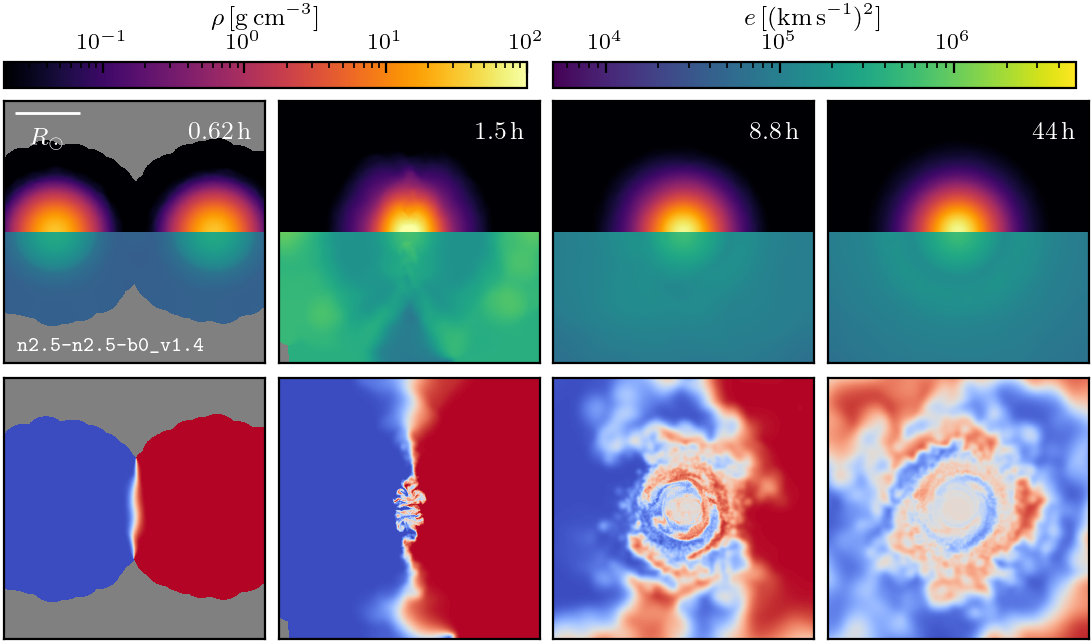}{0.49\textwidth}{(a) 1:1 mass-ratio, head-on ($b=0$) merger between $n=2.5$ stars. }
        \fig{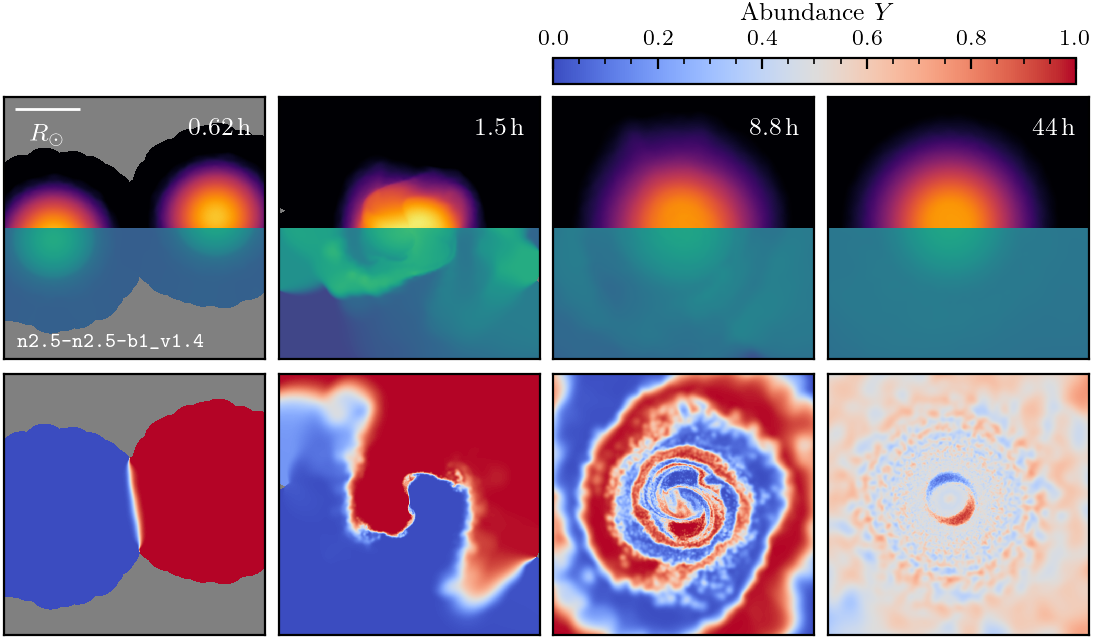}{0.49\textwidth}{(b) 1:1 mass-ratio, off-center ($b=R_\odot$) merger between $n=2.5$ stars.}
    }
    \vspace{-10pt}
    \gridline{
        \fig{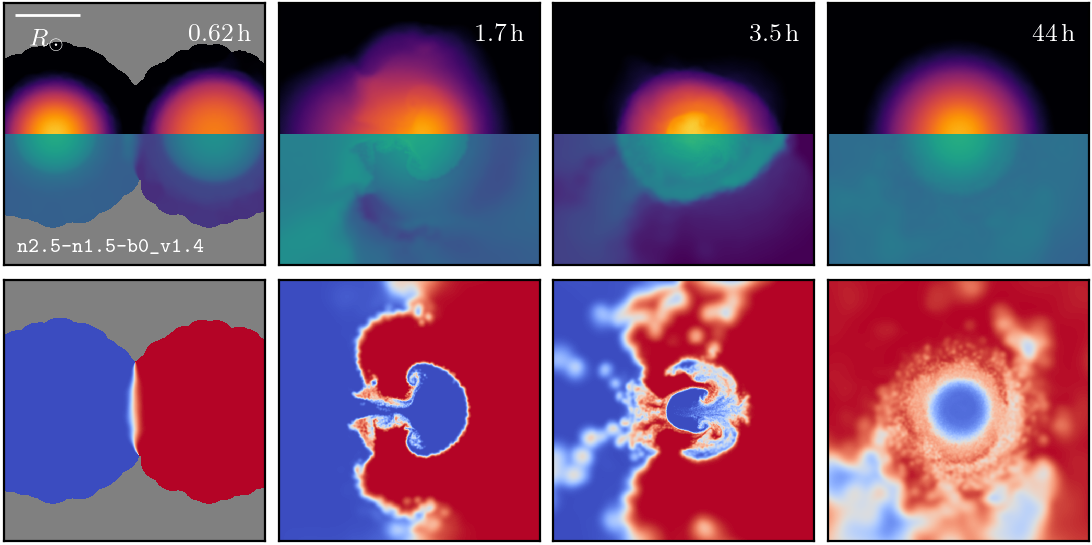}{0.49\textwidth}{(c) 1:1 mass-ratio, head-on ($b=0$) merger between $n=2.5$ and $n=1.5$ stars.}
        \fig{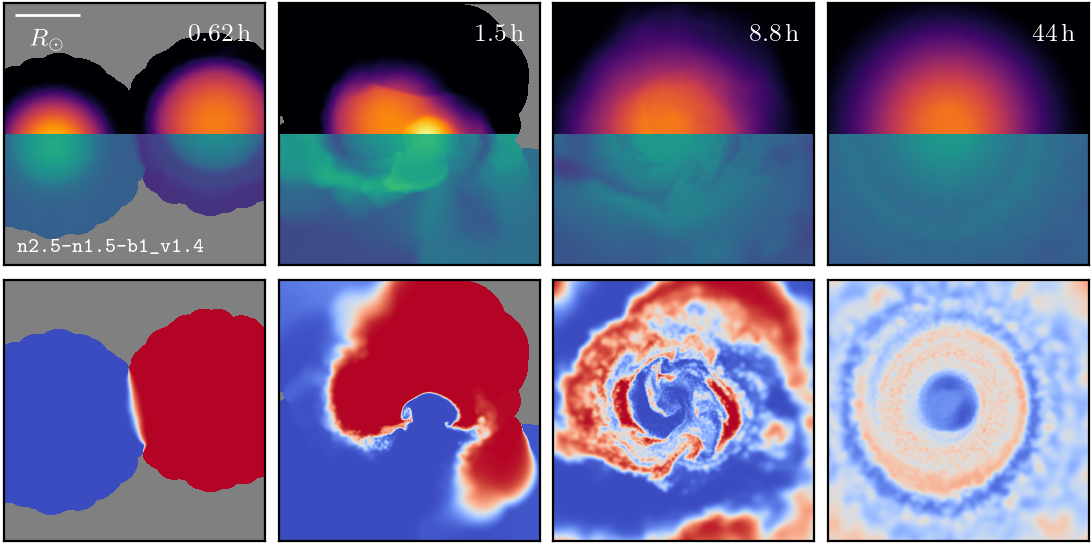}{0.49\textwidth}{(d) 1:1 mass-ratio, off-center ($b=R_\odot$) merger between $n=2.5$ and $n=1.5$ stars.}
    }
    \vspace{-10pt}
    \gridline{
        \fig{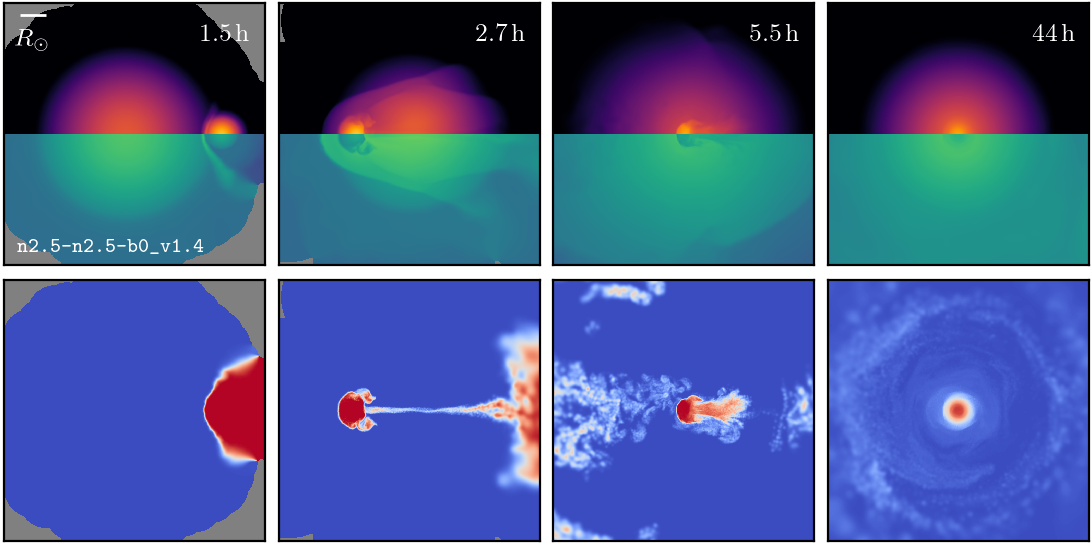}{0.49\textwidth}{(e) 10:1 mass-ratio, head-on ($b=0$) merger between $n=2.5$ stars.}
        \fig{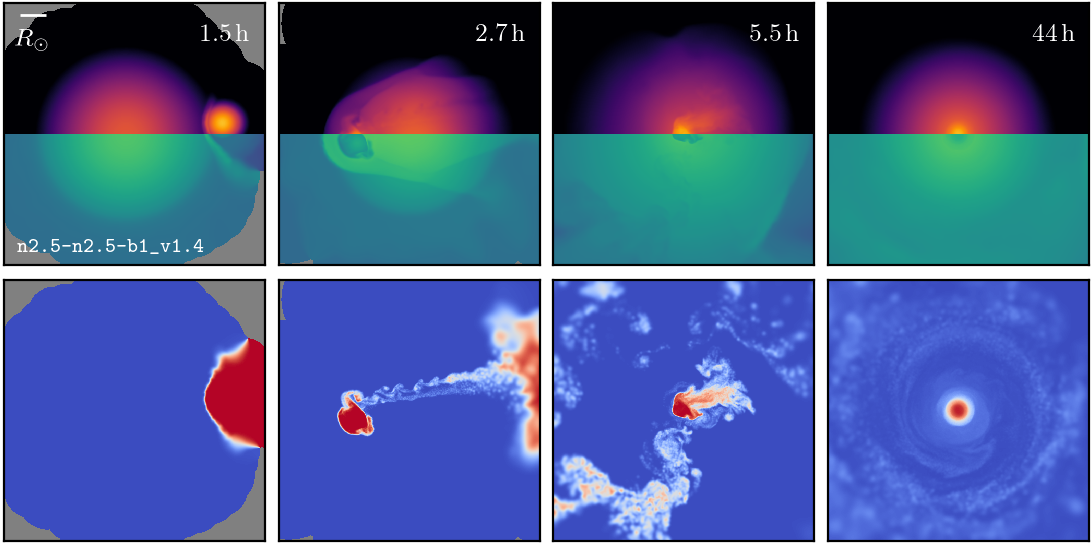}{0.49\textwidth}{(f) 10:1 mass-ratio, off-center ($b=R_\odot$) merger between $n=2.5$ stars.}
    }
    % \vspace{-10pt}
    % \gridline{
    %     \fig{figures/merger-n2.5_M10_R4_S0_Res138-n1.5_M1_R1_S0_Res64-b0_v1.4.png}{0.49\textwidth}{(b.3) 10:1 mass-ratio, head-on ($b=0$) merger between $n=2.5$ and $n=1.5$ stars.}
    %     \fig{figures/merger-n2.5_M10_R4_S0_Res138-n1.5_M1_R1_S0_Res64-b1_v1.4.png}{0.49\textwidth}{(b.4) 10:1 mass-ratio, off-center ($b=R_\odot$) merger between $n=2.5$ and $n=1.5$ stars.}
    % }
    \caption{Representative evolution of the polytropic stellar merger simulations. Each panel shows the mid-plane distributions of density $\rho$, specific internal energy $e$, and chemical abundance $Y$ at four stages of the merger. Panels (a)--(d) present $1:1$ mass-ratio mergers, while panels (e)--(f) show $10:1$ mergers. All simulations have $v_{\rm rel,0}=\sqrt{2G(M_1+M_2)/d_0}$ ($E_{\rm orbit}=0$); the stellar structures and impact parameters are varied as indicated in the sub-captions. The figure illustrates the characteristic shock evolution, turbulent mixing, and relaxation of the merger remnant across the explored parameter space.
}
    \label{fig:visualization}
\end{figure*}

% \begin{figure}
%     \centering
%     \includegraphics[width=\linewidth]{figures/mixing_evo-merger-n2.5_M1_R1_S0_Res64-n1.5_M1_R1_S0_Res64-b0_v0.pdf}
%     \caption{Caption}
%     \label{fig:placeholder}
% \end{figure}

\subsection{Evolution of the mergers}
\label{sec:res:mixing}

Fig.~\ref{fig:visualization} presents representative merger simulations spanning different mass ratios, stellar structures, and impact parameters. All runs shown have $v_{\rm rel,0}=\sqrt{2G(M_1+M_2)/d_0}$ ($E_{\rm orbit}=0$). Each simulation is displayed at four representative stages, illustrating the evolution of the mid-plane density $\rho$, specific internal energy $e$, and chemical abundance $Y$. While the detailed morphology varies with the merger parameters, all simulations follow the same overall sequence: first contact generates strong shocks and envelope expansion, followed by core interaction, turbulent mixing, and relaxation toward a quasi-hydrostatic remnant.

\paragraph{Equal-mass, identical stars}

Fig.~\ref{fig:visualization}a and \ref{fig:visualization}b show mergers between equal-mass, structurally identical ($n_1=n_2=2.5$) stars. In the head-on collision, the stellar cores merge shortly after first contact, producing a dense central region ($\gtrsim100\,\rm g\,cm^{-3}$). The release of gravitational energy drives strong turbulence and shock heating, leading to efficient chemical mixing in the core while the outer envelope largely retains the original composition of the progenitors. By $t\approx44\,\rm h$, the remnant has reached approximate hydrodynamical equilibrium with a nearly homogeneous core ($Y\approx0.5$).

For the off-center collision, the nonzero impact parameter introduces orbital angular momentum. During core coalescence, the remnant develops two trailing spiral arms that are gradually wound up by differential rotation. The resulting rotational shear enhances mixing throughout the central region, producing a largely homogeneous remnant except for a small ring-like region where the original cores co-rotate and remain only partially mixed.

% \begin{figure*}
%     \centering
%     \includegraphics[width=\linewidth]{figures/summary_plot.pdf}
%     \caption{The chemical element mixing patterns of the merger products for the parameter space explored in this study. The results are categorized into panels labeled by the mass ratio and stellar types, where we display the fates of the mergers of different $E_{\rm orbit}$ and $b$. The four values of $E_{\rm orbit}$, from left to right, correspond to that for $v_{\rm rel, 0} = (0, \sqrt{2}, 2, 4) v_{\rm circ,0}$ respectively.}
%     \label{fig:summary}
% \end{figure*}

\paragraph{Equal-mass, non-identical stars}

Fig.~\ref{fig:visualization}c and \ref{fig:visualization}d show mergers between equal-mass but structurally different stars ($n_1=2.5$ and $n_2=1.5$). Owing to its higher central density, the $n=2.5$ star survives the collision as a compact core that oscillates within the more diffuse $n=1.5$ star. The relative motion excites Rayleigh--Taylor instabilities and turbulent wakes, efficiently mixing the surrounding material while leaving the compact core largely intact. After the oscillations are damped, the remnant settles into a configuration consisting of an unmixed central core embedded within an extensively mixed inner envelope.

The off-center merger follows a similar evolution, except that differential rotation generates asymmetric spiral arms because of the different stellar structures. Although the spiral arms promote additional mixing in the envelope, the compact $n=2.5$ core remains largely unmixed. Consequently, the relaxed remnant exhibits a characteristic ``sandwiched'' abundance profile, with a low-abundance core surrounded by a high-abundance inner envelope and a low-abundance outer envelope.

\begin{figure*}
    \centering
    \includegraphics[width=\linewidth]{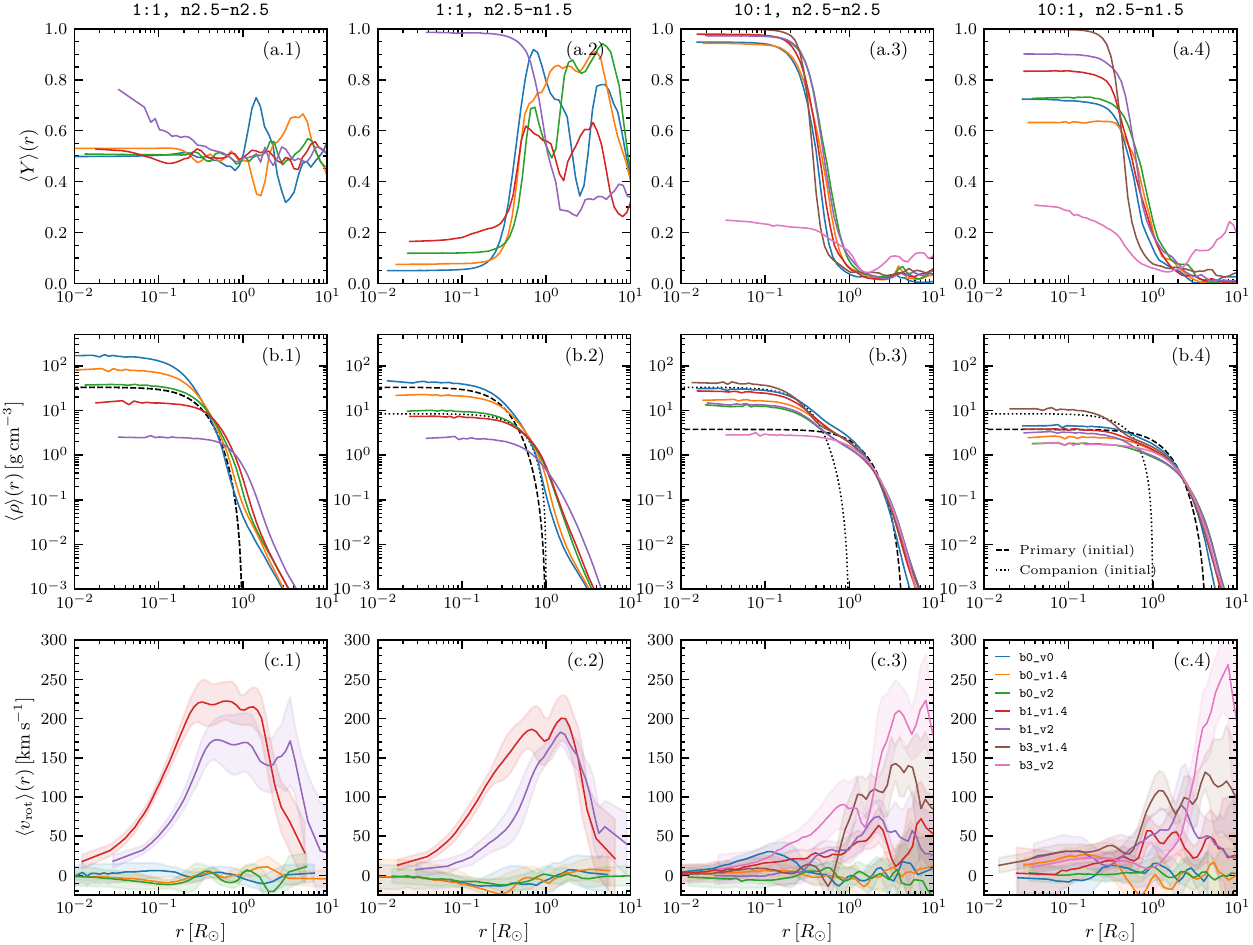}
    \caption{Radial profiles of chemical abundance $Y$ (\emph{top row}), density $\rho$ (\emph{middle row}), and rotational velocity $v_{\rm rot}$ (\emph{bottom row}) at the mid-plane ($z=0$) of the merger product. From left to right, columns are mergers of different mass ratios and stellar structures, including: (1) $1\,M_\odot:1\,M_\odot$, identical $n=2.5$ stars; (2) $1\,M_\odot:1\,M_\odot$, $n=2.5$ with $n=1.5$ stars; (3) $10\,M_\odot:1\,M_\odot$, identical $n=2.5$ stars; (4) $10\,M_\odot:1\,M_\odot$, $n=2.5$ with $n=1.5$ stars. 
    The colored lines in each panel (defined in the bottom-right 
    box) shows simulations with different impact parameter $b$ (in $R_\odot$) and $v_{\rm rel,0}$ (in $\sqrt{G(M_1+M_2)/d_0}$).
    }
    \label{fig:merger_product}
\end{figure*}

\begin{figure*}
    \centering
    \includegraphics[width=\linewidth]{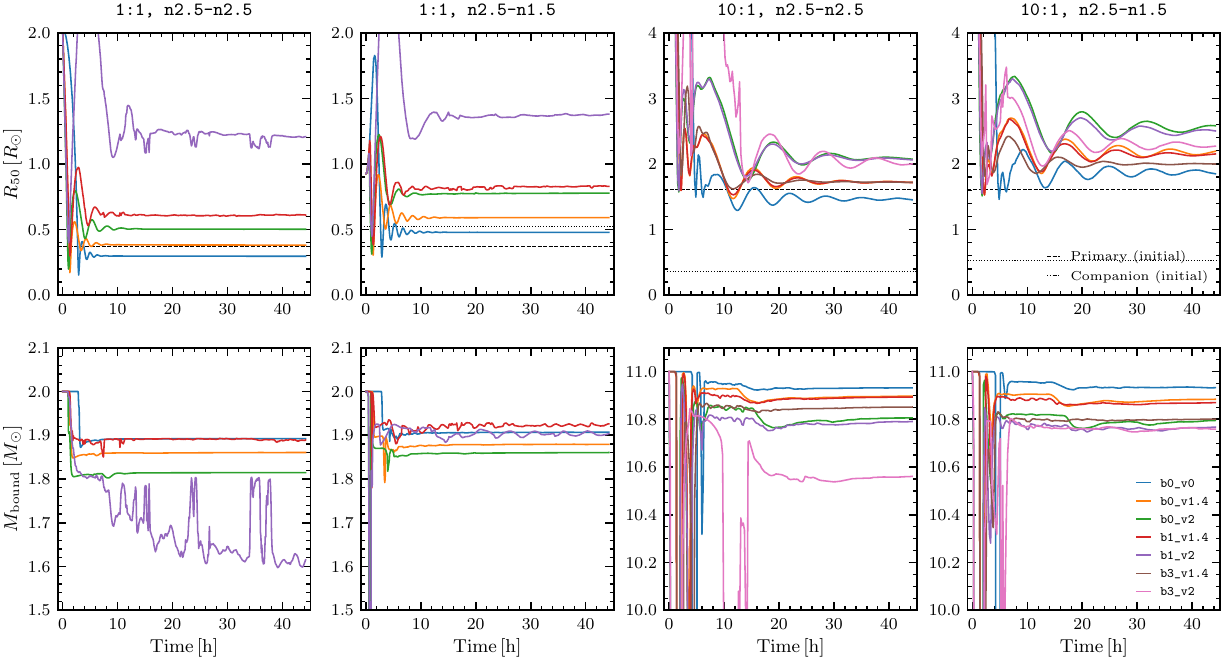}
    \caption{The time evolution of the half-mass radius ($R_{50}$, \emph{top}) and the gravitationally bound mass ($M_{\rm bound}$, \emph{bottom}). The layout of the panels is the same as Fig.~\ref{fig:merger_product}.
    }
    \label{fig:time_evo}
\end{figure*}

\paragraph{10:1 mass-ratio mergers}

Fig.~\ref{fig:visualization}e and \ref{fig:visualization}f show mergers with a mass ratio of $10:1$, where a $1\,M_\odot$ companion collides with a $10\,M_\odot$, $n=2.5$ primary. Because the companion is substantially denser than the primary, it survives the initial impact and plunges through the primary's envelope. Its motion drives strong bow shocks, strips material from the companion, and excites Rayleigh--Taylor instabilities along its trajectory. After several oscillations, the companion settles at the center of the remnant, producing a chemically enriched core while the extended envelope remains only weakly enriched.

The off-center merger evolves similarly, except that part of the orbital angular momentum is retained by the envelope, producing a mildly rotating remnant. Compared with the equal-mass mergers, however, the chemical redistribution is governed primarily by the penetration and survival of the companion core rather than by rotational mixing.

\subsection{Radial structures of the merger remnants}

By the end of the simulations, the merger remnants are approximately quasi-spherical, allowing their structures to be characterized by spherically averaged radial profiles. Fig.~\ref{fig:merger_product} summarizes the radial distributions of chemical abundance $Y$, density $\rho$, and rotational velocity $v_{\rm rot}$ measured from the final snapshots.

\subsubsection{Chemical mixing}

The first row of Fig.~\ref{fig:merger_product} shows the radial abundance profiles. For each remnant, we locate the density peak and compute the mean abundance within spherical shells. Despite the diversity of merger parameters, the remnants can be broadly classified into three categories according to their central chemical composition.

\paragraph{Well-mixed core.} The remnant develops a chemically homogeneous core with $Y\approx0.5$, indicating efficient turbulent mixing during core coalescence. This outcome is characteristic of mergers between identical stars (panel a.1), where neither progenitor core survives the collision.

\paragraph{Unenriched core.} The central abundance remains close to that of the primary ($Y_{\rm center}\approx0$), while the inner envelope is enriched by material from the companion ($Y_{\rm envelope}\approx1$). This configuration occurs in equal-mass mergers between structurally different stars (panel a.2), where the denser $n=2.5$ core survives the merger and remains largely unmixed. One exception is the high-velocity run \path{n2.5_n1.5_b1_v2}, which produces only a dilute remnant.

\paragraph{Enriched core.} The remnant develops a chemically enriched core ($Y_{\rm center}\approx1$) surrounded by an envelope with $Y_{\rm envelope}\approx0$. This outcome is typical of the $10:1$ mergers (panels a.3 and a.4), where the dense companion core sinks to the center of the primary. Again, the high-velocity mergers with $b=3\,R_\odot$ and $v_{\rm rel,0}=2\sqrt{G(M_1+M_2)/d_0}$ are exceptions because they fail to produce a compact remnant.

These three outcomes are primarily determined by the relative core densities of the progenitor stars. Since approximately $R\propto M^{0.6}$ \citep{ToutPolsEggleton_1996MNRAS.281..257T}, the mean stellar density scales as $\bar{\rho}\propto M^{-0.8}$, making the $1\,M_\odot$ star substantially denser than the $10\,M_\odot$ star. Consequently, in high mass-ratio mergers the companion core survives and replaces the original core of the primary. The distinction between $n=2.5$ and $n=1.5$ companions follows the same principle, as the latter possess less centrally concentrated cores and are therefore more susceptible to disruption and mixing.

{
This buoyancy argument is consistent with previous studies of stellar mergers and chemical mixing \citep[e.g.,][]{LombardiWarrenRasio_2002ApJ...568..939L,GaburovLombardiPortegiesZwart_2008MNRAS.383L...5G}, which characterize buoyancy using the specific entropy through the ``entropy-sorting'' prescription. We also examine the specific entropy, defined as $K=P/\rho^\gamma$, in our merger remnants. For all stable remnants, we find that $K$ generally increases with radius ($\dd K/\dd r>0$), consistent with a buoyantly stable configuration. 

Though entropy sorting predicts that the merger remnant should in general settle to an equilibrium with $\dd K/\dd r>0$, it does not predict chemical mixing when the stellar material is plunging or spiraling in, particularly turbulent mixing at the interfaces. Additionally, our simulations show prominent shocks (Fig.~\ref{fig:visualization}), which locally increase the specific entropy ($K$) rather than preserving it. Consequently, our simulations can be used to calibrate entropy-sorting algorithms to predict the redistribution of matter after a merger. 
}

\subsubsection{Density}

The second row of Fig.~\ref{fig:merger_product} compares the density profiles of the merger remnants with those of the isolated progenitors (dashed and dotted curves). In all cases, shock heating produces remnants that are more extended than the initial stars.

The central density depends on both the stellar structures and the collision parameters. For equal-mass mergers (panels b.1 and b.2), head-on collisions generally produce denser cores than off-center collisions because less orbital energy is retained as rotation. Increasing the initial relative velocity also reduces the central density by depositing more energy into the remnant.

The $10:1$ mergers exhibit a qualitatively different structure. For identical $n=2.5$ stars (panel b.3), the density profile contains two distinct components: a compact inner core resembling the original $1\,M_\odot$ companion embedded within an envelope dominated by the $10\,M_\odot$ primary. In contrast, when the companion has $n=1.5$ (panel b.4), its more diffuse core undergoes stronger disruption, producing a remnant whose density profile more closely resembles that of the original primary star.

\subsubsection{Rotation}

The third row of Fig.~\ref{fig:merger_product} shows the rotational velocity measured in the orbital mid-plane, where $\bm{v}_{\rm rot}\equiv\bm{v}_{\rm 2D}-(\bm{v}_{\rm 2D}\cdot\hat{\bm{r}}_{\rm 2D})\hat{\bm{r}}_{\rm 2D}$. As expected, head-on mergers ($b=0$) retain little net angular momentum and therefore exhibit negligible rotation. Off-center collisions, however, convert part of the orbital angular momentum into remnant rotation.

For the $1:1$ mergers (panels c.1 and c.2), the rotational velocity peaks near the original stellar surface ($r\sim0.5$--$1\,R_\odot$), reaching $\sim150$--$200\,\rm km\,s^{-1}$, well below the solar break-up velocity ($\sim440\,\rm km\,s^{-1}$). The extended envelope also retains substantial rotation.

The $10:1$ mergers (panels c.3 and c.4) exhibit a different rotational structure. The inner region remains nearly non-rotating, whereas most of the angular momentum is deposited in the extended envelope. The envelope rotation increases with both the impact parameter and the initial relative velocity, reaching $\sim100$--$200\,\rm km\,s^{-1}$ for the $b=3\,R_\odot$ runs.

This difference reflects how orbital angular momentum is redistributed during the merger. In equal-mass collisions, the two stellar cores interact directly, allowing angular momentum to be efficiently transferred into the central remnant. In contrast, for the $10:1$ mergers, the compact companion loses most of its orbital energy while traversing the primary's envelope before settling at the center, leaving the envelope as the primary reservoir of angular momentum.

% \begin{figure}
%     \gridline{
%         \fig{figures/merger-n3_M1_R1_S0_Res64-n1.5_M1_R1_S0_Res64-b0_v0.png}{\linewidth}{(a) 1:1 mass-ratio, head-on ($b=0$) merger between $n=3$ and $n=1.5$ stars. The initial relative velocity is 0 ($E_{\rm orbit}<0$).}
%     }
%     \vspace{-10pt}
%     \gridline{
%         \fig{figures/merger-n3_M1_R1_S0_Res64-n1.5_M1_R1_S0_Res64-b0_v1.4.png}{\linewidth}{(b) 1:1 mass-ratio, head-on ($b=0$) merger between $n=3$ and $n=1.5$ stars. The initial relative velocity is $\sqrt{2G(M_1+M_2)/d_0}$ ($E_{\rm orbit}=0$). }
%     }
%     \vspace{-10pt}
%     \gridline{
%         \fig{figures/merger-n3_M1_R1_S0_Res64-n3_M1_R1_S0_Res64-b0_v2.png}{\linewidth}{(c) 1:1 mass-ratio, head-on ($b=0$) merger between $n=3$ stars. The initial relative velocity is $2\sqrt{G(M_1+M_2)/d_0}$ ($E_{\rm orbit}>0$).}
%     }
%     \caption{Representative merger products with $n=3$ simulations. (a)~Driven by the approaching companion, the $n=3$ star becomes puffy enough so the companion sinks to form a chemical-abundant core. (b)~The $n=3$ expands but is still denser than the $n=1.5$ companion, leading to a well-mixed, slightly enriched core. (c)~The orbit energy is slightly positive, but the $n=3$ merger product is unstable.
%     }
%     \label{fig:n=3 runs}
% \end{figure}

\subsection{Oscillation, expansion, and mass loss}

Shock heating during the merger inflates the stellar envelope (Figs.~\ref{fig:visualization} and \ref{fig:merger_product}), producing an extended remnant that subsequently oscillates and relaxes toward hydrostatic equilibrium. We characterize the remnant size by the half-mass radius $R_{50}$. We also measure the gravitationally bound mass, $M_{\rm bound}$, defined as the total mass satisfying $v^2/2+\gamma e+\Phi<0$, where $\gamma e$ is the enthalpy and $\Phi$ is the gravitational potential.

The top panels of Fig.~\ref{fig:time_evo} show the evolution of $R_{50}$. All merger remnants undergo damped radial oscillations before settling into a quasi-steady state. For the $1:1$ mergers, the oscillations decay within $\sim5\,\rm h$, whereas the more extended $10:1$ remnants relax over longer timescales. The final remnant is generally more compact for lower relative velocities and head-on collisions. Typical equilibrium radii are $R_{50}\lesssim0.5\,R_\odot$ for $n=2.5$--$n=2.5$ mergers and $R_{50}\gtrsim0.5\,R_\odot$ for $n=2.5$--$n=1.5$ mergers, while the $10:1$ mergers produce substantially larger remnants with $R_{50}\sim1.5$--$2.5\,R_\odot$. The only notable exception is the high-velocity off-center run \texttt{b1\_v2}, which forms a diffuse remnant with $R_{50}>R_\odot$.

The oscillation period is approximately $\sim1.5\,\rm h$ for the $1:1$ mergers and $\sim8\,\rm h$ for the $10:1$ mergers, consistent with the expected fundamental radial mode ($\sim2\pi t_{\rm dyn}$) of the remnant. The oscillations are gradually damped, most likely through a turbulent cascade. Convergence tests presented in Appendix~\ref{app:convergence_tests} indicate that the damping is not dominated by SGS or numerical viscosity.

The bottom panels of Fig.~\ref{fig:time_evo} show the evolution of $M_{\rm bound}$. In all simulations, most of the mass remains gravitationally bound, with $M_{\rm bound}\approx1.8$--$1.9\,M_\odot$ for the $1:1$ mergers and $10.8$--$10.9\,M_\odot$ for the $10:1$ mergers. Mass loss increases with both the impact parameter and the initial relative velocity, reflecting the larger fraction of orbital energy converted into kinetic energy of the envelope. The dependence on stellar structure is comparatively weak, although the $1:1$ mergers with $n=2.5$ companions generally lose slightly more mass.

{
Overall, the prompt mass loss is modest, amounting to $\lesssim10\%$ for the $1:1$ mergers and $\lesssim2\%$ for the $10:1$ mergers, consistent with recent hydrodynamical studies \citep[e.g.,][]{RoseLombardiGonzalezPrieto_2026ApJ..1000..162R,GonzalezPrietoLombardiRose_2026ApJ..1005..131G} with $v_{\infty}\sim 0\,\rm km\,s^{-1}$. Our simulations, however, follow only the first $\sim 100$ dynamical times after the merger. Additional mass loss driven by the long-term thermal evolution of the remnant is therefore not captured. For example, one-dimensional stellar evolution models by \citet{Roman-GarzaFragosCharbonnel_2026A&A...707A.163R} predict that mergers between supermassive and massive stars may lose $\sim10$--30\% of their mass through pulsations over $\sim100\,\rm yr$.
}

% Part of the reason behind the \citet{Roman-GarzaFragosCharbonnel_2026A&A...707A.163R} result is that supermassive stars have $\gamma\approx 4/3$, which are dynamically unstable. So the $n=3$ polytropic stars, though not converging in relaxation runs (Fig.~\ref{fig:relaxation}), are still worth exploring, as developed in the next section.

{
\subsection{Additional numerical tests}

The results presented above adopt our fiducial numerical setup, including the SGS turbulent diffusivity and viscosity described in \S~\ref{sec:method}. To assess the robustness of our conclusions, we perform a suite of additional simulations that vary the SGS models and numerical resolution. The details are presented in Appendix~\ref{app:convergence_tests}. These tests demonstrate that the SGS turbulent mixing model is essential (at least for the MFM solver in \texttt{GIZMO}) and that our fiducial simulations are numerically converged. Furthermore, quantitative measurements of the mixing evolution confirm the picture inferred from the visualization plots: chemical mixing is initially dominated by large-scale shocks and bulk flows, followed by small-scale turbulent mixing during the subsequent relaxation.
}

{

\section{Mergers between AGN stars}
\label{sec:res_mesa}

\begin{figure}
    \centering
    \includegraphics[width=\linewidth]{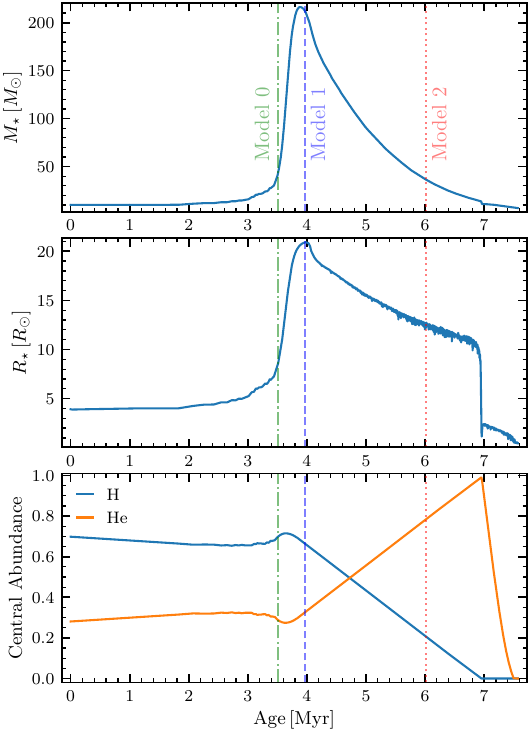}
    \caption{Evolutionary history of a ``metamorphic'' AGN star. From top to bottom, the panels show the stellar mass, radius, and central H/He abundances. We select two models (1, 2) from the main-sequence (H-burning) phase (labeled in the panels) and one model (0) from the accreting phase, which are later used as the initial conditions for the stellar merger simulations.}
    \label{fig:mesa_agn_star_evolution}
\end{figure}

In this section, we focus on stellar evolution in AGN disks. Unlike the more general, idealized simulations explored in \S~\ref{sec:res}, we construct three-dimensional stellar models based on the one-dimensional stellar 
evolutionary (\texttt{MESA}) models presented by \citet{XuChenLin_2026ApJ...997..206X}. An example is shown in Fig.~\ref{fig:mesa_agn_star_evolution}. The star is born as a $10\,M_\odot$ protostar that is embedded in the mid-plane of an AGN disk with an ambient density of $\rho_{\rm c}=10^{-16}\,\rm g\,cm^{-3}$ and temperature $T_{\rm c} \simeq 10^{5}\,\rm K$. The accretion rate depends on the stellar Eddington factor, $\lambda_\star \equiv L_\star/L_{\rm Edd}(M_\star)$, and is regulated by wind feedback following the prescription of \citet{ChenJiangGoodman_2024ApJ...974..106C}. In the \texttt{MESA} calculations, the accretion rate also depends on the feedback-transition parameter $\lambda_0$ defined by \citet{XuChenLin_2026ApJ...997..206X} (their Eq.~19). The evolutionary track shown adopts $\lambda_0=0.75$. The stellar mass grows rapidly after $\sim 3\,\rm Myr$, reaches a maximum at $\sim 3.8\,\rm Myr$, and subsequently declines.

The star accretes gas with a He abundance of $Y_{\rm disk}=0.2518$, together with $X_{\rm C,disk}=2.2\times 10^{-3}$, $X_{\rm N,disk}=7\times 10^{-4}$, and $X_{\rm O,disk}=6.3\times 10^{-3}$. During the accretion-dominated phase, the H and He abundances remain nearly pristine throughout the star, including in the core, because vigorous nuclear burning has not yet been established. After the stellar mass reaches its peak, the star enters the main-sequence phase, during which the central H abundance gradually decreases from $\sim 0.7$ to $\sim 0$, while the He abundance correspondingly increases. This evolution prescribes uniform diffusivity of typical values\footnote{$D_{\rm mix,rad}=10^5$--$10^7\,\rm cm^2\,s^{-1}$, but rotation is not explicitly modeled in their \texttt{MESA} simulations.} inferred from rotational mixing \citep[][]{Spruit_2002A&A...381..923S}, in which case mixing through the radiative layer between the stellar surface and the core is insufficient to replenish hydrogen at the rate it is consumed by the CNO cycle \citep{Xu_2025RAA....25k5013X}. Unlike the preceding accretion-dominated phase, the main-sequence evolution is feedback-dominated, as the large amount of energy released by the CNO cycle suppresses further accretion from the ambient gas.

Alternatively, \citet{XuChenLin_2026ApJ...997..206X} also found that, \textcolor{\revcolor}{if a
prescription for extra mixing is adopted}, hydrogen fuel can be continuously replenished into the stellar core, sustaining CNO burning indefinitely. Such stars may remain in a long-lived ``immortal'' state, retaining a main-sequence-like internal structure throughout their evolution \citep[][]{CantielloJermynLin_2021ApJ...910...94C, Ali-DibLin_2023MNRAS.526.5824A}.

The ``metamorphic'' main-sequence phase ends once the central hydrogen is depleted ($X_{\rm center}\sim 0$, $Y_{\rm center}\sim 1$), which occurs at $\sim 7\,\rm Myr$ in Fig.~\ref{fig:mesa_agn_star_evolution}. The star then enters the post-main-sequence phase, during which the onset of the triple-$\alpha$ process rapidly consumes the central helium, ultimately leading to core collapse and the formation of a BH remnant through a supernova explosion \citep{FryerHuangAli-Dib_2025MNRAS.537.1556F}.

Without hydrogen replenishment to the core, an old metamorphic star will eventually leave the main sequence. One possible rejuvenation channel, however, is a merger with a younger, hydrogen-rich metamorphic star. In the following, we describe the setup of our stellar merger simulations designed to explore this scenario.

\begin{deluxetable}{lcccc}
\tablecaption{Merger simulations of two AGN stars. The stellar models are selected from the evolutionary track shown in Fig.~\ref{fig:mesa_agn_star_evolution}. We consider both ``collisional'' simulations, in which the stars collide directly, and ``inspiral'' simulations, in which the stars merge through binary orbital evolution. Here $v_{\rm circ,0}=\sqrt{G(M_1+M_2)/d_0}$, where $d_0$ is the initial separation. We adopt $d_0=2(R_1+R_2)$ for the collisional simulations and $d_0=R_1+R_2$ for the inspiral simulations.
}
\label{tab:ics_mesa}
\tablehead{\colhead{Name} & Models & \colhead{$d_0\,[R_\odot]$} & \colhead{$b\,[R_\odot]$} & \colhead{$v_{\rm rel,0}/v_{\rm circ,0}$}}
\startdata
\path{Collision12-b0_v0} &1, 2 & 67.0 &  0 & 0  \\
\path{Collision12-b0_v1.4} & 1, 2 & 67.0 & 0 & $\sqrt{2}$  \\
\path{Collision12-b0_v2} & 1, 2 & 67.0 & 0 & 2  \\
\path{Collision12-b5_v1.4} & 1, 2 & 67.0 & 5 & $\sqrt{2}$  \\
\path{Collision12-b5_v2} & 1, 2 & 67.0 & 5 & 2  \\
\path{Collision12-b10_v1.4} & 1, 2 & 67.0 & 10 & $\sqrt{2}$  \\
\path{Collision12-b10_v2} & 1, 2 & 67.0 & 10 & 2  \\
\path{Inspiral12} & 1, 2 & 33.5 & 33.5 & 0.95  \\
\path{Inspiral02} & 0, 2 & 21.3 & 21.3 & 0.95  \\
\enddata
\end{deluxetable}

\begin{figure}
    \centering
    \gridline{
        \fig{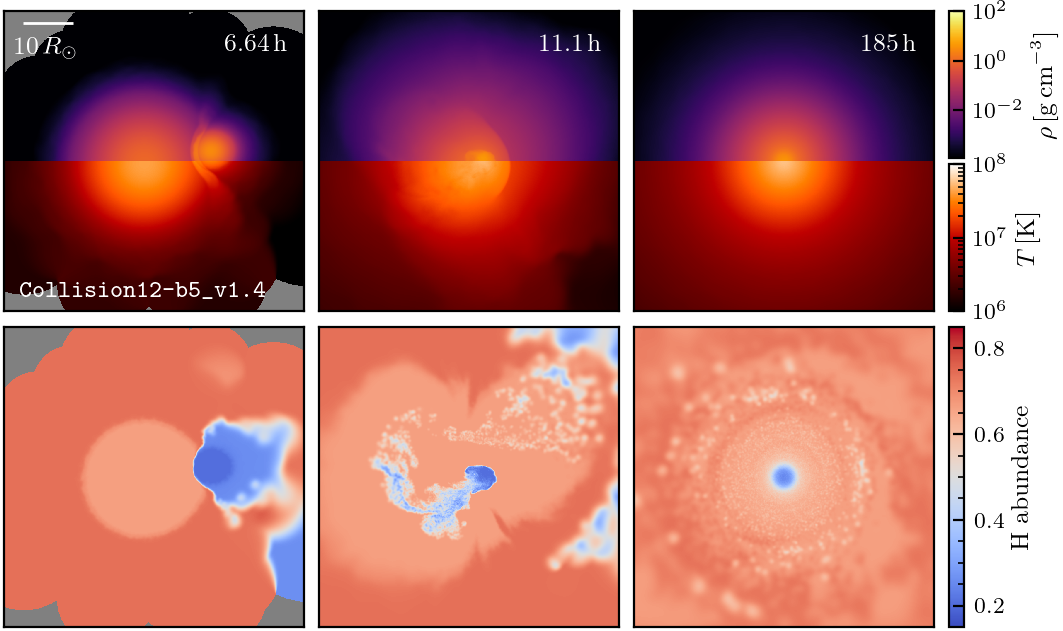}{\linewidth}{(a) Collision between models 1 and 2 (\path{Collision12-b5_v1.4}).}
    }
    \gridline{
        \fig{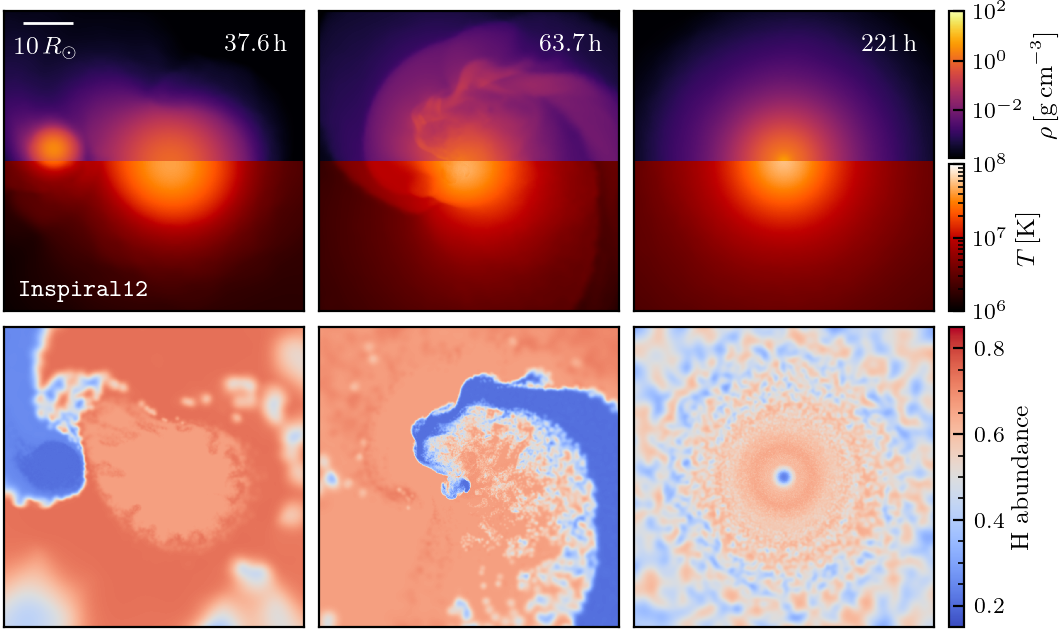}{\linewidth}{(b) Inspiral merger between models 1 and 2 (\path{Inspiral12}).}
    }
    \gridline{
        \fig{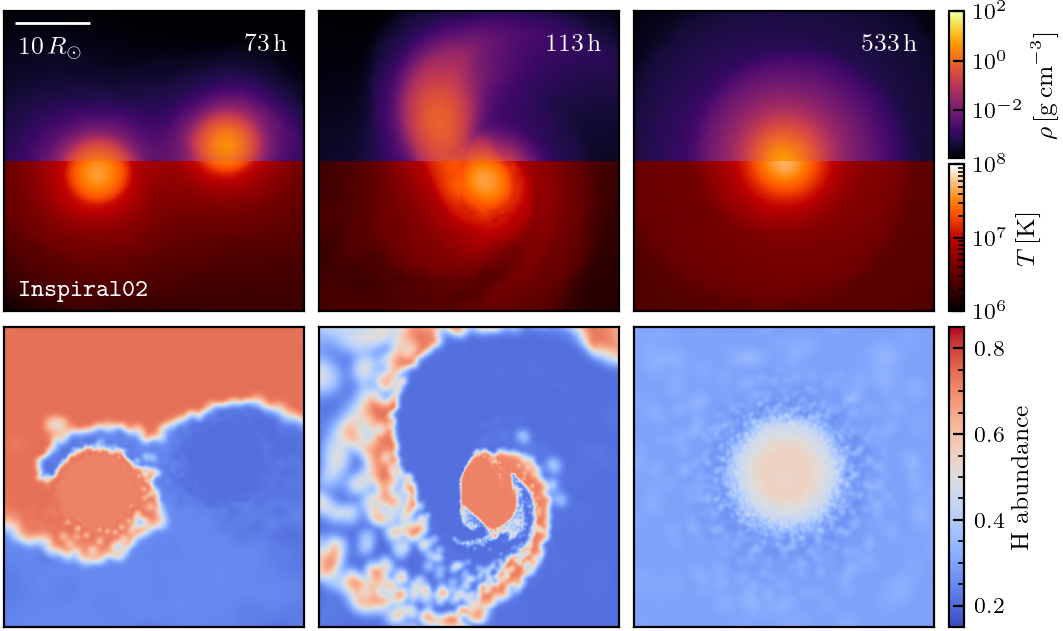}{\linewidth}{(c) Inspiral merger between models 0 and 2 (\path{Inspiral02}).}
    }
    \caption{Evolution of the AGN star mergers. Shown are the distributions of the mid-plane density $\rho$, temperature $T$, and hydrogen abundance at different stages of the simulations. From the full suite of simulations listed in Table~\ref{tab:ics_mesa}, we highlight three representative runs as labeled.}
    \label{fig:visualization_mesa}
\end{figure}

\subsection{Simulation setups}

We select two stellar models from the evolutionary track shown in Fig.~\ref{fig:mesa_agn_star_evolution}, as indicated in the figure. Both models are in the metamorphic main-sequence phase. ``Model~1'' is selected at $t\simeq 4\,\rm Myr$, shortly after the star enters the main sequence. At this stage, the star has $M_\star=213\,M_\odot$, $R_\star=20.9\,R_\odot$, and $X_{\rm center}\approx 0.67$. ``Model~2'' is selected at $t\simeq 6\,\rm Myr$, when the star is approaching the end of the main sequence, with $M_\star=36.3\,M_\odot$, $R_\star=12.6\,R_\odot$, and $X_{\rm center}\approx 0.2$. In addition, we consider an accreting star (``model~0'') at $t\simeq 3.5\,\rm Myr$, with $M_\star=44.9\,M_\odot$, $R_\star=8.73\,R_\odot$, and $X_{\rm center}\approx 0.7$. \textcolor{\revcolor}{We also note that ``model~0'' here only includes stellar structure, without any infalling disk material. }

As in the polytropic-star merger simulations presented in \S~\ref{sec:res}, we perform a suite of collisional simulations with different combinations of impact parameter $b$ and initial relative velocity $v_{\rm rel,0}$. Unless otherwise specified, we adopt the same initial configuration as in \S~\ref{sec:res}, with an initial separation of $d_0=2(R_1+R_2)$, such that the two stars undergo a direct geometric collision.

In addition, we perform an ``inspiral'' simulation to represent a different merger scenario, in which the two stars first form a binary and subsequently merge from a nearly circular orbit. In this configuration, the two stars are initially in contact, so we set $d_0=R_1+R_2$. The initial relative velocity is perpendicular to the line joining the stellar centers, giving $b=d_0$. We adopt an initial velocity of $v_{\rm rel,0}=0.95\sqrt{G(M_1+M_2)/d_0}$, which is slightly below the Keplerian value so that the binary gradually inspirals and merges.

We evolve each simulation for $\gtrsim 500\,t_{\rm dyn,\odot}$ ($220\,\rm h$), corresponding to approximately $100$ dynamical times of the more massive star. The mass resolution is $36.3/64^3\,M_\odot \approx 1.4\times10^{-4}\,M_\odot$. We employ the tabulated \texttt{MESA} EOS described in \S~\ref{sec:method:mesa_eos}. The abundances of all 21 chemical species in the \texttt{approx21} nuclear network are passively evolved according to Eq.~\eqref{equ:element_transport}.

\begin{figure*}
    \centering
    \includegraphics[width=\linewidth]{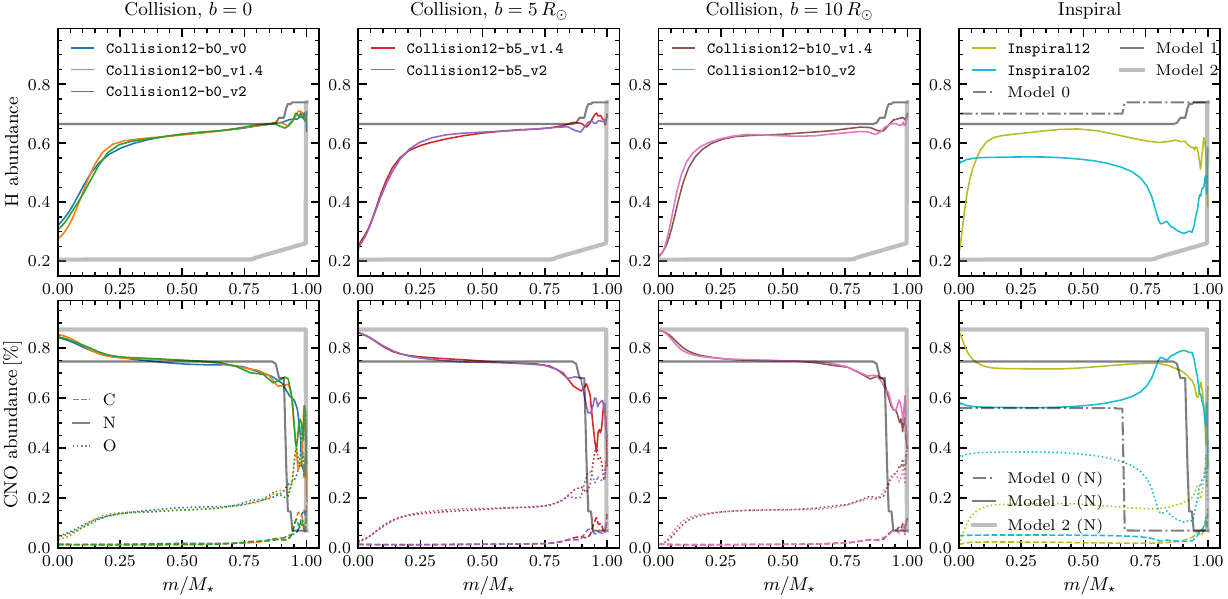}
    \caption{Radial chemical abundance profiles of the AGN-star merger remnants, including H (\emph{top panels}) and CNO elements (\emph{bottom panels}). The left and right columns show the collisional and inspiral simulations, respectively. For the H abundance, we also overplot the profiles of the two progenitor stars for comparison.}
    \label{fig:chemical_abundance_mesa}
\end{figure*}

\begin{figure*}
    \centering
    \includegraphics[width=\linewidth]{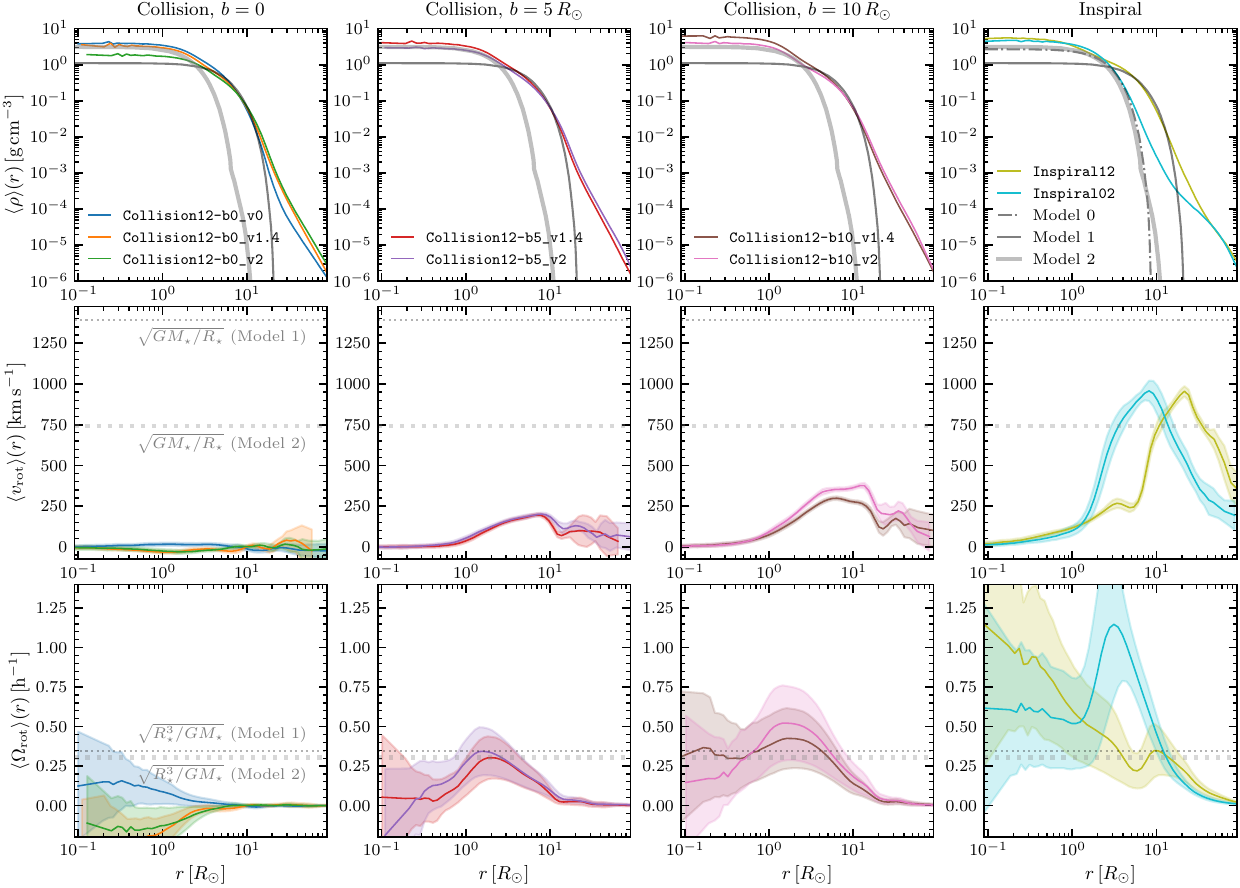}
    \caption{Density and rotational profiles of the AGN-star merger remnants. \emph{Top panels}: radial density profiles of the merger remnants and the two progenitor stars. \emph{Middle panels}: rotational velocity in the mid-plane. \emph{Bottom panels}: angular rotation rate $\Omega_{\rm rot}$. We also overplot the surface circular velocities and surface break-up rotation rates of the progenitor stars for comparison.}
    \label{fig:density_rotation_mesa}
\end{figure*}

\subsection{Chemical mixing}

\subsubsection{Merger with a young metamorphic star}

Fig.~\ref{fig:visualization_mesa} illustrates the evolution of the AGN-star merger simulations. Unlike the visualization of the polytropic-star mergers (Fig.~\ref{fig:visualization}), here we show the distributions of temperature and H abundance in the vicinity of the $z=0$ plane.

The overall evolution of the collisional simulation (\path{Collision12-b5_v1.4}; see Fig.~\ref{fig:visualization_mesa}) is broadly similar to that of the 10:1 mass-ratio polytropic-star mergers (sub-figures~e and f of Fig.~\ref{fig:visualization}). Although the secondary star is less massive, it has a higher central density, causing its core to sink toward the center of the primary star. Once the merger remnant reaches quasi-hydrostatic equilibrium, the central $r\lesssim R_\odot$ of the core is dominated by material from the secondary star, whereas the envelope is dominated by material from the primary. Although chemical mixing occurs in both regions, it does not alter this overall structure. Consequently, when an H-poor metamorphic star merges with an H-rich (and more massive) metamorphic star, the remnant retains an H-poor center while its envelope becomes H-rich. 

However, the merged star will continue to evolve toward a thermal equilibrium; since massive stars are highly convective, the small H-poor central core may well mix with the H-rich envelope, still leveraging the central H abundance.  

The inspiral simulation (\path{Inspiral12}; see Fig.~\ref{fig:visualization_mesa}) exhibits a markedly different evolutionary history. The binary first completes several orbits without merging, while both stars are tidally perturbed by their companion. These tidal interactions deform and expand the stellar envelopes. Around $t\sim 55.3\,\rm h$, the binary orbit begins to shrink rapidly, and by $t\sim 63.7\,\rm h$, the core of the secondary star has sunk to the center of the merger remnant, which subsequently relaxes toward equilibrium. Compared with the collisional case, the inspiral simulation converts a much larger fraction of the orbital angular momentum into the rotation of the merger remnant. As the secondary spirals inward, it leaves behind a trailing stream of H-poor material, producing additional mixing within the envelope. Although the overall H distribution remains qualitatively similar to that of the collisional case: the core remains H-poor, while the envelope has a lower H abundance than in the collisional simulation (\path{Collision12-b5_v1.4}).

Fig.~\ref{fig:chemical_abundance_mesa} further quantifies the radial distributions of H, C, N, and O abundances. In all simulations, the central H abundance is closer to that of model~2 (the less massive but more compact star), with $X_{\rm center}\gtrsim X_{\rm center,2}\approx 0.2$. Among the collisional runs, the central H abundance is highest for head-on collisions ($X_{\rm center}\approx 0.3$) and decreases for larger impact parameters ($X_{\rm center}\approx 0.2$), suggesting more efficient chemical mixing when the secondary core sinks directly toward the center of the primary. For a fixed impact parameter, the impact velocity plays a secondary role, as the resulting H-abundance profiles are very similar. The inspiral simulation likewise produces a core with $X_{\rm center}\approx 0.2$, although the envelope H abundance is lower than in the collisional runs because of the enhanced mixing that occurs while the secondary spirals inward.

These results suggest that, for an old metamorphic star approaching the end of the main sequence, merging with a younger metamorphic star does not significantly replenish the hydrogen abundance {at the very center of the star}. This is because the older metamorphic star is more compact, causing its core composition to dominate that of the merger remnant. This conclusion is robust across the different merger scenarios considered here, including both collisional and inspiral mergers.  {Nevertheless, a substantial amount of hydrogen can be replenished to the envelope, which may be engulfed by the convection zone 
during the subsequent thermal and nuclear evolution.}

We additionally examine the CNO abundances. The merger transports nitrogen synthesized in the stellar cores to the surface of the merger remnant, producing surface nitrogen abundances that exceed those of either progenitor. The exact degree of N enrichment depends on the merger parameters, particularly the impact parameter $b$. Taking the surface N/O ratio as an example (Fig.~\ref{fig:chemical_abundance_mesa}), we find ${\rm N/O}\sim 1$ for $b=0$, ${\rm N/O}\sim 2$ for $b=5\,R_\odot$, and ${\rm N/O}\sim 1.5$ for $b=10\,R_\odot$. The inspiral simulation of the same binary yields an even higher surface ratio, ${\rm N/O}\sim 3$. All merger remnants have surface N/O ratios well above that of the AGN disk ($X_{\rm N,disk}/X_{\rm O,disk}\approx 0.1$). Likewise, the merger remnants have surface ${\rm N/C}\gtrsim 5$, substantially exceeding the disk value of $X_{\rm N,disk}/X_{\rm C,disk}\approx 0.3$. 

These results are also consistent with the merger interpretation of SN~1987A, whose circumstellar nebula is likewise nitrogen enriched, with ${\rm N/O}\sim 1.1$--1.7 and ${\rm N/C}\sim 5$--6.1 \citep[depending on the measurement;][]{LundqvistFransson_1996ApJ...464..924L,SonnebornFranssonLundqvist_1997ApJ...477..848S,MattilaLundqvistGroningsson_2010ApJ...717.1140M}. Moreover, because the merger remnant becomes enriched in nitrogen and other CNO elements at its surface, its stellar wind may be further enhanced by the increased metal content. Such merger remnants may therefore contribute to the nitrogen enrichment of AGN disks \citep[][]{HuangLinShields_2023MNRAS.525.5702H,IsobeMaiolinoD'Eugenio_2025MNRAS.541L..71I}.

\subsubsection{Merger with an accreting star}

We visualize the inspiral merger between model~0 and model~2 (\path{Inspiral02}) in Fig.~\ref{fig:visualization_mesa}c. Unlike the mergers between models~1 and 2 (e.g., \path{Collision12-b5_v1.4} and \path{Inspiral12}), this merger produces an H-rich core ($X_{\rm center}\approx 0.55$) surrounded by an H-poor envelope. This difference arises because models~0 and 2 have comparable masses and similar core radii (see the density distributions in Fig.~\ref{fig:visualization_mesa}). Consequently, the merger resembles the 1:1 mass-ratio, $n=2.5$--$n=2.5$ polytropic mergers studied in \S~\ref{sec:res} (also see Fig.~\ref{fig:visualization}), in which the stellar cores mix efficiently to form a well-mixed central region. This contrasts with the higher mass-ratio mergers between models~1 and 2, which preserve the H-poor core of the older metamorphic star in the merger remnant.

The efficient mixing is also reflected in the surface nitrogen abundance (rightmost column of Fig.~\ref{fig:chemical_abundance_mesa}), with ${\rm N/O}\sim 2$ and ${\rm N/C}\sim 6$. Both ratios are substantially higher than the corresponding abundances in the AGN disk.

\subsection{Density and rotation}

Fig.~\ref{fig:density_rotation_mesa} shows the density and rotational profiles of the merger remnants. The density profiles retain clear imprints of both progenitor stars. The central density is $\sim 3\,\rm g\,cm^{-3}$, comparable to the core density of the secondary star (model~2), whereas the density profile of the envelope closely resembles that of the more massive primary star (model~1). Beyond $r\gtrsim 20\,R_\odot$, corresponding to $\rho\lesssim 10^{-3}\,\rm g\,cm^{-3}$, the remnant develops an extended outer envelope rather than exhibiting a sharp density cutoff. This extended structure is produced by shock heating during the merger, which inflates the stellar surface layers. Overall, the density profiles are remarkably similar across the different merger configurations.

The middle panels show the rotational velocity in the mid-plane. Unlike the density profiles, the rotational properties of the merger remnants depend strongly on the impact parameter, reflecting the conversion of orbital angular momentum into stellar rotation. As expected, the head-on collisions ($b=0$) produce essentially non-rotating remnants. For off-center collisions, the rotational velocity increases with increasing impact parameter. Nevertheless, for all collisional runs explored here, the rotational velocity remains below the break-up velocity ($\sqrt{GM_\star/R_\star}$) of both progenitor stars.

The inspiral case is substantially different because the binary initially carries much more orbital angular momentum. Consequently, the merger remnant is rapidly rotating, with a peak rotational velocity that exceeds the break-up velocity of the secondary star (model~2), while remaining below that of the primary star (model~1). In addition, the rotational profile exhibits a local maximum near $r\sim 3\,R_\odot$, approximately corresponding to the transition between the core (dominated by material from the secondary star) and the envelope (dominated by material from the primary star). The inspiral merger between models~0 and 2 is also rapidly rotating.

The bottom panels show the angular rotation rate, $\Omega_{\rm rot}(r)$. This quantity is computed in radial shells. \textcolor{\revcolor}{For all particles within a shell $[r,r+\Delta r]$, we measure the mass-weighted mean rotational angular velocity as $\Omega_{\rm rot}=\sum_i m_i j_{z,i}/\sum_i m_i(x_i^2+y_i^2)$, where $j_{z,i}=x_i v_{y,i}-y_i v_{x,i}$ is the specific angular momentum of particle $i$ about the rotation axis. We characterize the dispersion in the angular-momentum contributions by $\Delta\Omega_{\rm rot}=\sqrt{\left\langle\left(m_i j_{z,i}-\left\langle m_i j_{z,i}\right\rangle\right)^2\right\rangle}/\sum_i m_i(x_i^2+y_i^2)$, which provides a measure of the fluctuations in the rotational velocity due to turbulent motions. The shaded regions indicate $\Omega_{\rm rot}\pm\Delta\Omega_{\rm rot}$.
} All merger remnants, except those produced by head-on collisions, exhibit significant differential rotation. In particular, the inspiral remnant possesses a rapidly rotating core whose angular velocity exceeds the surface break-up rate of both progenitor stars ($\sqrt{GM_\star/R_\star^3}$). 

This result suggests that rotation could play an important role in the subsequent long-term evolution of the merger remnant \citep[][]{MaederMeynet_2000ARA&A..38..143M,HegerLangerWoosley_2000ApJ...528..368H,BrottdeMinkCantiello_2011A&A...530A.115B,deMinkLangerIzzard_2013ApJ...764..166D}. In particular, the high rotation rate found in the inspiral simulations may lead to strong rotational mixing \citep[][]{Spruit_2002A&A...381..923S}, that efficiently transfer the surface H-rich fuels to the core, which is exactly what was assumed for the immortal star models \citep[][]{CantielloJermynLin_2021ApJ...910...94C,JermynDittmannMcKernan_2022ApJ...929..133J}.

\section{Implications for AGN stars}
\label{sec:implications}

Once the two AGN stars merge, there are four evolutionary stages with increasing timescales after collisions between stars embedded in an AGN disk.
\begin{enumerate}
    \item \emph{Dynamical evolution towards a hydrostatic equilibrium}. This equilibrium is established in $\sim 100$ stellar dynamical times, which is captured with our simulations in \S~\ref{sec:res_mesa}. 
    \item \emph{Internal structure adjustment towards a thermal equilibrium}. This requires  the Kelvin--Helmholtz timescale, $t_{\rm KH}\sim GM_\star^2/(2R_\star L_\star)$, usually $10^4\,\rm yr$, which is beyond our simulations. At the end of this stage, the merged star's interior is composed of (a)~a nuclear burning core, (b)~a convective envelope, (c)~a radiative layer, and (d)~a mass exchange zone with the ambient medium. Convection leads to efficient mixing in zones a and b, which contain most of the star's mass.
    \item \emph{Mass exchange with the disk towards an accretion--wind equilibrium}. This is because the stars in the AGN context are in an accreting boundary condition. 
    \item \emph{Onward nuclear burning}. Once the AGN star reaches accretion--wind equilibrium, its internal chemical composition changes due to nuclear burning.
\end{enumerate}

\begin{figure*}
    \centering
    \includegraphics[width=\linewidth]{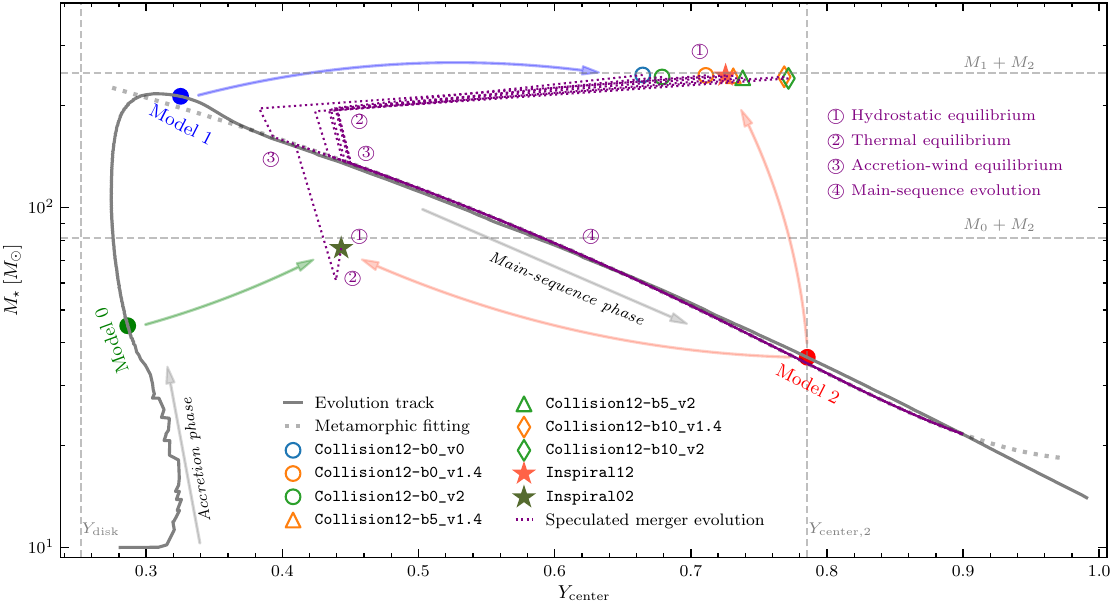}
    \caption{Main-sequence evolution of accreting and metamorphic stars in the $(Y_{\rm center}, M_\star)$ plane. Each evolutionary track begins with the background He abundance of the AGN disk ($Y_{\rm disk}$), undergoes a rapid accretion phase followed by a metamorphic main-sequence phase, and eventually enters the He-burning stage \citep[][]{XuChenLin_2026ApJ...997..206X}. We mark the three stellar models (0, 1, and 2) selected for the merger simulations, together with the corresponding merger remnants (see also Table~\ref{tab:ics_mesa}). The purple dotted lines illustrate the evolution of these merger remnants, which undergo four stages as annotated in the figure (purple text). The evolution beyond stage 1 (hydrostatic equilibrium) remains uncertain, but is extensively discussed in \S~\ref{sec:implications}.
    }
    \label{fig:metamorphic_merger}
\end{figure*}

The long-term evolution of the merger remnants can ultimately be investigated using a stellar evolution code such as \texttt{MESA}. One possible approach is to map the density and chemical abundance profiles of the merger remnants onto a one-dimensional \texttt{MESA} model and continue the stellar evolution. Without performing such calculations here, we first outline the evolutionary pathway with the aid of a schematic figure, Fig.~\ref{fig:metamorphic_merger}, which shows the same stellar track as in Fig.~\ref{fig:mesa_agn_star_evolution}, but in the $(Y_{\rm center}, M_\star)$ plane. For the metamorphic star in main-sequence evolution, \citet{XuChenLin_2026ApJ...997..206X} found that the Eddington ratio is a constant, more precisely, $\lambda_\star\approx \lambda_0$, where $\lambda_0$ is the feedback-transition parameter (also see \S~\ref{sec:res_mesa}). {\color{\revcolor}Once $\lambda_\star$ is close to unity, the star shades mass significantly \citep[][]{owocki2004,OwockiTownsendQuataert_2017MNRAS.472.3749O}, leading to an anti-correlation between $M_\star$ and $Y_{\rm center}$. For the fiducial value $\lambda_0=0.75$ and disk central density $\rho_{\rm c} = 10^{-16}\rm g\,cm^{-3}$ adopted by \citet[][]{XuChenLin_2026ApJ...997..206X}, the main-sequence evolution can be described by the fitting formula
\begin{align}
    M_\star\,[M_\odot] = 403.6\, (1-Y_{\rm center})^{2.07}+18.1.
    \label{equ:fitting}
\end{align}
\citet{XuChenLin_2026ApJ...997..206X} also found that, for the fiducial $\lambda_0$ and $\rho_{\rm c}$, stars converge to the same main-sequence relation given by Eq.~\eqref{equ:fitting}, even when the disk He abundance $Y_{\rm disk}$ varies. }

Fig.~\ref{fig:metamorphic_merger} also presents the position of the AGN stellar mergers in the $(Y_{\rm center}, M_{\star})$ plane at the end of our simulations in \S~\ref{sec:res_mesa}, which covers stage~1 of the merger evolution. For mergers, $M_\star$ is $M_{\rm merger}$, which is the self-gravitating remnant mass estimated using the criterion $e+p/\rho+v^2/2+\Phi<0$, where $e+p/\rho$ is the specific enthalpy. We find that all simulations retain nearly the entire mass of the two progenitors, yielding $M_{\rm merger}\approx M_1+M_2$. The central He abundance $Y_{\rm center}$ can also be obtained directly (see Fig.~\ref{fig:chemical_abundance_mesa}).

Below, we discuss the evolution of the merger remnant after the hydrostatic equilibrium (i.e., stage 2 and beyond).

\subsection{Stage 2: towards a thermal equilibrium}

\subsubsection{Core convection and rejuvenation of an old metamorphic star}

We first consider the change of $Y_{\rm center}$ after the thermal relaxation. 

\paragraph{Merger with a metamorphic star}
\label{sec:metamorphicmergers}

These are represented by mergers between models~1 and 2. From Fig.~\ref{fig:chemical_abundance_mesa}, the central He abundance of the merger remnants lies in the range $Y_{\rm center}\sim 0.65$--0.75, only slightly lower than that of the older progenitor ($Y_{\rm center,2}=0.79$). This is primarily because model~2 is always less massive but more compact than model~1. 

However, since massive stars are highly convective at the core, this small region ($r\lesssim R_\odot$, see Fig.~\ref{fig:visualization_mesa}) with high He abundance can mix with the outer regions throughout the thermal relaxation. As a result, the post-relaxation central He abundance can be estimated with $\bar Y_{\rm center}$, defined as the mass-weighted average He abundance inside $m\le M_{\rm merger}/2$, to consider convective mixing. We find that for mergers between models 1 and 2, $\bar Y_{\rm center} \sim 0.375$ (for the inspiral simulation) or $\bar Y_{\rm center} \sim 0.43$ (for multiple collisional simulations). This suggests that merging with a younger metamorphic star rejuvenates the older star, shifting it to an earlier evolutionary stage with a lower central He abundance.

\paragraph{Merger with an accreting star}
\label{sec:accretingmergers}

We next consider mergers between an old metamorphic star and a young and rapidly accreting star. Our simulation \path{Inspiral02} provides such an example. Unlike model~1, model~0 has a mass comparable to that of model~2. This allows the cores of the two stars to mix efficiently, substantially reducing the central He abundance of the merger remnant. We find that the remnant of \path{Inspiral02} has $Y_{\rm center} \approx \bar Y_{\rm center}\approx 0.45$, corresponding to a similar level of rejuvenation as that produced by the inspiral merger with model~1 (\path{Inspiral12}).

\subsubsection{Mass loss during thermal relaxation}

During this relaxation phase, the total luminosity of the merger remnant is
\begin{align}
    L_{\rm tot} = L_\star + L_{\rm acc,th} + L_{\rm acc,KE} + L_{\rm grav}.
\end{align}
Here, $L_\star$ is the luminosity generated by nuclear burning, $L_{\rm acc,th}\sim \dot{M}_{\rm acc} c_{\rm s,disk}^2\sim \dot M_{\rm Bondi} c_{\rm s,disk}^2$ is the thermal energy released through accretion (where $c_{\rm s,disk}$ is the sound speed in the ambient disk), and $L_{\rm acc,KE}\sim \dot{M}_{\rm acc} v_{\star}^2 \sim \dot{M}_{\rm acc} GM_\star/R_\star$ is the kinetic energy released by accretion. These three terms contribute to the luminosity of an AGN star that is already in thermal equilibrium \citep{XuChenLin_2026ApJ...997..206X}. In addition, $L_{\rm grav}$ accounts for the energy released by gravitational contraction. Averaged over the thermal timescale, $L_{\rm grav}$ can be approximated as $L_{\rm grav} \sim f_{\rm grav} G M_\star^2/(2R_\star t_{\rm KH}) \sim f_{\rm grav} L_\star$, where $f_{\rm grav}$ parametrizes the difference in the stellar binding energy before and after thermal relaxation.

The total luminosity of the merger remnant is dominated by $L_\star$ and $L_{\rm grav}$. From \citet{XuChenLin_2026ApJ...997..206X}, typical values are $L_{\rm acc,th} \sim 0.2\, (M_\star/M_\odot)^2\,L_\odot$ and $L_{\rm acc,KE} \sim 200\, (M_\star/M_\odot)^3/(R_\star/R_\odot)\,L_\odot$ (additional parameter dependencies are ignored here), whereas $L_\star =\lambda_\star L_{\rm Edd,\star}=3\times10^4\lambda_\star (M_\star/M_\odot)\,L_\odot$. Although the merger 
increases the stellar mass, the hierarchy $L_\star\gg L_{\rm acc,th},
L_{\rm acc,KE}$ remains valid. {Moreover, $\dot{M}_{\rm acc}$ is quenched by the radiation pressure as $\lambda_\star \rightarrow 1$.}

For a merger remnant whose hydrogen is not yet depleted, $L_\star = \int \dd m\,\epsilon_{\rm CNO} \sim \int \dd m\,\epsilon_{\rm CNO,0}\rho X X_{\rm CNO} f(T)$, where $\epsilon_{\rm CNO}$ is the energy generation rate of the CNO cycle, $\epsilon_{\rm CNO,0}$ is a normalization constant, and $f(T)$ describes the temperature dependence. Immediately after the merger, we find that $X$, $\rho$, and $T$ can all be elevated relative to those of the progenitors (cf. Figs.~\ref{fig:visualization_mesa}, \ref{fig:chemical_abundance_mesa}, and \ref{fig:density_rotation_mesa}), suggesting that the merger remnant may initially be more luminous than either progenitor.

{\color{\revcolor} The excess luminosity of the merger product can drive mass loss during thermal relaxation. A self-consistent calculation would require, e.g., a \texttt{MESA} simulation; here we give a rough estimate based on the super-Eddington wind models of \citet{OwockiTownsendQuataert_2017MNRAS.472.3749O}. Steady wind solutions in their model require the total luminosity to exceed the (electron-scattering) Eddington luminosity, $\lambda_{\rm tot}\equiv L_{\rm tot}/L_{\rm Edd,\star}>1$, and the mass-loss rate to satisfy $m\equiv \dot M_{\rm wind}GM_\star/(R_\star L_{\rm tot})\le 1-1/\lambda_{\rm tot}$, which keeps the emergent luminosity at or above $L_{\rm Edd,\star}$. Here $L_{\rm tot}\approx L_\star+L_{\rm grav}$, with $L_{\rm grav}\equiv f_{\rm grav}L_\star$ the luminosity released by contraction, and $R_\star$ is the radius at which the wind is launched. We adopt $\dot M_{\rm wind}=\eta\max[(L_{\rm tot}-L_{\rm Edd,\star})R_\star/GM_\star,\,0]$ with $\eta\le1$, i.e., $m=\eta(1-1/\lambda_{\rm tot})$. Writing $\lambda_\star\equiv L_\star/L_{\rm Edd,\star}$ and $t_{\rm KH}\equiv GM_\star^2/(2R_\star L_\star)$, this gives $\dot M_\star\simeq-\dot M_{\rm wind}\simeq-\eta(1+f_{\rm grav}-1/\lambda_\star)\,M_\star/2t_{\rm KH}$. Integrating over the relaxation time $\Delta t$ yields $M_{\rm relaxed}/M_{\rm merger}\sim\exp[-\eta(1+f_{\rm grav}-1/\lambda_\star)\,\Delta t/2t_{\rm KH}]$. We adopt $\lambda_\star=1$, $f_{\rm grav}=0.5$ (so $\lambda_{\rm tot}=1.5$), and $\eta=0.5$. The last choice corresponds to $m=1/6$, for which the terminal-speed scaling of \citet[][their Eq.~44]{OwockiTownsendQuataert_2017MNRAS.472.3749O}, $v_\infty^2/v_{\rm esc}^2\approx\lambda_{\rm tot}/(1+m\lambda_{\rm tot})-1$, gives $v_\infty\approx0.45\,v_{\rm esc}$, where $v_{\rm esc}=\sqrt{2GM_\star/R_\star}$. With $\Delta t\sim t_{\rm KH}$--$2t_{\rm KH}$, we obtain $M_{\rm relaxed}/M_{\rm merger}\sim0.8$--$0.9$. The estimate is sensitive to $\lambda_\star$: no wind is driven for $\lambda_\star<1/(1+f_{\rm grav})$. It should therefore be regarded as an order-of-magnitude approximation.

}

\subsection{Stage 3: Towards an accretion--wind equilibrium}

In thermal equilibrium, the newly acquired $M_{\rm merger}$ and $Y_{\rm merger}$ determine the nuclear reaction rate and therefore the stellar luminosity $L_\star$ and Eddington ratio $\lambda_\star$, and both $L_\star$ and $\lambda_\star$ increase with $M_\star$ for fixed $Y_{\rm center}$ \citep[][]{KippenhahnWeigertWeiss_2013sse..book.....K}.  Embedded in the dense disk gas,
the merged stars would gain mass through accretion of disk gas 
with $Y_{\rm disk}$ if their $\lambda_\star < \lambda_0$.
But, stars with $\lambda_\star > \lambda_0$ would shed mass 
with $Y_{\rm center}$ through radiation-pressure-driven winds. These 
stellar response leads to an accretion--wind equilibrium 
with $\lambda_\star \sim \lambda_0$ in stage 3.\footnote{{\color{\revcolor} Note that $\lambda_0$ is the characteristic Eddington factor defined in \citet{XuChenLin_2026ApJ...997..206X}, above which the wind mass loss is significant, which is fundamentally controlled by their prescription on radiation-driven winds (their Eq.~22). It can be generalized to other prescription: e.g., with the super-Eddington wind model in \citet{OwockiTownsendQuataert_2017MNRAS.472.3749O}, $\lambda_0$ should be treated as unity.  }}

\paragraph{Mergers with a metamorphic star}

The merged star's $M_{\rm relaxed}-Y_{\rm center}$ values fall above the original $M_{\star} (\lambda_0, Y_{\rm center})-Y_{\rm center} (\lambda_0)$ track in Fig.~\ref{fig:metamorphic_merger}. This suggests the star is over-luminous with $\lambda_\star > \lambda_0$ such that its accretion is suppressed. During its evolution towards an accretion--wind equilibrium, the merged star loses mass, without significant changes in $Y_{\rm center}$, while lowering the Eddington factor to $\lambda_\star \to \lambda_0$.  In the $(Y_{\rm center},M_\star)$ plane, the merger remnant therefore evolves approximately along a vertical trajectory from $(\bar Y_{\rm center},M_{\rm relaxed})$ to $(\bar Y_{\rm center},M_{\star}(\lambda_0, \bar Y_{\rm center}))$.

\paragraph{Mergers with an accreting star}
After reaching a thermal equilibrium (at the end of stage 2), 
the mass $M_\star$ of the merger remnants of metamorphic and 
accreting stars grows through accretion of disk gas with 
$Y_{\rm disk} < Y_{\rm center}$.  Consequently, their core and envelope are replenished with a fresh supply of H-rich disk gas.  Their modified Helium abundance is 
\begin{equation}
Y_{\rm center} \sim (M_{\rm merger} Y_{\rm merger} + (M_{\star}-M_{\rm merger}) 
Y_{\rm disk}) / M_{\star},
\label{eq:addedy}
\end{equation}
satisfying $Y_{\rm disk} < Y_{\rm center} < Y_{\rm merger}$. The merger remnant is also below the universal main sequence, suggesting that it is not massive (luminous) enough such that accretion will dominate, leading to mass growth.

In the $(Y_{\rm center}, M_\star)$ plane, the merger remnant therefore evolves upwards and inwards from $(Y_{\rm center},M_{\rm merger})$.
At the end of stage 3, the star will reach an accretion--wind equilibrium with $\lambda_\star\sim \lambda_0$, such that the $M_\star-Y_{\rm center}$ values satisfy both Eq.~\eqref{equ:fitting}.

\subsection{Stage 4: main-sequence evolution}

Once the stellar merger remnants reach accretion--wind equilibrium, they may converge to the universal main sequence calculated by \citet{XuChenLin_2026ApJ...997..206X} (Eq.~\ref{equ:fitting}). In their further evolution, they can evolve with $\lambda_\star\rightarrow\lambda_0 (\sim 1)$, just like normal metamorphic main-sequence stars. They may also follow the same mass--radius relation. From the evolutionary track shown in Fig.~\ref{fig:mesa_agn_star_evolution}, we find that main-sequence stars of different ages approximately satisfy $R_\star \sim 10\,(M_\star/10\,M_\odot)^{0.25}\,R_\odot$. 

Furthermore, Fig.~\ref{fig:mesa_agn_star_evolution} shows that $Y_{\rm center}$ increases nearly linearly with time. Related to this fact, the merger is expected to prolong the main-sequence lifetime of the old metamorphic star, following the argument of \citet{XuChenLin_2026ApJ...997..206X}. For a hydrogen-burning star, $\eta \dot M_{\rm H} c^2 \sim \lambda_0 L_{\rm Edd,\star}$, where $\dot M_{\rm H}$ is the hydrogen consumption rate and $\eta\sim0.007$ is the energy conversion efficiency of the CNO cycle. The main-sequence lifetime is then $\tau_{\rm MS} \sim X_{\rm center} M_{\rm core}/\dot M_{\rm H} \sim X_{\rm center}\eta/\lambda_0 \cdot M_{\rm core}/M_\star \cdot \tau_{\rm Sal}$, where $\tau_{\rm Sal}=M_\star c^2/L_{\rm Edd,\star}$ is the Salpeter timescale. Since $M_{\rm core}\sim M_\star$ for very massive stars, $\tau_{\rm MS}\sim 3(X_{\rm center}/\lambda_0)\,\rm Myr$, in reasonable agreement with the stellar evolution models of \citet{XuChenLin_2026ApJ...997..206X}. Consequently, for mergers between models~1 and 2, $X_{\rm center}$ increases by $\sim0.35$--0.4 {after thermal relaxation}, extending the main-sequence lifetime by $\Delta\tau_{\rm MS}\sim 1.4$--$1.6 \,\rm Myr$. For the inspiral merger between models~0 and 2, $\Delta X_{\rm center}\sim0.35$, also giving $\Delta\tau_{\rm MS}\sim1.4\,\rm Myr$.

We emphasize that this speculative long-term evolution requires verification with dedicated stellar evolution calculations. Several important caveats remain. First, the luminosities before and after thermal relaxation are only estimated approximately. Second, the chemical composition of the merger remnant differs substantially from that of ordinary metamorphic main-sequence stars. Third, rapid rotation inferred from inspiral simulations may significantly affect the subsequent evolution: now rotational mixing can be efficient in bringing accreted hydrogen across the radiative layer to the core \citep[][]{Spruit_2002A&A...381..923S}, potentially changing the metamorphic scenario \citep[][]{CantielloJermynLin_2021ApJ...910...94C,JermynDittmannMcKernan_2022ApJ...929..133J,Ali-DibLin_2023MNRAS.526.5824A,XuChenLin_2026ApJ...997..206X}. These complexities are beyond the scope of the present analytic treatment. A quantitative investigation of the long-term evolution with \texttt{MESA} will be presented elsewhere (Xu et al. in preparation).

\begin{figure*}
    \centering
    \includegraphics[width=\linewidth]{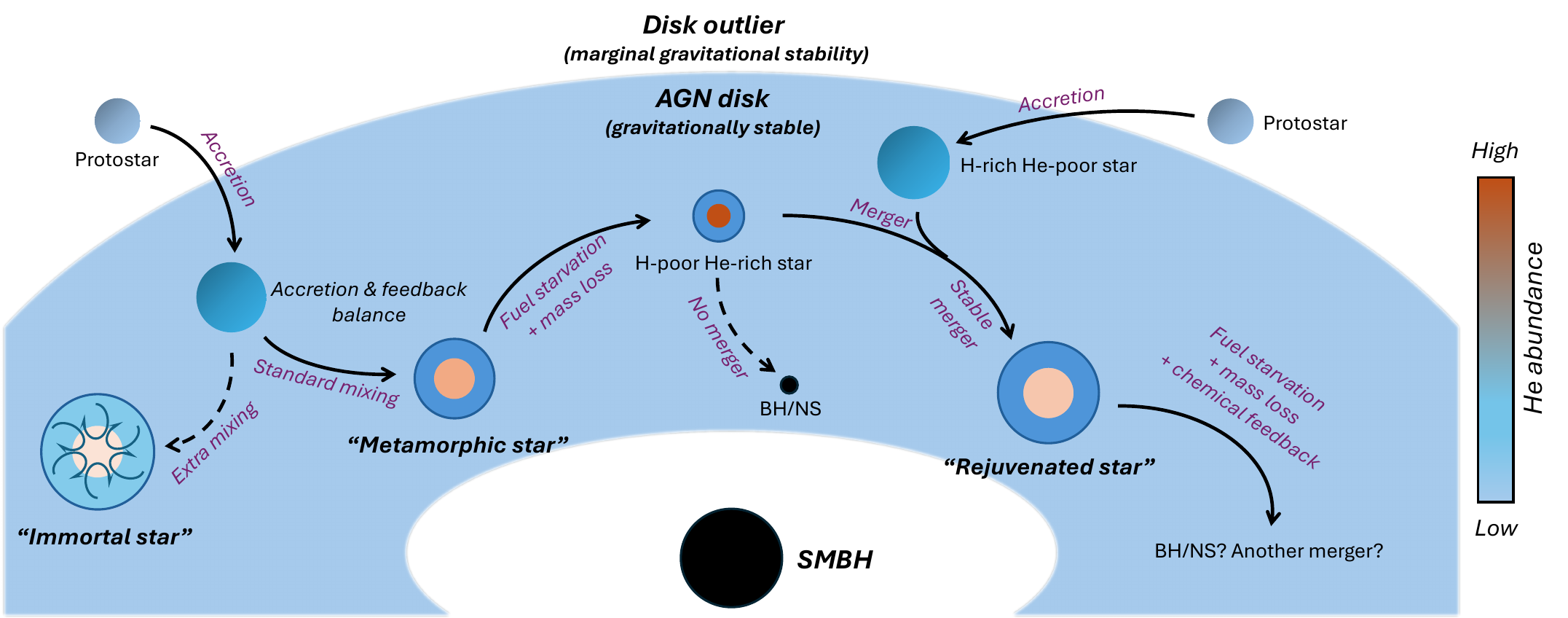}
    \caption{Illustration: evolution channels of massive stars in AGN disks. If the merger rate is high enough (e.g., $\gtrsim 10^{-6}\,\rm yr^{-1}$), such merger events may also contribute to the chemical enrichment of the AGN disk. }
    \label{fig:merger_in_agns}
\end{figure*}

\subsection{The merger rates of metamorphic stars}

We revisit the merger-rate estimate of \citet{ChenLin_2024ApJ...967...88C} and examine whether its underlying assumptions remain applicable to metamorphic stars embedded in AGN disks. Stellar mergers may occur either through direct physical collisions or through the inspiral of binaries formed after close gravitational encounters.

For a stellar surface density $s_\star$ and velocity dispersion $\Delta v$, the stellar volume density is $n_\star\sim s_\star/H\sim s_\star/(\Delta v/\Omega)$, where $\Omega=(GM_{\rm BH}/R^3)^{1/2}$ is the local Keplerian frequency. The encounter rate for each star is then $\Gamma\sim\pi R_{\rm eff}^2n_\star\Delta v\sim\pi R_{\rm eff}^2s_\star\Omega$, where $R_{\rm eff}$ is the effective interaction radius. Following \citet{ChenLin_2024ApJ...967...88C}, $R_{\rm eff}=\min[(1+\Theta_\star)^{1/2}R_\star,R_{\rm H}]$, 
where the Safronov number 
$\Theta_\star= G M_\star/(R_\star \Delta v ^2)$ accounts for gravitational focusing and $R_{\rm H}\equiv(M_\star/3M_{\rm BH})^{1/3}R$ is the Hill radius.

If $R_{\rm eff}$ is set by the stellar radius, corresponding to direct geometric collisions, the merger rate is extremely small. For {pessimistic} parameters ($M_{\rm BH}=10^8\,M_\odot$, $R=1\,{\rm pc}$, $s_\star=10^4\,{\rm pc}^{-2}$, $R_\star=10\,R_\odot$, and $\Theta_\star=1$), one obtains $\Gamma_{\rm col}\sim10^{-9}\,{\rm yr^{-1}}$, implying that direct collisions are negligible. 
{But, gas drag and collisions limit the magnitude of \textcolor{\revcolor}{$\Delta v$} such that $\Theta \gg 1$ (Wang et al. in preparation).}  In this case, the 
encounter rate becomes orders of magnitude larger if $R_{\rm eff}=R_{\rm H}$, because $R_{\rm H}\gg R_\star$. In this picture, close encounters inside the Hill sphere are efficiently converted into bound binaries by gas dissipation, yielding a characteristic binary-capture timescale of $\tau_{\rm cap}\sim10^4$--$10^6\,{\rm yr}$ \citep[][]{ChenLin_2024ApJ...967...88C}.

The subsequent binary evolution is considerably more uncertain. Based on numerical simulations, \citet{ChenLin_2024ApJ...967...88C} adopted an inspiral timescale $\tau_{\rm ins}\sim a/\dot a\sim M_\star/\dot M_\star$ \citep[][]{LaiMunoz_2023ARA&A..61..517L}, giving $\tau_{\rm ins}\sim10^4/\lambda_\star\,{\rm yr}$ after neglecting its weaker parameter dependence. This estimate is appropriate for rapidly accreting stars, whose circumstellar and circumbinary gas efficiently removes orbital angular momentum.

Metamorphic stars, however, are expected to be feedback-dominated and therefore accrete much more slowly. The reduced gas supply naturally lengthens $\tau_{\rm ins}$, making binary hardening, rather than binary capture, the bottleneck for stellar mergers. Furthermore, feedback may evacuate gas from the vicinity of the binary, suppressing the formation of mini-disks or even opening a circumbinary cavity. In this regime, the gravitational torques responsible for orbital decay become substantially more uncertain, and binaries captured within the Hill sphere may not necessarily shrink to stellar contact \citep[][]{DempseyLiMishra_2022ApJ...940..155D}. Consequently, the merger rates predicted by \citet{ChenLin_2024ApJ...967...88C} should likely be regarded as optimistic upper limits when applied to metamorphic stars.

Given these uncertainties, we consider two limiting scenarios. In the optimistic case, binary hardening remains efficient, so the merger rate approaches the Hill-sphere encounter rate of $\sim10^{-6}$--$10^{-4}\,{\rm yr^{-1}}$. A metamorphic star may then experience $\sim1$--100 mergers over its main-sequence lifetime, substantially affecting both stellar evolution and the chemical enrichment of the AGN disk (Fig.~\ref{fig:merger_in_agns}). In the opposite limit, binary hardening is strongly suppressed, and only a small fraction of captured binaries eventually merge. The merger rate could then approach that of direct geometric collisions ($\sim10^{-9}\,{\rm yr^{-1}}$), in which case mergers would have only a minor influence on the stellar population and the chemical evolution of the AGN disk.

{A single AGN disk may host both limits. The upper limit may be applicable for the inner regions ($\lesssim 10^4 R_\bullet$, the SMBH's gravitational radius) where embedded stars are densely populated. The lower limit may be more appropriate to the less populated outer regions of the AGN disks where $s_\star$ is relatively small. The coexistence of both populations may introduce observable signatures on the metallicity and its gradient in AGN disks \citep{HuangLinShields_2023MNRAS.525.5702H}.}

}

\section{Summary}
\label{sec:summary}

{
We perform a suite of binary stellar merger simulations to study the mixing of chemical elements during stellar mergers, first using an idealized polytropic EOS and then a more realistic, composition-dependent EOS.
}

With the polytropic EOS and stellar models (\S~\ref{sec:res}), we explore a broad parameter space, including stellar structure (polytropic index $n$), mass ratio ($M_1/M_2$), initial relative velocity ($v_{\rm rel,0}$), and impact parameter ($b$). The simulations produce a wide variety of post-merger outcomes. We summarize the main results below.

\begin{itemize}
    \item The dominant mixing mechanism depends on the merger stage. During the initial coalescence phase, shock-driven splashing and spirals of material govern the mixing. Gravity subsequently promotes the formation of Rayleigh--Taylor interfaces that further enhance chemical mixing. As the remnant relaxes toward hydrodynamic equilibrium, turbulent diffusion smooths the abundance distribution. 

    \item The merger remnant can become more compact and centrally concentrated than either progenitor, particularly for equal-mass mergers with low relative velocities. In contrast, mergers with high relative velocities produce more diffuse remnants and may become dynamically unstable if the impact energy is sufficiently large. For off-center collisions, the non-zero orbital angular momentum spins up the merger remnant. 

    \item Although stellar mergers eject material to radii much larger than the initial stellar sizes, the majority ($\gtrsim90\%$) of the mass remains gravitationally bound for the parameter space explored here. Additional mass loss is expected during the subsequent thermal relaxation phase.

    \item Immediately after coalescence, the merger remnant oscillates at a frequency close to its fundamental mode. These oscillations damp before the end of the simulation, likely through the transfer of energy to smaller-scale, higher-frequency modes.
\end{itemize}

{

In \S~\ref{sec:res_mesa}, we perform merger simulations using a realistic EOS and stellar models generated with \texttt{MESA}. Specifically, we select three models along the metamorphic stellar evolutionary track of \citet{XuChenLin_2026ApJ...997..206X}: an old H-burning metamorphic star, a young H-burning metamorphic star, and an even younger accreting star (cf. Fig.~\ref{fig:mesa_agn_star_evolution}). With this more realistic setup, we find that the qualitative merger outcomes, including chemical mixing, density structure, rotation, and mass retention, are broadly consistent with those obtained from the parameter survey using polytropic models. We further find that mergers can rejuvenate old H-poor/He-rich metamorphic stars to earlier evolutionary stages. Fig.~\ref{fig:merger_in_agns} summarizes these results.

\begin{itemize}
    \item A merger with a younger H-rich/He-poor metamorphic star increases the central hydrogen abundance by only $\Delta X_{\rm center}\sim0.05$--0.1 while preserving $\gtrsim95\%$ of the total progenitor mass. This limited increase is primarily because old metamorphic stars are substantially more compact than younger ones. Whether the merger proceeds through a direct collision or an inspiral does not qualitatively change this conclusion. However, convection may mix the compact core with the H-rich material outside, still efficiently rejuvenating the star by $\Delta X_{\rm center}\sim0.3$ after thermal relaxation. 

    \item In contrast, a merger with an H-rich accreting star of comparable mass and radius can already produce a well-mixed core in $\sim 100$ dynamical times, increasing the central hydrogen abundance by $\Delta X_{\rm center}\sim0.3$. Therefore, the main-sequence lifespan can be extended by $\sim 1.2\,\rm Myr$.

    \item The surface nitrogen abundance is enhanced in all simulations, yielding ${\rm N/O}\sim1$--3 and ${\rm N/C}\gtrsim5$. These abundance ratios are significantly higher than those of the AGN disk material, where ${\rm N/O}\sim0.1$ and ${\rm N/C}\gtrsim0.3$, and are comparable to those observed in the nebula surrounding SN~1987A \citep[][]{LundqvistFransson_1996ApJ...464..924L,SonnebornFranssonLundqvist_1997ApJ...477..848S,MattilaLundqvistGroningsson_2010ApJ...717.1140M}.

    \item Compared with collisional mergers, remnants produced through quasi-circular inspirals exhibit much more rapid differential rotation, which is expected to play an important role in their subsequent evolution \citep[][]{MaederMeynet_2000ARA&A..38..143M}. The high rotation rate may enhance internal mixing \citep[][]{Spruit_2002A&A...381..923S}, which is required by immortal stars \citep[][]{CantielloJermynLin_2021ApJ...910...94C,JermynDittmannMcKernan_2022ApJ...929..133J}.
\end{itemize}

We also speculate on the long-term evolution of the merger remnants (Fig.~\ref{fig:metamorphic_merger}). The merger remnant is expected to become temporarily over-luminous and may lose a fraction of its retained mass during thermal relaxation, while its central H and He abundances remain nearly unchanged. The remnant then evolves towards an accretion--wind equilibrium, and it may lose mass if it is over-luminous with $\lambda_\star>\lambda_0$, or accrete mass if it is less luminous with $\lambda_\star<\lambda_0$. This modulation leads to a convergence towards a nearly universal main-sequence evolutionary track that isolated metamorphic stars follow (see \S~\ref{sec:implications}). Throughout this evolution, the remnant is expected to shed a substantial amount of chemically enriched material through stellar winds, potentially contributing to the chemical enrichment of the AGN disk.

Several caveats apply, e.g., we do not include radiative transfer \citep[][]{HatfullIvanovaLombardi_2021MNRAS.507..385H,HatfullIvanova_2025ApJ...982...83H} or magnetic fields \citep{SchneiderOhlmannPodsiadlowski_2019Natur.574..211S,RyuAmaroSeoaneTaylor_2024MNRAS.528.6193R,RyuSillsPakmor_2025ApJ...980L..38R,VynatheyaRyuWang_2026ApJ...999...64V}. We plan to address these effects in future work. In addition, the long-term evolution of merger remnants remains uncertain and will ultimately be investigated using detailed stellar evolution calculations (Xu et al. in preparation).
}

%% Also note that the akcnowlodgment environment does not support long amounts of text. If you have a lot of people and institutions to acknowledge, do not use this command. Instead, create a new \section{Acknowledgments}.
\begin{acknowledgments}
We thank the anonymous referee for important comments to improve this manuscript. We also thank Zhenghao Xu for providing the AGN stellar models, and Michela Mapelli for useful comments. YS and NM acknowledge the support of the Natural Sciences and Engineering Research Council of Canada (NSERC) under the funding reference No. 568580. The related computation is performed on the University of Toronto cluster ``Trillium,'' supported by SciNet (scinethpc.ca) and the Digital Research Alliance of Canada (alliancecan.ca).
\end{acknowledgments}


\vspace{5mm}
% \facilities{HST(STIS), Swift(XRT and UVOT), AAVSO, CTIO:1.3m,


\software{\texttt{GIZMO} \citep{Hopkins_2015MNRAS.450...53H}
          }



\appendix
\section{Effective Polytropic Index Approximation}
\label{app:gas_mixing}

%need some motivation to validate the current method
%mixing is a complicated problem, hows the stars internal energy compares to kinetic energy? 
%pressure in both star plays important role in setting the mixing
%we study a simplified problem, the goal is to understand wheater adoping gamma-eff provides a reasonable approximation to 
% \xiaoshan{Condense this Section, discuss with Yanlong, Doug. }

\begin{figure*}
    \centering
    \includegraphics[width=0.495\linewidth]{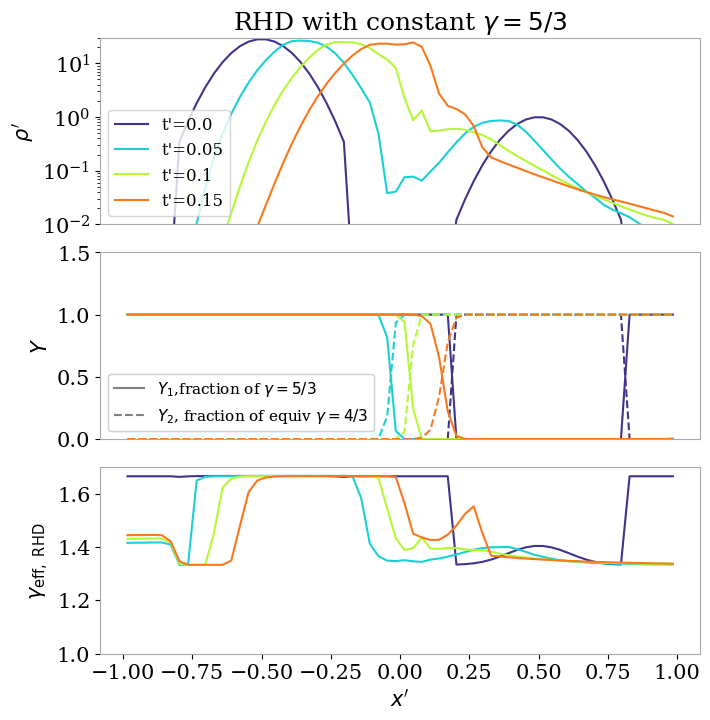}
    \includegraphics[width=0.495\linewidth]{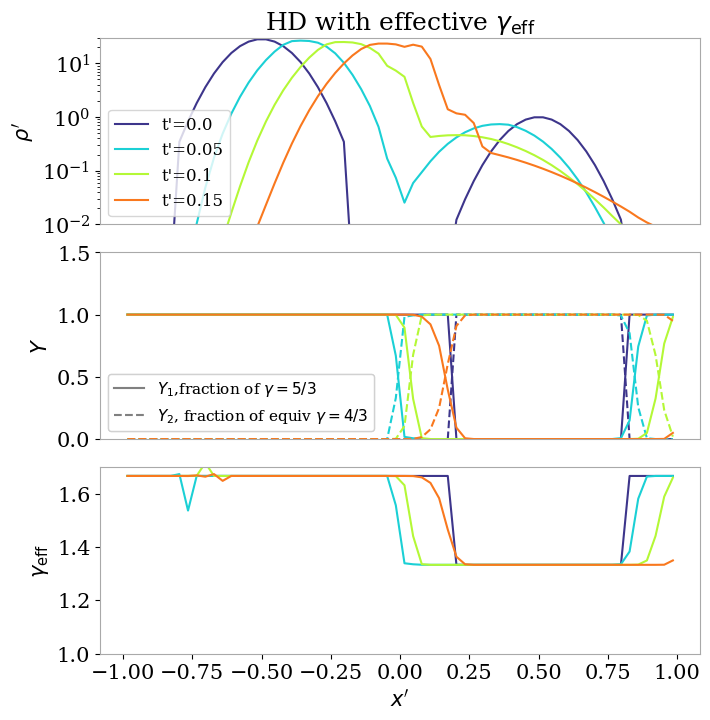}
    \caption{The RHD and HD one-dimensional mix test (left and right panels, respectively). From top to bottom row in each panel: gas density, passive scalar concentration (solid for equivalent $\gamma_{\rm eff}\approx5/3$ gas, dashed for equivalent $\gamma_{\rm eff}\approx4/3$ gas), effective polytropic index $\gamma_{\rm eff}$. }
    \label{fig:appendix_primitive_timeseries}
\end{figure*}

In this section, we validate the way we evaluate $\gamma_{\rm eff}$ of the polytropic gas mixing (\S~\ref{sec:method:mixing_of_gas}), by testing it against the more realistic RHD simulation of the radiation-pressure-dominated gas ($\gamma
 \approx 4/3$) mixed with the thermal-pressure-dominated gas ($\gamma\approx 5/3$). 

We set up a one-dimensional RHD simulation of such a mixing, which solves the radiation transfer equation to account for the coupling between radiation and gas; alongside, a counterpart hydrodynamic (HD) simulation of polytropic gas mixing following $\gamma=\gamma_{\rm eff} = \sum_i \rho_i \gamma_i / \rho$. Both the RHD and HD simulations are performed with Athena++ \citep{StoneTomidaWhite_2020ApJS..249....4S,Jiang_2022ApJS..263....4J}, which we use to solve unit-less equations assuming density, temperature, and length scaling $\rho_{0},~T_{0},~l_{0}$, or their equivalents. While the hydrodynamic test is scale-free, the radiation hydrodynamic test is uniquely set by the choice of scaling. For simplicity, we assume the same scaling in both tests. We label the unit-less variables in code with prime $X'$, the variables with physical units by $X=X_{0}X'$. In the rest of the section, we report unit-less variables unless explicitly specified.

We adopt density scaling $\rho_{0}=10^{-5}\rm g~cm^{-3}$, length scaling $l_{0}=6.96\times10^{10}{\rm cm}=R_{\odot}$ and temperature unit $T_{0}=10^{6}\rm K$, giving velocity scaling of $v_{0}=1.17\times10^{7}\rm cm\, s^{-1}$. We adopt 80 angles for the angular resolution in the radiation test.

The one-dimensional simulation domain spans $x'\in[-1,~1]$ with 64 uniformly-spaced grids. The density and pressure floor for the hydrodynamic solver are $\rho_{\rm floor}'=10^{-7},~P_{\rm floor}'=10^{-9}$. The gas clouds are modeled as Gaussian density profiles $\rho'=\rho_{0}'\exp(-(x'-x_{0}')^{2}/2\sigma'^{2})$ and assuming a width of $\sigma'=0.1$. We assume the first and second Gaussian centers at $x_{1}'=-0.5$ and $x_{2}'=0.5$, with $\rho_{0}'=\rho_{1}'$ and $\rho_{0}'=\rho_{2}'$ respectively. Gas within clouds $|x'-x_{0}'|<3\sigma'$ is initialized with velocities $v_{1}'=v_{x}',~v_{2}'=-v_{x}'$, so that the two clouds collide and mix with each other. If we estimating $c_{s}=\sqrt{\gamma P_{\rm tot}/\rho}$, the collision is supersonic with moderate Mach number $\mathcal{M}=v_{x}'/c_{s}'\approx 4$.

Similar to \S~\ref{sec:method:mixing_of_gas}, we dye gas within each cloud with passive scalar $Y_{1}$ and $Y_{2}$, and background low-density gas is assumed with $Y_{1}=1.0$. These passive scalars are evolved by $\partial_t (\rho Y) + \nabla \cdot (\rho Y \bm{v}) =0$, same as Eq.~\eqref{equ:element_transport} except for the absence of the SGS diffusivity. The boundary conditions for all variables are set to outflow. We specify other numerical setups for each test in the following subsections.

\subsection{Radiation-hydrodynamic simulation}

We assume the cloud on the left/right (labeled by subscript 1/2) is gas/radiation pressure dominated. In the second cloud central density is $\rho_{2}'=1.0$ ($\rho_{2}=10^{-5} \rm g~cm^{-3}$) and its temperature is $T_{2}'=1.0$ ($T_{2}=10^{6}$K). The estimated ratio of radiation pressure to gas pressure $P_{\rm rad,2}/P_{\rm gas,2}=a_{R}T_{2}^{4}/3.0/(k_{B}\rho_{2}T_{2}/\mu m_{p})\approx2$. 

We assume the first cloud temperature is $T_{1}'=0.1$ ($T_{1}=10^{5}$K), and the gas pressure matches the total pressure of the second cloud. Solving for its density $(k_{B}\rho_{2}T_{1}/\mu m_{p})=P_{\rm rad,2}+P_{\rm gas,2}$, we have $\rho'_{1}=28.3$. The ratio of radiation pressure to gas pressure is $P_{\rm rad,1}/P_{\rm gas,1}=a_{R}T_{1}^{4}/3.0/(k_{B}\rho_{1}T_{1}/\mu m_{p})\approx6.47\times10^{-5}$. 

We set the scatter opacity to be $\kappa_{s}=0.32\rm cm^{2}~g^{-1}$, absorption opacity $\kappa_{R}=0.0$ (Rosseland mean opacity), $\kappa_{P}=0.0$ (Planck mean opacity) for simplicity. Both clouds are optically-thick $\tau\sim\rho\sigma\kappa_{s}>1$. When initializing the gas, we initialize isotropic intensity corresponding to the local gas temperature, so that for each angle $\textbf{n}$, $I(\textbf{n})'=J(\textbf{n})'=B(T_{\rm gas}')$, where $B$ is the Planck function. 

The gas equation of state assumes a polytropic $\gamma=5/3$. In the analysis, we define an effective $\gamma$ for radiative fluid:
\begin{equation}\label{eq:gamma_rhd}
    \gamma_{\rm eff,~RHD}=1+\frac{P_{\rm rad}+P_{\rm gas}}{E_{\rm rad}+E_{\rm IE,gas}},~c_{\rm s}^{2}=\gamma(P_{\rm rad}+P_{\rm gas})/\rho,
\end{equation}
where $P_{\rm rad}$ and $E_{\rm rad}$ is radiation pressure tensor and energy density respectively. The radiation field is approximately isotropic in the simulation, so that $P_{\rm rad}\approx E_{\rm rad}/3.0$.

The left panel of Fig.~\ref{fig:appendix_primitive_timeseries} shows the density, passive scalar $Y$ and $\gamma_{\rm eff,~RHD}$ (Equation~\ref{eq:gamma_rhd}) evolution. Initially, the second cloud $\gamma_{\rm eff,RHD}$ is slightly higher than $4/3$, because radiation pressure is only larger than gas pressure by a factor of a few $P_{\rm rad,2}/P_{\rm gas,2}\approx2$. At $t'=0.05$, as the two clouds touch, a thin layer of shock forms at the contact surface $x'\sim 0.0$ with increased entropy. From $t'=0.1$ to $t'=0.15$, the two Gaussians merge. The momentum of the gas-pressure-dominated cloud 1 is higher due to its higher gas density; the merged cloud density distribution shifts toward the positive $x'$ direction. The background gas becomes radiation pressure dominated due to its low density $\rho'=10^{-3}$, with $\gamma_{\rm eff,~RHD}\approx4/3$. 

\subsection{Hydrodynamic simulation with $\gamma_{\rm eff}$}

In this test, we adopt a polytropic equation of state, but calculate the effective adiabatic index as $\gamma_{\rm eff}  \approx \sum_i \rho_i \gamma_i / \rho\equiv\sum_i Y_i \gamma_i$. Other dynamical setups are identical to the RHD counterpart.

The evolution of density, passive scalar concentration $Y$, and $\gamma_{\rm eff}$ are shown in the right panel of Fig.~\ref{fig:appendix_primitive_timeseries}. The density structure during mixing is similar to the RHD runs, with moderate differences at the contact surface at $t'=0.05$. The final mixed density distribution at $t'=0.15$ between the two tests is similar.

From this test, we conclude that the approach of adopting $\gamma_{\rm eff}$ as described in the main text approximates the mix between fluids with different energy density content reasonably well, in the limit of such simplified opacity structure. More realistic physical problems of mixing will depend on factors such as dimensionality, turbulence or diffusivity prescription, and opacity function; these will be explored in future work.

\begin{figure*}
    \centering
    \includegraphics[width=\linewidth]{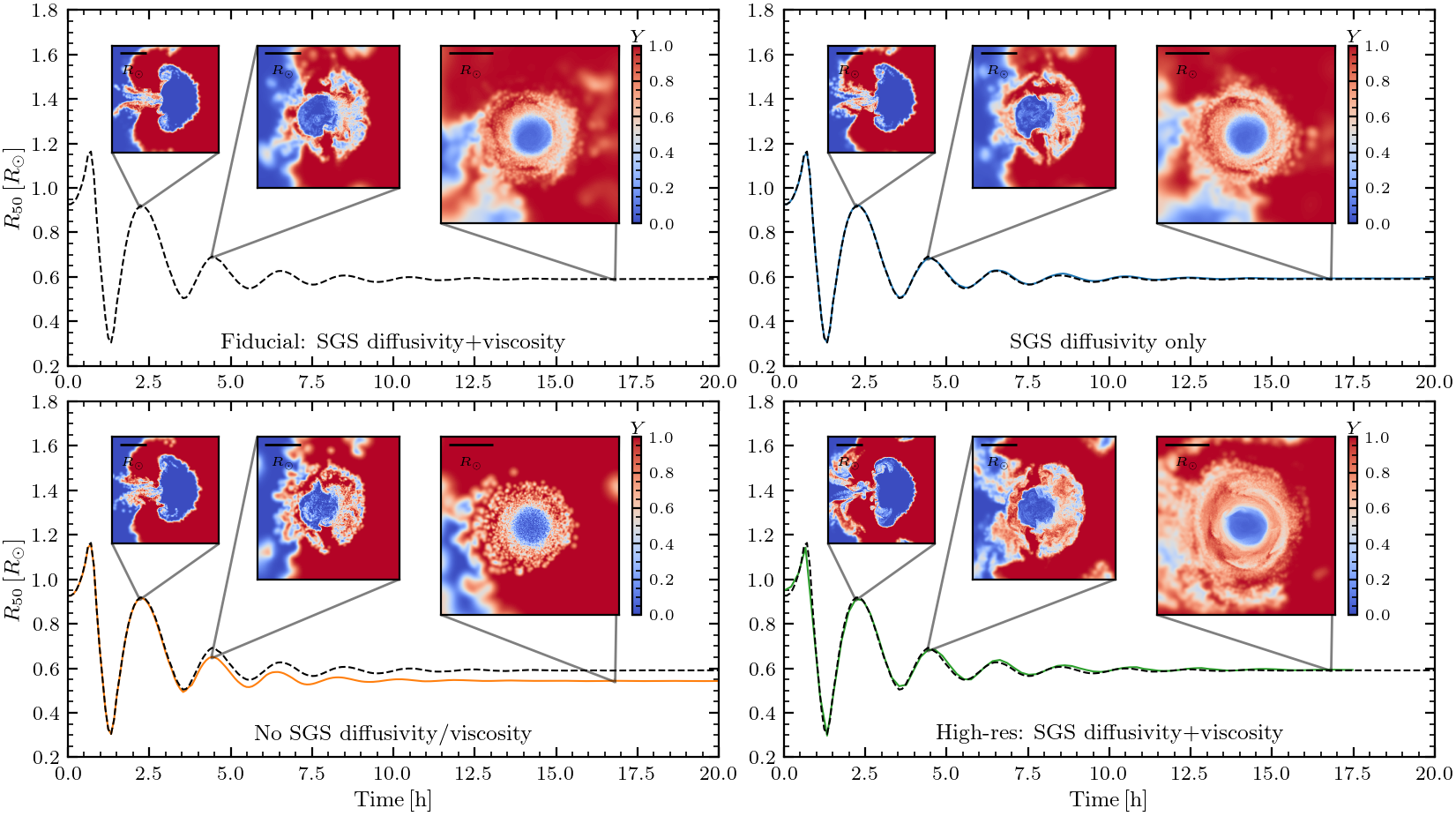}
    \caption{Test simulations based on the 1:1 mass ratio merger \texttt{n2.5-n-2.5\_b0\_v1.4}. Each panel shows the oscillation and damping in the half-mass radius ($R_{50}$) of the merger, and the mid-plane abundance maps at different stages. Four simulations are: (1) with SGS diffusivity+viscosity (fiducial; \emph{upper left}); (2) with SGS diffusivity but without SGS viscosity (\emph{upper right}); (3) without SGS diffusivity/viscosity (\emph{lower left}); (4) with fiducial setups but at a higher ($8\times$) resolution (\emph{lower right}).
    }
    \label{fig:convergence_tests}
\end{figure*}

\section{Numerical and convergence tests}
\label{app:convergence_tests}

The governing equations (Eqs.~\ref{equ:conintuity}--\ref{equ:element_transport}) include additional sub-grid diffusion and viscosity terms, which introduce extra dissipation compared to simulations that omit these terms. To assess their impact, we perform a set of additional simulations to quantify and justify the effects of these sub-grid contributions.

\subsection{Post-merger oscillation and damping}

Fig.~\ref{fig:convergence_tests} presents a suite of simulations based on the 1:1 mass-ratio merger run \texttt{n2.5-n1.5\_b0\_v1.4}. We consider four different numerical setups:
(1) a fiducial run that solves Eqs.~\eqref{equ:conintuity}--\eqref{equ:element_transport}, including both sub-grid-scale (SGS) diffusivity and viscosity;
(2) a run that retains the SGS diffusivity term in Eq.~\eqref{equ:element_transport} but removes SGS viscosity (i.e., all $\nu_{\rm sgs}$-associated terms in Eqs.~\ref{equ:momentum} and \ref{equ:energy});
(3) a run that further removes SGS viscosity entirely, eliminating all $\nu_{\rm sgs}$-associated terms in Eqs.~\ref{equ:momentum}--\ref{equ:element_transport};
and (4) a higher-resolution run that retains both SGS diffusivity and viscosity, but with an $8\times$ increase in mass resolution, corresponding to $m_{\rm gas}=1/128^3\,M_\odot \approx 5\times 10^{-7}\,M_\odot$.

We plot the half-mass radius, $R_{50}$, of the merger remnant (Fig.~\ref{fig:time_evo}), which exhibits oscillatory behavior followed by damping. This damping may arise from sub-grid-scale (SGS) viscosity, from numerical viscosity (if the SGS viscosity is subdominant), or from resolved turbulence that transfers energy from this low-frequency mode to higher-frequency fluctuations. If the damping is dominated by numerical viscosity, the simulation results are not physically robust. Each panel of Fig.~\ref{fig:convergence_tests} also shows mid-plane abundance maps at three representative evolutionary stages. 

Comparing the fiducial run with the SGS-diffusion-only run, we find good agreement in the oscillatory behavior of $R_{50}$, which rules out SGS viscosity as the primary cause of the observed damping. The corresponding abundance maps are also qualitatively similar between the two runs.

\begin{figure*}
    \centering
    \includegraphics[width=\linewidth]{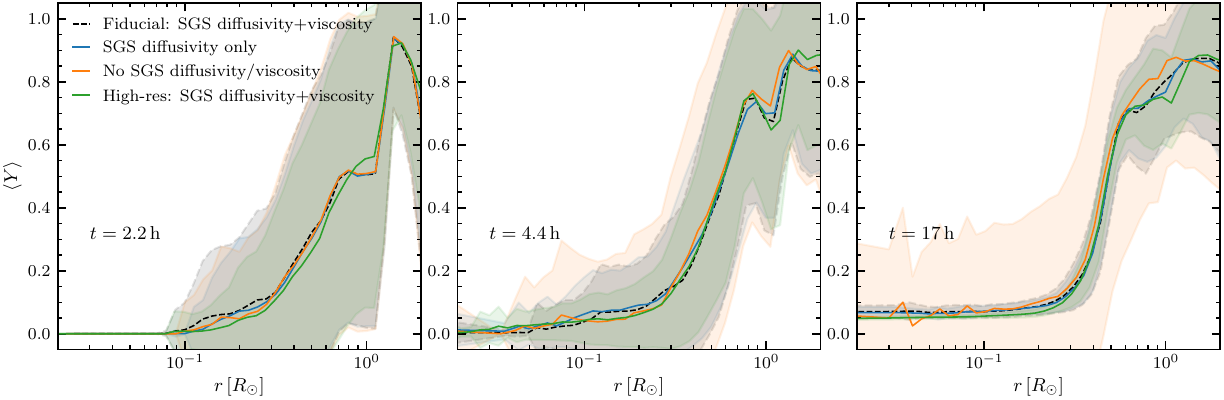}
    \caption{The chemical abundance ($Y$) profile at different evolutionary stages for the four test simulations. Here the panels correspond to $t=2.2\,\rm h$, $4.4\,\rm h$, and $17\,\rm h$, identical to that of the inset panels in Fig.~\ref{fig:convergence_tests}. In eahc panel, we plot the $Y$ profiles, including their standard deviations (shaded regions).} 
    \label{fig:convergence_tests_chemical}
\end{figure*}

Comparing the fiducial run with the run without SGS diffusivity and viscosity, the oscillations of $R_{50}$ agree well at early times ($t \lesssim 4\,\rm h$) but diverge thereafter. In the absence of SGS terms, $R_{50}$ eventually settles to a slightly smaller value. The corresponding abundance maps are also similar to the fiducial case at $t \lesssim 4\,\rm h$, but become significantly noisier at later times. This behavior arises because, without SGS diffusivity, passive scalars are frozen in gas cells without inter-cell exchange. While large-scale splashing of material can mix gas with different $Y$, it produces sharp discontinuities between cells. In contrast, simulations with SGS diffusivity permit continuous mixing, resulting in smoother $Y$ distributions.

In the higher-resolution run that includes both SGS diffusivity and viscosity, the evolution of $R_{50}$ closely matches that of the fiducial run, and the abundance maps are qualitatively identical. So the damping in $R_{50}$ is due to physical reasons rather than numerical effects. One possibility is that the low-frequency oscillations transfer their energy to higher-frequency modes, likely through a turbulent cascade. These high-frequency modes may eventually reach the resolution limit and be damped by numerical viscosity; however, this regime lies beyond the scope of the present study.

{

\subsection{Chemical abundance profiles}

Fig.~\ref{fig:convergence_tests_chemical} further quantifies the mixing through the radial chemical abundance profiles, each corresponding to the inset abundance maps in Fig.~\ref{fig:convergence_tests}. To construct these radial profiles, we take the density peak of the merger remnant as the center. As shown by the abundance maps, the chemical distributions at $t=2.2\,\rm h$ and $4.4\,\rm h$ are highly non-spherical, resulting in large standard deviations. Nevertheless, both the average profiles and their standard deviations are very similar among the four test simulations, indicating that the SGS diffusivity and viscosity have little effect on the early stages of chemical mixing.

By $t=17\,\rm h$, however, significant differences emerge between the simulations with and without SGS diffusivity/viscosity. When SGS diffusivity is disabled (orange lines), the average $Y$ profile remains similar to those of the other runs, but the standard deviation becomes much larger ($\sigma_Y\sim0.2$ at the center) than in the other simulations ($\sigma_Y\sim0.02$ at the center), consistent with the visual comparison in Fig.~\ref{fig:convergence_tests}. Comparing the early and late stages of the evolution, we conclude that large-scale material splashing dominates the mixing initially, whereas turbulent diffusion becomes increasingly important after the large-scale oscillations begin to damp.

Comparing the fiducial setup with the SGS-diffusivity-only setup, we find only minor differences in the $Y$ profiles, and the overall evolution remains qualitatively unchanged. Likewise, simulations with the same fiducial physics but different numerical resolutions produce very similar $Y$ profiles. By $t=17\,\rm h$, the high-resolution simulation exhibits a smaller standard deviation in $Y$, as expected from its improved spatial resolution. This convergence study demonstrates that our fiducial resolution ($m_{\rm gas}=1/64^3\,M_\odot$) is sufficient to capture the radial abundance distribution of the merger remnant and that our results are numerically robust.
}


\bibliography{bh,star,lin}{}

@ARTICLE{TomarHopkinsKremer_2026PhRvD.113f3036T,
       author = {{Tomar}, Yashvardhan and {Hopkins}, Philip F. and {Kremer}, Kyle},
        title = "{Suppressed capture and merger rates in AGN}",
      journal = {\prd},
         year = 2026,
        month = mar,
       volume = {113},
       number = {6},
          eid = {063036},
        pages = {063036},
          doi = {10.1103/gcwt-rmml},
archivePrefix = {arXiv},
       eprint = {2601.02487},
 primaryClass = {astro-ph.HE},
       adsurl = {https://ui.adsabs.harvard.edu/abs/2026PhRvD.113f3036T}
}

@ARTICLE{ShiDaiMurray_2026ApJ...997..309S,
       author = {{Shi}, Yanlong and {Dai}, Liang and {Murray}, Norman and {Ye}, Claire S. and {Matzner}, Christopher D. and {Pascale}, Massimo},
        title = "{Very Massive Stars and High N/O: A Tale of the Nitrogen-enriched Super Star Cluster in the Sunburst Arc}",
      journal = {\apj},
         year = 2026,
        month = feb,
       volume = {997},
       number = {2},
          eid = {309},
        pages = {309},
          doi = {10.3847/1538-4357/ae3150},
archivePrefix = {arXiv},
       eprint = {2510.15823},
 primaryClass = {astro-ph.GA},
       adsurl = {https://ui.adsabs.harvard.edu/abs/2026ApJ...997..309S}
}

@ARTICLE{YuLai_2025ApJ...993...88Y,
       author = {{Yu}, Fangyuan and {Lai}, Dong},
        title = "{Binary Stars Approaching Supermassive Black Holes: Hydrodynamics of Stellar Collisions, Mass Fallback, and Partial Tidal Disruption Events}",
      journal = {\apj},
         year = 2025,
        month = nov,
       volume = {993},
       number = {1},
          eid = {88},
        pages = {88},
          doi = {10.3847/1538-4357/ae032b},
archivePrefix = {arXiv},
       eprint = {2504.14146},
 primaryClass = {astro-ph.HE},
       adsurl = {https://ui.adsabs.harvard.edu/abs/2025ApJ...993...88Y}
}

@ARTICLE{Schneider_2025arXiv250918421S,
       author = {{Schneider}, Fabian R.~N.},
        title = "{Theory, Simulations and Observations of Stellar Mergers}",
      journal = {arXiv e-prints},
         year = 2025,
        month = sep,
          eid = {arXiv:2509.18421},
        pages = {arXiv:2509.18421},
          doi = {10.48550/arXiv.2509.18421},
archivePrefix = {arXiv},
       eprint = {2509.18421},
 primaryClass = {astro-ph.SR},
       adsurl = {https://ui.adsabs.harvard.edu/abs/2025arXiv250918421S}
}

@ARTICLE{RantalaNaabLahen_2024MNRAS.531.3770R,
       author = {{Rantala}, Antti and {Naab}, Thorsten and {Lah{\'e}n}, Natalia},
        title = "{FROST-CLUSTERS - I. Hierarchical star cluster assembly boosts intermediate-mass black hole formation}",
      journal = {\mnras},
         year = 2024,
        month = jul,
       volume = {531},
       number = {3},
        pages = {3770-3799},
          doi = {10.1093/mnras/stae1413},
archivePrefix = {arXiv},
       eprint = {2403.10602},
 primaryClass = {astro-ph.GA},
       adsurl = {https://ui.adsabs.harvard.edu/abs/2024MNRAS.531.3770R}
}

@ARTICLE{HuangLinShields_2023MNRAS.525.5702H,
       author = {{Huang}, Jiamu and {Lin}, Douglas N.~C. and {Shields}, Gregory},
        title = "{Metal enrichment due to embedded stars in AGN discs}",
      journal = {\mnras},
         year = 2023,
        month = nov,
       volume = {525},
       number = {4},
        pages = {5702-5718},
          doi = {10.1093/mnras/stad2642},
archivePrefix = {arXiv},
       eprint = {2308.15761},
 primaryClass = {astro-ph.GA},
       adsurl = {https://ui.adsabs.harvard.edu/abs/2023MNRAS.525.5702H}
}

@ARTICLE{LaiMunoz_2023ARA&A..61..517L,
       author = {{Lai}, Dong and {Mu{\~n}oz}, Diego J.},
        title = "{Circumbinary Accretion: From Binary Stars to Massive Binary Black Holes}",
      journal = {\araa},
         year = 2023,
        month = aug,
       volume = {61},
        pages = {517-560},
          doi = {10.1146/annurev-astro-052622-022933},
archivePrefix = {arXiv},
       eprint = {2211.00028},
 primaryClass = {astro-ph.HE},
       adsurl = {https://ui.adsabs.harvard.edu/abs/2023ARA&A..61..517L}
}

@ARTICLE{DempseyLiMishra_2022ApJ...940..155D,
       author = {{Dempsey}, Adam M. and {Li}, Hui and {Mishra}, Bhupendra and {Li}, Shengtai},
        title = "{Contracting and Expanding Binary Black Holes in 3D Low-mass AGN Disks: The Importance of Separation}",
      journal = {\apj},
         year = 2022,
        month = dec,
       volume = {940},
       number = {2},
          eid = {155},
        pages = {155},
          doi = {10.3847/1538-4357/ac9d92},
archivePrefix = {arXiv},
       eprint = {2203.06534},
 primaryClass = {astro-ph.HE},
       adsurl = {https://ui.adsabs.harvard.edu/abs/2022ApJ...940..155D}
}

@ARTICLE{FordMcKernan_2022MNRAS.517.5827F,
       author = {{Ford}, K.~E. Saavik and {McKernan}, Barry},
        title = "{Binary black hole merger rates in AGN discs versus nuclear star clusters: loud beats quiet}",
      journal = {\mnras},
         year = 2022,
        month = dec,
       volume = {517},
       number = {4},
        pages = {5827-5834},
          doi = {10.1093/mnras/stac2861},
archivePrefix = {arXiv},
       eprint = {2109.03212},
 primaryClass = {astro-ph.HE},
       adsurl = {https://ui.adsabs.harvard.edu/abs/2022MNRAS.517.5827F}
}

@ARTICLE{Jiang_2022ApJS..263....4J,
       author = {{Jiang}, Yan-Fei},
        title = "{Multigroup Radiation Magnetohydrodynamics Based on Discrete Ordinates including Compton Scattering}",
      journal = {\apjs},
         year = 2022,
        month = nov,
       volume = {263},
       number = {1},
          eid = {4},
        pages = {4},
          doi = {10.3847/1538-4365/ac9231},
archivePrefix = {arXiv},
       eprint = {2209.06240},
 primaryClass = {astro-ph.IM},
       adsurl = {https://ui.adsabs.harvard.edu/abs/2022ApJS..263....4J}
}

@ARTICLE{DiCarloMapelliPasquato_2021MNRAS.507.5132D,
       author = {{Di Carlo}, Ugo N. and {Mapelli}, Michela and {Pasquato}, Mario and {Rastello}, Sara and {Ballone}, Alessandro and {Dall'Amico}, Marco and {Giacobbo}, Nicola and {Iorio}, Giuliano and {Spera}, Mario and {Torniamenti}, Stefano and et al.},
        title = "{Intermediate-mass black holes from stellar mergers in young star clusters}",
      journal = {\mnras},
         year = 2021,
        month = nov,
       volume = {507},
       number = {4},
        pages = {5132-5143},
          doi = {10.1093/mnras/stab2390},
archivePrefix = {arXiv},
       eprint = {2105.01085},
 primaryClass = {astro-ph.GA},
       adsurl = {https://ui.adsabs.harvard.edu/abs/2021MNRAS.507.5132D}
}

@ARTICLE{Rennehan_2021MNRAS.506.2836R,
       author = {{Rennehan}, Douglas},
        title = "{Mixing matters}",
      journal = {\mnras},
         year = 2021,
        month = sep,
       volume = {506},
       number = {2},
        pages = {2836-2852},
          doi = {10.1093/mnras/stab1813},
archivePrefix = {arXiv},
       eprint = {2104.07673},
 primaryClass = {astro-ph.GA},
       adsurl = {https://ui.adsabs.harvard.edu/abs/2021MNRAS.506.2836R}
}

@ARTICLE{ShiGrudicHopkins_2021MNRAS.505.2753S,
       author = {{Shi}, Yanlong and {Grudi{\'c}}, Michael Y. and {Hopkins}, Philip F.},
        title = "{The mass budget for intermediate-mass black holes in dense star clusters}",
      journal = {\mnras},
         year = 2021,
        month = aug,
       volume = {505},
       number = {2},
        pages = {2753-2763},
          doi = {10.1093/mnras/stab1470},
archivePrefix = {arXiv},
       eprint = {2008.12290},
 primaryClass = {astro-ph.GA},
       adsurl = {https://ui.adsabs.harvard.edu/abs/2021MNRAS.505.2753S}
}

@ARTICLE{KremerSperaBecker_2020ApJ...903...45K,
       author = {{Kremer}, Kyle and {Spera}, Mario and {Becker}, Devin and {Chatterjee}, Sourav and {Di Carlo}, Ugo N. and {Fragione}, Giacomo and {Rodriguez}, Carl L. and {Ye}, Claire S. and {Rasio}, Frederic A.},
        title = "{Populating the Upper Black Hole Mass Gap through Stellar Collisions in Young Star Clusters}",
      journal = {\apj},
         year = 2020,
        month = nov,
       volume = {903},
       number = {1},
          eid = {45},
        pages = {45},
          doi = {10.3847/1538-4357/abb945},
archivePrefix = {arXiv},
       eprint = {2006.10771},
 primaryClass = {astro-ph.HE},
       adsurl = {https://ui.adsabs.harvard.edu/abs/2020ApJ...903...45K}
}

@ARTICLE{TagawaHaimanKocsis_2020ApJ...898...25T,
       author = {{Tagawa}, Hiromichi and {Haiman}, Zolt{\'a}n and {Kocsis}, Bence},
        title = "{Formation and Evolution of Compact-object Binaries in AGN Disks}",
      journal = {\apj},
         year = 2020,
        month = jul,
       volume = {898},
       number = {1},
          eid = {25},
        pages = {25},
          doi = {10.3847/1538-4357/ab9b8c},
archivePrefix = {arXiv},
       eprint = {1912.08218},
 primaryClass = {astro-ph.GA},
       adsurl = {https://ui.adsabs.harvard.edu/abs/2020ApJ...898...25T}
}

@ARTICLE{StoneTomidaWhite_2020ApJS..249....4S,
       author = {{Stone}, James M. and {Tomida}, Kengo and {White}, Christopher J. and {Felker}, Kyle G.},
        title = "{The Athena++ Adaptive Mesh Refinement Framework: Design and Magnetohydrodynamic Solvers}",
      journal = {\apjs},
         year = 2020,
        month = jul,
       volume = {249},
       number = {1},
          eid = {4},
        pages = {4},
          doi = {10.3847/1538-4365/ab929b},
archivePrefix = {arXiv},
       eprint = {2005.06651},
 primaryClass = {astro-ph.IM},
       adsurl = {https://ui.adsabs.harvard.edu/abs/2020ApJS..249....4S}
}

@ARTICLE{RennehanBabulHopkins_2019MNRAS.483.3810R,
       author = {{Rennehan}, Douglas and {Babul}, Arif and {Hopkins}, Philip F. and {Dav{\'e}}, Romeel and {Moa}, Belaid},
        title = "{Dynamic localized turbulent diffusion and its impact on the galactic ecosystem}",
      journal = {\mnras},
         year = 2019,
        month = mar,
       volume = {483},
       number = {3},
        pages = {3810-3831},
          doi = {10.1093/mnras/sty3376},
archivePrefix = {arXiv},
       eprint = {1807.11509},
 primaryClass = {astro-ph.GA},
       adsurl = {https://ui.adsabs.harvard.edu/abs/2019MNRAS.483.3810R}
}

@ARTICLE{HopkinsWetzelKeres_2018MNRAS.480..800H,
       author = {{Hopkins}, Philip F. and {Wetzel}, Andrew and {Kere{\v{s}}}, Du{\v{s}}an and {Faucher-Gigu{\`e}re}, Claude-Andr{\'e} and {Quataert}, Eliot and {Boylan-Kolchin}, Michael and {Murray}, Norman and {Hayward}, Christopher C. and {Garrison-Kimmel}, Shea and {Hummels}, Cameron and et al.},
        title = "{FIRE-2 simulations: physics versus numerics in galaxy formation}",
      journal = {\mnras},
         year = 2018,
        month = oct,
       volume = {480},
       number = {1},
        pages = {800-863},
          doi = {10.1093/mnras/sty1690},
archivePrefix = {arXiv},
       eprint = {1702.06148},
 primaryClass = {astro-ph.GA},
       adsurl = {https://ui.adsabs.harvard.edu/abs/2018MNRAS.480..800H}
}

@ARTICLE{EscalaWetzelKirby_2018MNRAS.474.2194E,
       author = {{Escala}, Ivanna and {Wetzel}, Andrew and {Kirby}, Evan N. and {Hopkins}, Philip F. and {Ma}, Xiangcheng and {Wheeler}, Coral and {Kere{\v{s}}}, Du{\v{s}}an and {Faucher-Gigu{\`e}re}, Claude-Andr{\'e} and {Quataert}, Eliot},
        title = "{Modelling chemical abundance distributions for dwarf galaxies in the Local Group: the impact of turbulent metal diffusion}",
      journal = {\mnras},
         year = 2018,
        month = feb,
       volume = {474},
       number = {2},
        pages = {2194-2211},
          doi = {10.1093/mnras/stx2858},
archivePrefix = {arXiv},
       eprint = {1710.06533},
 primaryClass = {astro-ph.GA},
       adsurl = {https://ui.adsabs.harvard.edu/abs/2018MNRAS.474.2194E}
}

@ARTICLE{ColbrookMaHopkins_2017MNRAS.467.2421C,
       author = {{Colbrook}, Matthew J. and {Ma}, Xiangcheng and {Hopkins}, Philip F. and {Squire}, Jonathan},
        title = "{Scaling laws of passive-scalar diffusion in the interstellar medium}",
      journal = {\mnras},
         year = 2017,
        month = may,
       volume = {467},
       number = {2},
        pages = {2421-2429},
          doi = {10.1093/mnras/stx261},
archivePrefix = {arXiv},
       eprint = {1610.06590},
 primaryClass = {astro-ph.GA},
       adsurl = {https://ui.adsabs.harvard.edu/abs/2017MNRAS.467.2421C}
}

@ARTICLE{BartosKocsisHaiman_2017ApJ...835..165B,
       author = {{Bartos}, Imre and {Kocsis}, Bence and {Haiman}, Zolt{\'a}n and {M{\'a}rka}, Szabolcs},
        title = "{Rapid and Bright Stellar-mass Binary Black Hole Mergers in Active Galactic Nuclei}",
      journal = {\apj},
         year = 2017,
        month = feb,
       volume = {835},
       number = {2},
          eid = {165},
        pages = {165},
          doi = {10.3847/1538-4357/835/2/165},
archivePrefix = {arXiv},
       eprint = {1602.03831},
 primaryClass = {astro-ph.HE},
       adsurl = {https://ui.adsabs.harvard.edu/abs/2017ApJ...835..165B}
}

@ARTICLE{Mapelli_2016MNRAS.459.3432M,
       author = {{Mapelli}, Michela},
        title = "{Massive black hole binaries from runaway collisions: the impact of metallicity}",
      journal = {\mnras},
         year = 2016,
        month = jul,
       volume = {459},
       number = {4},
        pages = {3432-3446},
          doi = {10.1093/mnras/stw869},
archivePrefix = {arXiv},
       eprint = {1604.03559},
 primaryClass = {astro-ph.GA},
       adsurl = {https://ui.adsabs.harvard.edu/abs/2016MNRAS.459.3432M}
}

@ARTICLE{Hopkins_2015MNRAS.450...53H,
       author = {{Hopkins}, Philip F.},
        title = "{A new class of accurate, mesh-free hydrodynamic simulation methods}",
      journal = {\mnras},
         year = 2015,
        month = jun,
       volume = {450},
       number = {1},
        pages = {53-110},
          doi = {10.1093/mnras/stv195},
archivePrefix = {arXiv},
       eprint = {1409.7395},
 primaryClass = {astro-ph.CO},
       adsurl = {https://ui.adsabs.harvard.edu/abs/2015MNRAS.450...53H}
}

@ARTICLE{ShenWadsleyStinson_2010MNRAS.407.1581S,
       author = {{Shen}, S. and {Wadsley}, J. and {Stinson}, G.},
        title = "{The enrichment of the intergalactic medium with adiabatic feedback - I. Metal cooling and metal diffusion}",
      journal = {\mnras},
         year = 2010,
        month = sep,
       volume = {407},
       number = {3},
        pages = {1581-1596},
          doi = {10.1111/j.1365-2966.2010.17047.x},
archivePrefix = {arXiv},
       eprint = {0910.5956},
 primaryClass = {astro-ph.CO},
       adsurl = {https://ui.adsabs.harvard.edu/abs/2010MNRAS.407.1581S}
}

@ARTICLE{GurkanFreitagRasio_2004ApJ...604..632G,
       author = {{G{\"u}rkan}, M. Atakan and {Freitag}, Marc and {Rasio}, Frederic A.},
        title = "{Formation of Massive Black Holes in Dense Star Clusters. I. Mass Segregation and Core Collapse}",
      journal = {\apj},
         year = 2004,
        month = apr,
       volume = {604},
       number = {2},
        pages = {632-652},
          doi = {10.1086/381968},
archivePrefix = {arXiv},
       eprint = {astro-ph/0308449},
 primaryClass = {astro-ph},
       adsurl = {https://ui.adsabs.harvard.edu/abs/2004ApJ...604..632G}
}

@ARTICLE{Goodman_2003MNRAS.339..937G,
       author = {{Goodman}, Jeremy},
        title = "{Self-gravity and quasi-stellar object discs}",
      journal = {\mnras},
         year = 2003,
        month = mar,
       volume = {339},
       number = {4},
        pages = {937-948},
          doi = {10.1046/j.1365-8711.2003.06241.x},
archivePrefix = {arXiv},
       eprint = {astro-ph/0201001},
 primaryClass = {astro-ph},
       adsurl = {https://ui.adsabs.harvard.edu/abs/2003MNRAS.339..937G}
}

@ARTICLE{PortegiesZwartMcMillan_2002ApJ...576..899P,
       author = {{Portegies Zwart}, Simon F. and {McMillan}, Stephen L.~W.},
        title = "{The Runaway Growth of Intermediate-Mass Black Holes in Dense Star Clusters}",
      journal = {\apj},
         year = 2002,
        month = sep,
       volume = {576},
       number = {2},
        pages = {899-907},
          doi = {10.1086/341798},
archivePrefix = {arXiv},
       eprint = {astro-ph/0201055},
 primaryClass = {astro-ph},
       adsurl = {https://ui.adsabs.harvard.edu/abs/2002ApJ...576..899P}
}

@ARTICLE{ToutPolsEggleton_1996MNRAS.281..257T,
       author = {{Tout}, Christopher A. and {Pols}, Onno R. and {Eggleton}, Peter P. and {Han}, Zhanwen},
        title = "{Zero-age main-seqence radii and luminosities as analytic functions of mass and metallicity}",
      journal = {\mnras},
         year = 1996,
        month = jul,
       volume = {281},
       number = {1},
        pages = {257-262},
          doi = {10.1093/mnras/281.1.257},
       adsurl = {https://ui.adsabs.harvard.edu/abs/1996MNRAS.281..257T}
}

@ARTICLE{GermanoPiomelliMoin_1991PhFlA...3.1760G,
       author = {{Germano}, Massimo and {Piomelli}, Ugo and {Moin}, Parviz and {Cabot}, William H.},
        title = "{A dynamic subgrid-scale eddy viscosity model}",
      journal = {Physics of Fluids A},
         year = 1991,
        month = jul,
       volume = {3},
       number = {7},
        pages = {1760-1765},
          doi = {10.1063/1.857955},
       adsurl = {https://ui.adsabs.harvard.edu/abs/1991PhFlA...3.1760G}
}

@ARTICLE{Smagorinsky_1963MWRv...91...99S,
       author = {{Smagorinsky}, J.},
        title = "{General Circulation Experiments with the Primitive Equations}",
      journal = {Monthly Weather Review},
         year = 1963,
        month = jan,
       volume = {91},
       number = {3},
        pages = {99},
          doi = {10.1175/1520-0493(1963)091<0099:GCEWTP>2.3.CO;2},
       adsurl = {https://ui.adsabs.harvard.edu/abs/1963MWRv...91...99S}
}

@ARTICLE{ida2026,
       author = {{Ida}, Shigeru and {Li}, Ya-Ping and {Pan}, Jun-Peng and {Chen}, Yi-Xian and {Lin}, Douglas N.~C.},
        title = "{Outward Migration of a Gas Accreting Planet: A Semianalytical Formula}",
      journal = {\apj},
         year = 2026,
        month = feb,
       volume = {997},
       number = {2},
          eid = {160},
        pages = {160},
          doi = {10.3847/1538-4357/ae2612},
archivePrefix = {arXiv},
       eprint = {2512.00304},
 primaryClass = {astro-ph.EP},
       adsurl = {https://ui.adsabs.harvard.edu/abs/2026ApJ...997..160I}
}

@ARTICLE{wu2024,
       author = {{Wu}, Yinhao and {Chen}, Yi-Xian and {Lin}, Douglas N.~C.},
        title = "{Chaotic Type I migration in turbulent discs}",
      journal = {\mnras},
         year = 2024,
        month = feb,
       volume = {528},
       number = {1},
        pages = {L127-L132},
          doi = {10.1093/mnrasl/slad183},
archivePrefix = {arXiv},
       eprint = {2311.15747},
 primaryClass = {astro-ph.EP},
       adsurl = {https://ui.adsabs.harvard.edu/abs/2024MNRAS.528L.127W}
}

@ARTICLE{kleynelson2012,
       author = {{Kley}, W. and {Nelson}, R.~P.},
        title = "{Planet-Disk Interaction and Orbital Evolution}",
      journal = {\araa},
         year = 2012,
        month = sep,
       volume = {50},
        pages = {211-249},
          doi = {10.1146/annurev-astro-081811-125523},
archivePrefix = {arXiv},
       eprint = {1203.1184},
 primaryClass = {astro-ph.EP},
       adsurl = {https://ui.adsabs.harvard.edu/abs/2012ARA&A..50..211K}
}

@ARTICLE{linpap1986,
       author = {{Lin}, D.~N.~C. and {Papaloizou}, John},
        title = "{On the Tidal Interaction between Protoplanets and the Protoplanetary Disk. III. Orbital Migration of Protoplanets}",
      journal = {\apj},
         year = 1986,
        month = oct,
       volume = {309},
        pages = {846},
          doi = {10.1086/164653},
       adsurl = {https://ui.adsabs.harvard.edu/abs/1986ApJ...309..846L}
}

@ARTICLE{ward1997,
       author = {{Ward}, William R.},
        title = "{Protoplanet Migration by Nebula Tides}",
      journal = {\icarus},
         year = 1997,
        month = apr,
       volume = {126},
       number = {2},
        pages = {261-281},
          doi = {10.1006/icar.1996.5647},
       adsurl = {https://ui.adsabs.harvard.edu/abs/1997Icar..126..261W}
}

@ARTICLE{paardekooper2011,
       author = {{Paardekooper}, S.-J. and {Baruteau}, C. and {Kley}, W.},
        title = "{A torque formula for non-isothermal Type I planetary migration - II. Effects of diffusion}",
      journal = {\mnras},
         year = 2011,
        month = jan,
       volume = {410},
       number = {1},
        pages = {293-303},
          doi = {10.1111/j.1365-2966.2010.17442.x},
archivePrefix = {arXiv},
       eprint = {1007.4964},
 primaryClass = {astro-ph.EP},
       adsurl = {https://ui.adsabs.harvard.edu/abs/2011MNRAS.410..293P}
}

@ARTICLE{artymowicz1993,
       author = {{Artymowicz}, Pawel and {Lin}, D.~N.~C. and {Wampler}, E.~J.},
        title = "{Star Trapping and Metallicity Enrichment in Quasars and Active Galactic Nuclei}",
      journal = {\apj},
         year = 1993,
        month = jun,
       volume = {409},
        pages = {592},
          doi = {10.1086/172690},
       adsurl = {https://ui.adsabs.harvard.edu/abs/1993ApJ...409..592A}
}

@ARTICLE{davies2020,
       author = {{Davies}, Melvyn B. and {Lin}, Doug N.~C.},
        title = "{Making massive stars in the Galactic Centre via accretion on to low-mass stars within an accretion disc}",
      journal = {\mnras},
         year = 2020,
        month = nov,
       volume = {498},
       number = {3},
        pages = {3452-3456},
          doi = {10.1093/mnras/staa2590},
archivePrefix = {arXiv},
       eprint = {2008.10033},
 primaryClass = {astro-ph.GA},
       adsurl = {https://ui.adsabs.harvard.edu/abs/2020MNRAS.498.3452D}
}

@ARTICLE{owocki2004,
       author = {{Owocki}, Stanley P. and {Gayley}, Kenneth G. and {Shaviv}, Nir J.},
        title = "{A Porosity-Length Formalism for Photon-Tiring-limited Mass Loss from Stars above the Eddington Limit}",
      journal = {\apj},
         year = 2004,
        month = nov,
       volume = {616},
       number = {1},
        pages = {525-541},
          doi = {10.1086/424910},
archivePrefix = {arXiv},
       eprint = {astro-ph/0409573},
 primaryClass = {astro-ph},
       adsurl = {https://ui.adsabs.harvard.edu/abs/2004ApJ...616..525O}
}

@ARTICLE{GonzalezPrietoLombardiRose_2026ApJ..1005..131G,
       author = {{Gonz{\'a}lez Prieto}, Elena and {Lombardi}, Jr., James C. and {Rose}, Sanaea C. and {Gibson}, Charles F.~A. and {O'Connor}, Christopher E. and {Starkenburg}, Tjitske and {K{\i}ro{\u{g}}lu}, Fulya and {Kremer}, Kyle and {Sand}, Tristan C. and {Rasio}, Frederic A.},
        title = "{Machine Learning Methods for Stellar Collisions. I. Predicting Outcomes of SPH Simulations}",
      journal = {\apj},
         year = 2026,
        month = jul,
       volume = {1005},
       number = {2},
          eid = {131},
        pages = {131},
          doi = {10.3847/1538-4357/ae6585},
archivePrefix = {arXiv},
       eprint = {2602.10191},
 primaryClass = {astro-ph.HE},
       adsurl = {https://ui.adsabs.harvard.edu/abs/2026ApJ..1005..131G}
}

@ARTICLE{RoseLombardiGonzalezPrieto_2026ApJ..1000..162R,
       author = {{Rose}, Sanaea C. and {Lombardi}, Jr., James C. and {Gonz{\'a}lez Prieto}, Elena and {K{\i}ro{\u{g}}lu}, Fulya and {Rasio}, Frederic A.},
        title = "{Modeling Stellar Collisions in Galactic Nuclei Using Hydrodynamic Simulations and Machine Learning}",
      journal = {\apj},
         year = 2026,
        month = apr,
       volume = {1000},
       number = {2},
          eid = {162},
        pages = {162},
          doi = {10.3847/1538-4357/ae459b},
archivePrefix = {arXiv},
       eprint = {2511.01811},
 primaryClass = {astro-ph.GA},
       adsurl = {https://ui.adsabs.harvard.edu/abs/2026ApJ..1000..162R}
}

@ARTICLE{VynatheyaRyuWang_2026ApJ...999...64V,
       author = {{Vynatheya}, Pavan and {Ryu}, Taeho and {Wang}, Chen and {Sills}, Alison and {Pakmor}, R{\"u}diger},
        title = "{The Collision and Merger Products of Stars Do Not Look Alike: A Magnetohydrodynamics Comparison}",
      journal = {\apj},
         year = 2026,
        month = mar,
       volume = {999},
       number = {1},
          eid = {64},
        pages = {64},
          doi = {10.3847/1538-4357/ae40ba},
archivePrefix = {arXiv},
       eprint = {2510.13736},
 primaryClass = {astro-ph.SR},
       adsurl = {https://ui.adsabs.harvard.edu/abs/2026ApJ...999...64V}
}

@ARTICLE{Roman-GarzaFragosCharbonnel_2026A&A...707A.163R,
       author = {{Roman-Garza}, J. and {Fragos}, T. and {Charbonnel}, C. and {Ram{\'\i}rez-Galeano}, L. and {Kruckow}, M. and {Farag}, E.},
        title = "{Massive stellar cannibals: How stellar mergers drive mass loss in extremely massive stars}",
      journal = {\aap},
         year = 2026,
        month = mar,
       volume = {707},
          eid = {A163},
        pages = {A163},
          doi = {10.1051/0004-6361/202557899},
archivePrefix = {arXiv},
       eprint = {2602.02141},
 primaryClass = {astro-ph.SR},
       adsurl = {https://ui.adsabs.harvard.edu/abs/2026A&A...707A.163R}
}

@ARTICLE{XuChenLin_2026ApJ...997..206X,
       author = {{Xu}, Zheng-Hao and {Chen}, Yi-Xian and {Lin}, Douglas N.~C.},
        title = "{Stellar Evolution with Radiative Feedback in AGN Disks}",
      journal = {\apj},
         year = 2026,
        month = feb,
       volume = {997},
       number = {2},
          eid = {206},
        pages = {206},
          doi = {10.3847/1538-4357/ae2271},
archivePrefix = {arXiv},
       eprint = {2511.03904},
 primaryClass = {astro-ph.GA},
       adsurl = {https://ui.adsabs.harvard.edu/abs/2026ApJ...997..206X}
}

@ARTICLE{Xu_2025RAA....25k5013X,
       author = {{Xu}, Zheng-Hao},
        title = "{The Fate of Stars Embedded in AGN Disks is Determined by an Internal Mixing Threshold}",
      journal = {Research in Astronomy and Astrophysics},
         year = 2025,
        month = nov,
       volume = {25},
       number = {11},
          eid = {115013},
        pages = {115013},
          doi = {10.1088/1674-4527/adfeb9},
       adsurl = {https://ui.adsabs.harvard.edu/abs/2025RAA....25k5013X}
}

@ARTICLE{HatfullIvanova_2025ApJ...982...83H,
       author = {{Hatfull}, Roger W.~M. and {Ivanova}, Natalia},
        title = "{Simulating a Stellar Binary Merger. II. Obtaining a Light Curve}",
      journal = {\apj},
         year = 2025,
        month = apr,
       volume = {982},
       number = {2},
          eid = {83},
        pages = {83},
          doi = {10.3847/1538-4357/ada6b8},
archivePrefix = {arXiv},
       eprint = {2412.06583},
 primaryClass = {astro-ph.SR},
       adsurl = {https://ui.adsabs.harvard.edu/abs/2025ApJ...982...83H}
}

@ARTICLE{FabjDittmannCantiello_2025ApJ...981...16F,
       author = {{Fabj}, Gaia and {Dittmann}, Alexander J. and {Cantiello}, Matteo and {Perna}, Rosalba and {Samsing}, Johan},
        title = "{Mapping the Outcomes of Stellar Evolution in the Disks of Active Galactic Nuclei}",
      journal = {\apj},
         year = 2025,
        month = mar,
       volume = {981},
       number = {1},
          eid = {16},
        pages = {16},
          doi = {10.3847/1538-4357/ada896},
archivePrefix = {arXiv},
       eprint = {2408.16050},
 primaryClass = {astro-ph.GA},
       adsurl = {https://ui.adsabs.harvard.edu/abs/2025ApJ...981...16F}
}

@ARTICLE{RyuSillsPakmor_2025ApJ...980L..38R,
       author = {{Ryu}, Taeho and {Sills}, Alison and {Pakmor}, Ruediger and {de Mink}, Selma and {Mathieu}, Robert},
        title = "{Magnetic Field Amplification during Stellar Collisions between Low-mass Stars}",
      journal = {\apjl},
         year = 2025,
        month = feb,
       volume = {980},
       number = {2},
          eid = {L38},
        pages = {L38},
          doi = {10.3847/2041-8213/adaf94},
archivePrefix = {arXiv},
       eprint = {2410.00148},
 primaryClass = {astro-ph.SR},
       adsurl = {https://ui.adsabs.harvard.edu/abs/2025ApJ...980L..38R}
}

@ARTICLE{FryerHuangAli-Dib_2025MNRAS.537.1556F,
       author = {{Fryer}, Chris L. and {Huang}, Jiamu and {Ali-Dib}, Mohamad and {Andrews}, Amaya and {Xu}, Zhenghao and {Lin}, Douglas N.~C.},
        title = "{Stellar population and metal production in AGN discs}",
      journal = {\mnras},
         year = 2025,
        month = feb,
       volume = {537},
       number = {2},
        pages = {1556-1570},
          doi = {10.1093/mnras/staf130},
archivePrefix = {arXiv},
       eprint = {2501.06973},
 primaryClass = {astro-ph.HE},
       adsurl = {https://ui.adsabs.harvard.edu/abs/2025MNRAS.537.1556F}
}

@ARTICLE{DittmannCantiello_2025ApJ...979..245D,
       author = {{Dittmann}, Alexander J. and {Cantiello}, Matteo},
        title = "{A Semi-analytical Model for Stellar Evolution in AGN Disks}",
      journal = {\apj},
         year = 2025,
        month = feb,
       volume = {979},
       number = {2},
          eid = {245},
        pages = {245},
          doi = {10.3847/1538-4357/ad9e92},
archivePrefix = {arXiv},
       eprint = {2409.02981},
 primaryClass = {astro-ph.GA},
       adsurl = {https://ui.adsabs.harvard.edu/abs/2025ApJ...979..245D}
}

@ARTICLE{ChenJiangGoodman_2024ApJ...974..106C,
       author = {{Chen}, Yi-Xian and {Jiang}, Yan-Fei and {Goodman}, Jeremy and {Lin}, Douglas N.~C.},
        title = "{Radiation Hydrodynamic Simulations of Massive Stars in Gas-rich Environments: Accretion of AGN Stars Suppressed by Thermal Feedback}",
      journal = {\apj},
         year = 2024,
        month = oct,
       volume = {974},
       number = {1},
          eid = {106},
        pages = {106},
          doi = {10.3847/1538-4357/ad6dd4},
archivePrefix = {arXiv},
       eprint = {2408.12017},
 primaryClass = {astro-ph.HE},
       adsurl = {https://ui.adsabs.harvard.edu/abs/2024ApJ...974..106C}
}

@ARTICLE{ChenLin_2024ApJ...967...88C,
       author = {{Chen}, Yi-Xian and {Lin}, Douglas N.~C.},
        title = "{The Population of Massive Stars in Active Galactic Nuclei Disks}",
      journal = {\apj},
         year = 2024,
        month = jun,
       volume = {967},
       number = {2},
          eid = {88},
        pages = {88},
          doi = {10.3847/1538-4357/ad3c3a},
archivePrefix = {arXiv},
       eprint = {2404.08780},
 primaryClass = {astro-ph.GA},
       adsurl = {https://ui.adsabs.harvard.edu/abs/2024ApJ...967...88C}
}

@ARTICLE{RyuAmaroSeoaneTaylor_2024MNRAS.528.6193R,
       author = {{Ryu}, Taeho and {Amaro Seoane}, Pau and {Taylor}, Andrew M. and {Ohlmann}, Sebastian T.},
        title = "{Collisions of red giants in galactic nuclei}",
      journal = {\mnras},
         year = 2024,
        month = mar,
       volume = {528},
       number = {4},
        pages = {6193-6209},
          doi = {10.1093/mnras/stae396},
archivePrefix = {arXiv},
       eprint = {2307.07338},
 primaryClass = {astro-ph.HE},
       adsurl = {https://ui.adsabs.harvard.edu/abs/2024MNRAS.528.6193R}
}

@ARTICLE{HennecoSchneiderLaplace_2024A&A...682A.169H,
       author = {{Henneco}, J. and {Schneider}, F.~R.~N. and {Laplace}, E.},
        title = "{Contact tracing of binary stars: Pathways to stellar mergers}",
      journal = {\aap},
         year = 2024,
        month = feb,
       volume = {682},
          eid = {A169},
        pages = {A169},
          doi = {10.1051/0004-6361/202347893},
archivePrefix = {arXiv},
       eprint = {2311.12124},
 primaryClass = {astro-ph.SR},
       adsurl = {https://ui.adsabs.harvard.edu/abs/2024A&A...682A.169H}
}

@ARTICLE{Ali-DibLin_2023MNRAS.526.5824A,
       author = {{Ali-Dib}, Mohamad and {Lin}, Douglas N.~C.},
        title = "{The impermanent fate of massive stars in AGN discs}",
      journal = {\mnras},
         year = 2023,
        month = dec,
       volume = {526},
       number = {4},
        pages = {5824-5838},
          doi = {10.1093/mnras/stad2774},
archivePrefix = {arXiv},
       eprint = {2309.04392},
 primaryClass = {astro-ph.GA},
       adsurl = {https://ui.adsabs.harvard.edu/abs/2023MNRAS.526.5824A}
}

@ARTICLE{JermynBauerSchwab_2023ApJS..265...15J,
       author = {{Jermyn}, Adam S. and {Bauer}, Evan B. and {Schwab}, Josiah and {Farmer}, R. and {Ball}, Warrick H. and {Bellinger}, Earl P. and {Dotter}, Aaron and {Joyce}, Meridith and {Marchant}, Pablo and {Mombarg}, Joey S.~G. and et al.},
        title = "{Modules for Experiments in Stellar Astrophysics (MESA): Time-dependent Convection, Energy Conservation, Automatic Differentiation, and Infrastructure}",
      journal = {\apjs},
         year = 2023,
        month = mar,
       volume = {265},
       number = {1},
          eid = {15},
        pages = {15},
          doi = {10.3847/1538-4365/acae8d},
archivePrefix = {arXiv},
       eprint = {2208.03651},
 primaryClass = {astro-ph.SR},
       adsurl = {https://ui.adsabs.harvard.edu/abs/2023ApJS..265...15J}
}

@ARTICLE{JermynDittmannMcKernan_2022ApJ...929..133J,
       author = {{Jermyn}, Adam S. and {Dittmann}, Alexander J. and {McKernan}, B. and {Ford}, K.~E.~S. and {Cantiello}, Matteo},
        title = "{Effects of an Immortal Stellar Population in AGN Disks}",
      journal = {\apj},
         year = 2022,
        month = apr,
       volume = {929},
       number = {2},
          eid = {133},
        pages = {133},
          doi = {10.3847/1538-4357/ac5d40},
archivePrefix = {arXiv},
       eprint = {2203.06187},
 primaryClass = {astro-ph.GA},
       adsurl = {https://ui.adsabs.harvard.edu/abs/2022ApJ...929..133J}
}

@ARTICLE{HatfullIvanovaLombardi_2021MNRAS.507..385H,
       author = {{Hatfull}, Roger W.~M. and {Ivanova}, Natalia and {Lombardi}, James C.},
        title = "{Simulating a stellar contact binary merger - I. Stellar models}",
      journal = {\mnras},
         year = 2021,
        month = oct,
       volume = {507},
       number = {1},
        pages = {385-397},
          doi = {10.1093/mnras/stab2140},
archivePrefix = {arXiv},
       eprint = {2107.11480},
 primaryClass = {astro-ph.SR},
       adsurl = {https://ui.adsabs.harvard.edu/abs/2021MNRAS.507..385H}
}

@ARTICLE{DittmannCantielloJermyn_2021ApJ...916...48D,
       author = {{Dittmann}, Alexander J. and {Cantiello}, Matteo and {Jermyn}, Adam S.},
        title = "{Accretion onto Stars in the Disks of Active Galactic Nuclei}",
      journal = {\apj},
         year = 2021,
        month = jul,
       volume = {916},
       number = {1},
          eid = {48},
        pages = {48},
          doi = {10.3847/1538-4357/ac042c},
archivePrefix = {arXiv},
       eprint = {2102.12484},
 primaryClass = {astro-ph.GA},
       adsurl = {https://ui.adsabs.harvard.edu/abs/2021ApJ...916...48D}
}

@ARTICLE{JermynSchwabBauer_2021ApJ...913...72J,
       author = {{Jermyn}, Adam S. and {Schwab}, Josiah and {Bauer}, Evan and {Timmes}, F.~X. and {Potekhin}, Alexander Y.},
        title = "{Skye: A Differentiable Equation of State}",
      journal = {\apj},
         year = 2021,
        month = may,
       volume = {913},
       number = {1},
          eid = {72},
        pages = {72},
          doi = {10.3847/1538-4357/abf48e},
archivePrefix = {arXiv},
       eprint = {2104.00691},
 primaryClass = {astro-ph.SR},
       adsurl = {https://ui.adsabs.harvard.edu/abs/2021ApJ...913...72J}
}

@ARTICLE{CantielloJermynLin_2021ApJ...910...94C,
       author = {{Cantiello}, Matteo and {Jermyn}, Adam S. and {Lin}, Douglas N.~C.},
        title = "{Stellar Evolution in AGN Disks}",
      journal = {\apj},
         year = 2021,
        month = apr,
       volume = {910},
       number = {2},
          eid = {94},
        pages = {94},
          doi = {10.3847/1538-4357/abdf4f},
archivePrefix = {arXiv},
       eprint = {2009.03936},
 primaryClass = {astro-ph.SR},
       adsurl = {https://ui.adsabs.harvard.edu/abs/2021ApJ...910...94C}
}

@ARTICLE{SchneiderOhlmannPodsiadlowski_2019Natur.574..211S,
       author = {{Schneider}, Fabian R.~N. and {Ohlmann}, Sebastian T. and {Podsiadlowski}, Philipp and {R{\"o}pke}, Friedrich K. and {Balbus}, Steven A. and {Pakmor}, R{\"u}diger and {Springel}, Volker},
        title = "{Stellar mergers as the origin of magnetic massive stars}",
      journal = {\nat},
         year = 2019,
        month = oct,
       volume = {574},
       number = {7777},
        pages = {211-214},
          doi = {10.1038/s41586-019-1621-5},
archivePrefix = {arXiv},
       eprint = {1910.14058},
 primaryClass = {astro-ph.SR},
       adsurl = {https://ui.adsabs.harvard.edu/abs/2019Natur.574..211S}
}

@ARTICLE{OwockiTownsendQuataert_2017MNRAS.472.3749O,
       author = {{Owocki}, Stanley P. and {Townsend}, Richard H.~D. and {Quataert}, Eliot},
        title = "{Super-Eddington stellar winds: unifying radiative-enthalpy versus flux-driven models}",
      journal = {\mnras},
         year = 2017,
        month = dec,
       volume = {472},
       number = {3},
        pages = {3749-3760},
          doi = {10.1093/mnras/stx2251},
archivePrefix = {arXiv},
       eprint = {1708.07790},
 primaryClass = {astro-ph.SR},
       adsurl = {https://ui.adsabs.harvard.edu/abs/2017MNRAS.472.3749O}
}

@ARTICLE{OhlmannRopkePakmor_2017A&A...599A...5O,
       author = {{Ohlmann}, Sebastian T. and {R{\"o}pke}, Friedrich K. and {Pakmor}, R{\"u}diger and {Springel}, Volker},
        title = "{Constructing stable 3D hydrodynamical models of giant stars}",
      journal = {\aap},
         year = 2017,
        month = mar,
       volume = {599},
          eid = {A5},
        pages = {A5},
          doi = {10.1051/0004-6361/201629692},
archivePrefix = {arXiv},
       eprint = {1612.00008},
 primaryClass = {astro-ph.SR},
       adsurl = {https://ui.adsabs.harvard.edu/abs/2017A&A...599A...5O}
}

@ARTICLE{SchneiderPodsiadlowskiLanger_2016MNRAS.457.2355S,
       author = {{Schneider}, F.~R.~N. and {Podsiadlowski}, Ph. and {Langer}, N. and {Castro}, N. and {Fossati}, L.},
        title = "{Rejuvenation of stellar mergers and the origin of magnetic fields in massive stars}",
      journal = {\mnras},
         year = 2016,
        month = apr,
       volume = {457},
       number = {3},
        pages = {2355-2365},
          doi = {10.1093/mnras/stw148},
archivePrefix = {arXiv},
       eprint = {1601.05084},
 primaryClass = {astro-ph.SR},
       adsurl = {https://ui.adsabs.harvard.edu/abs/2016MNRAS.457.2355S}
}

@ARTICLE{GlebbeekGaburovPortegiesZwart_2013MNRAS.434.3497G,
       author = {{Glebbeek}, Evert and {Gaburov}, Evghenii and {Portegies Zwart}, Simon and {Pols}, Onno R.},
        title = "{Structure and evolution of high-mass stellar mergers}",
      journal = {\mnras},
         year = 2013,
        month = oct,
       volume = {434},
       number = {4},
        pages = {3497-3510},
          doi = {10.1093/mnras/stt1268},
archivePrefix = {arXiv},
       eprint = {1307.2445},
 primaryClass = {astro-ph.SR},
       adsurl = {https://ui.adsabs.harvard.edu/abs/2013MNRAS.434.3497G}
}

@ARTICLE{deMinkLangerIzzard_2013ApJ...764..166D,
       author = {{de Mink}, S.~E. and {Langer}, N. and {Izzard}, R.~G. and {Sana}, H. and {de Koter}, A.},
        title = "{The Rotation Rates of Massive Stars: The Role of Binary Interaction through Tides, Mass Transfer, and Mergers}",
      journal = {\apj},
         year = 2013,
        month = feb,
       volume = {764},
       number = {2},
          eid = {166},
        pages = {166},
          doi = {10.1088/0004-637X/764/2/166},
archivePrefix = {arXiv},
       eprint = {1211.3742},
 primaryClass = {astro-ph.SR},
       adsurl = {https://ui.adsabs.harvard.edu/abs/2013ApJ...764..166D}
}

@BOOK{KippenhahnWeigertWeiss_2013sse..book.....K,
       author = {{Kippenhahn}, Rudolf and {Weigert}, Alfred and {Weiss}, Achim},
        title = "{Stellar Structure and Evolution}",
         year = 2013,
          doi = {10.1007/978-3-642-30304-3},
       adsurl = {https://ui.adsabs.harvard.edu/abs/2013sse..book.....K}
}

@software{Irwin_2012ascl.soft11002I,
       author = {{Irwin}, Alan W.},
        title = "{FreeEOS: Equation of State for stellar interiors calculations}",
 howpublished = {Astrophysics Source Code Library, record ascl:1211.002},
         year = 2012,
        month = nov,
          eid = {ascl:1211.002},
archivePrefix = {ascl},
       eprint = {1211.002},
       adsurl = {https://ui.adsabs.harvard.edu/abs/2012ascl.soft11002I}
}

@ARTICLE{BrottdeMinkCantiello_2011A&A...530A.115B,
       author = {{Brott}, I. and {de Mink}, S.~E. and {Cantiello}, M. and {Langer}, N. and {de Koter}, A. and {Evans}, C.~J. and {Hunter}, I. and {Trundle}, C. and {Vink}, J.~S.},
        title = "{Rotating massive main-sequence stars. I. Grids of evolutionary models and isochrones}",
      journal = {\aap},
         year = 2011,
        month = jun,
       volume = {530},
          eid = {A115},
        pages = {A115},
          doi = {10.1051/0004-6361/201016113},
archivePrefix = {arXiv},
       eprint = {1102.0530},
 primaryClass = {astro-ph.SR},
       adsurl = {https://ui.adsabs.harvard.edu/abs/2011A&A...530A.115B}
}

@ARTICLE{MattilaLundqvistGroningsson_2010ApJ...717.1140M,
       author = {{Mattila}, Seppo and {Lundqvist}, Peter and {Gr{\"o}ningsson}, Per and {Meikle}, Peter and {Stathakis}, Raylee and {Fransson}, Claes and {Cannon}, Russell},
        title = "{Abundances and Density Structure of the Inner Circumstellar Ring Around SN 1987A}",
      journal = {\apj},
         year = 2010,
        month = jul,
       volume = {717},
       number = {2},
        pages = {1140-1156},
          doi = {10.1088/0004-637X/717/2/1140},
archivePrefix = {arXiv},
       eprint = {1002.4195},
 primaryClass = {astro-ph.SR},
       adsurl = {https://ui.adsabs.harvard.edu/abs/2010ApJ...717.1140M}
}

@ARTICLE{GlebbeekPolsHurley_2008A&A...488.1007G,
       author = {{Glebbeek}, E. and {Pols}, O.~R. and {Hurley}, J.~R.},
        title = "{Evolution of stellar collision products in open clusters. I. Blue stragglers in N-body models of M 67}",
      journal = {\aap},
         year = 2008,
        month = sep,
       volume = {488},
       number = {3},
        pages = {1007-1015},
          doi = {10.1051/0004-6361:200809930},
archivePrefix = {arXiv},
       eprint = {0806.0863},
 primaryClass = {astro-ph},
       adsurl = {https://ui.adsabs.harvard.edu/abs/2008A&A...488.1007G}
}

@ARTICLE{GlebbeekPols_2008A&A...488.1017G,
       author = {{Glebbeek}, E. and {Pols}, O.~R.},
        title = "{Evolution of stellar collision products in open clusters. II. A grid of low-mass collisions}",
      journal = {\aap},
         year = 2008,
        month = sep,
       volume = {488},
       number = {3},
        pages = {1017-1025},
          doi = {10.1051/0004-6361:200809931},
archivePrefix = {arXiv},
       eprint = {0806.0865},
 primaryClass = {astro-ph},
       adsurl = {https://ui.adsabs.harvard.edu/abs/2008A&A...488.1017G}
}

@ARTICLE{GaburovLombardiPortegiesZwart_2008MNRAS.383L...5G,
       author = {{Gaburov}, E. and {Lombardi}, J.~C. and {Portegies Zwart}, S.},
        title = "{Mixing in massive stellar mergers}",
      journal = {\mnras},
         year = 2008,
        month = jan,
       volume = {383},
       number = {1},
        pages = {L5-L9},
          doi = {10.1111/j.1745-3933.2007.00399.x},
archivePrefix = {arXiv},
       eprint = {0707.3021},
 primaryClass = {astro-ph},
       adsurl = {https://ui.adsabs.harvard.edu/abs/2008MNRAS.383L...5G}
}

@ARTICLE{CollinZahn_2008A&A...477..419C,
       author = {{Collin}, S. and {Zahn}, J.-P.},
        title = "{Star formation in accretion discs: from the Galactic center to active galactic nuclei}",
      journal = {\aap},
         year = 2008,
        month = jan,
       volume = {477},
       number = {2},
        pages = {419-435},
          doi = {10.1051/0004-6361:20078191},
archivePrefix = {arXiv},
       eprint = {0709.3772},
 primaryClass = {astro-ph},
       adsurl = {https://ui.adsabs.harvard.edu/abs/2008A&A...477..419C}
}

@ARTICLE{SillsAdamsDavies_2005MNRAS.358..716S,
       author = {{Sills}, Alison and {Adams}, Tim and {Davies}, Melvyn B.},
        title = "{Blue stragglers as stellar collision products: the angular momentum question}",
      journal = {\mnras},
         year = 2005,
        month = apr,
       volume = {358},
       number = {3},
        pages = {716-725},
          doi = {10.1111/j.1365-2966.2005.08809.x},
archivePrefix = {arXiv},
       eprint = {astro-ph/0501142},
 primaryClass = {astro-ph},
       adsurl = {https://ui.adsabs.harvard.edu/abs/2005MNRAS.358..716S}
}

@ARTICLE{KlessenLin_2003PhRvE..67d6311K,
       author = {{Klessen}, Ralf S. and {Lin}, Douglas N.},
        title = "{Diffusion in supersonic turbulent compressible flows}",
      journal = {\pre},
         year = 2003,
        month = apr,
       volume = {67},
       number = {4},
          eid = {046311},
        pages = {046311},
          doi = {10.1103/PhysRevE.67.046311},
archivePrefix = {arXiv},
       eprint = {astro-ph/0302527},
 primaryClass = {astro-ph},
       adsurl = {https://ui.adsabs.harvard.edu/abs/2003PhRvE..67d6311K}
}

@ARTICLE{RogersNayfonov_2002ApJ...576.1064R,
       author = {{Rogers}, F.~J. and {Nayfonov}, A.},
        title = "{Updated and Expanded OPAL Equation-of-State Tables: Implications for Helioseismology}",
      journal = {\apj},
         year = 2002,
        month = sep,
       volume = {576},
       number = {2},
        pages = {1064-1074},
          doi = {10.1086/341894},
       adsurl = {https://ui.adsabs.harvard.edu/abs/2002ApJ...576.1064R}
}

@ARTICLE{LombardiWarrenRasio_2002ApJ...568..939L,
       author = {{Lombardi}, Jr., James C. and {Warren}, Jessica S. and {Rasio}, Frederic A. and {Sills}, Alison and {Warren}, Aaron R.},
        title = "{Stellar Collisions and the Interior Structure of Blue Stragglers}",
      journal = {\apj},
         year = 2002,
        month = apr,
       volume = {568},
       number = {2},
        pages = {939-953},
          doi = {10.1086/339060},
archivePrefix = {arXiv},
       eprint = {astro-ph/0107388},
 primaryClass = {astro-ph},
       adsurl = {https://ui.adsabs.harvard.edu/abs/2002ApJ...568..939L}
}

@ARTICLE{Spruit_2002A&A...381..923S,
       author = {{Spruit}, H.~C.},
        title = "{Dynamo action by differential rotation in a stably stratified stellar interior}",
      journal = {\aap},
         year = 2002,
        month = jan,
       volume = {381},
        pages = {923-932},
          doi = {10.1051/0004-6361:20011465},
archivePrefix = {arXiv},
       eprint = {astro-ph/0108207},
 primaryClass = {astro-ph},
       adsurl = {https://ui.adsabs.harvard.edu/abs/2002A&A...381..923S}
}

@ARTICLE{WellsteinLangerBraun_2001A&A...369..939W,
       author = {{Wellstein}, S. and {Langer}, N. and {Braun}, H.},
        title = "{Formation of contact in massive close binaries}",
      journal = {\aap},
         year = 2001,
        month = apr,
       volume = {369},
        pages = {939-959},
          doi = {10.1051/0004-6361:20010151},
archivePrefix = {arXiv},
       eprint = {astro-ph/0102244},
 primaryClass = {astro-ph},
       adsurl = {https://ui.adsabs.harvard.edu/abs/2001A&A...369..939W}
}

@ARTICLE{TimmesSwesty_2000ApJS..126..501T,
       author = {{Timmes}, F.~X. and {Swesty}, F. Douglas},
        title = "{The Accuracy, Consistency, and Speed of an Electron-Positron Equation of State Based on Table Interpolation of the Helmholtz Free Energy}",
      journal = {\apjs},
         year = 2000,
        month = feb,
       volume = {126},
       number = {2},
        pages = {501-516},
          doi = {10.1086/313304},
       adsurl = {https://ui.adsabs.harvard.edu/abs/2000ApJS..126..501T}
}

@ARTICLE{MaederMeynet_2000ARA&A..38..143M,
       author = {{Maeder}, Andr{\'e} and {Meynet}, Georges},
        title = "{The Evolution of Rotating Stars}",
      journal = {\araa},
         year = 2000,
        month = jan,
       volume = {38},
        pages = {143-190},
          doi = {10.1146/annurev.astro.38.1.143},
archivePrefix = {arXiv},
       eprint = {astro-ph/0004204},
 primaryClass = {astro-ph},
       adsurl = {https://ui.adsabs.harvard.edu/abs/2000ARA&A..38..143M}
}

@ARTICLE{HegerLangerWoosley_2000ApJ...528..368H,
       author = {{Heger}, A. and {Langer}, N. and {Woosley}, S.~E.},
        title = "{Presupernova Evolution of Rotating Massive Stars. I. Numerical Method and Evolution of the Internal Stellar Structure}",
      journal = {\apj},
         year = 2000,
        month = jan,
       volume = {528},
       number = {1},
        pages = {368-396},
          doi = {10.1086/308158},
archivePrefix = {arXiv},
       eprint = {astro-ph/9904132},
 primaryClass = {astro-ph},
       adsurl = {https://ui.adsabs.harvard.edu/abs/2000ApJ...528..368H}
}

@ARTICLE{SillsLombardiBailyn_1997ApJ...487..290S,
       author = {{Sills}, Alison and {Lombardi}, Jr., James C. and {Bailyn}, Charles D. and {Demarque}, Pierre and {Rasio}, Frederic A. and {Shapiro}, Stuart L.},
        title = "{Evolution of Stellar Collision Products in Globular Clusters. I. Head-on Collisions}",
      journal = {\apj},
         year = 1997,
        month = sep,
       volume = {487},
       number = {1},
        pages = {290-303},
          doi = {10.1086/304588},
archivePrefix = {arXiv},
       eprint = {astro-ph/9705019},
 primaryClass = {astro-ph},
       adsurl = {https://ui.adsabs.harvard.edu/abs/1997ApJ...487..290S}
}

@ARTICLE{SonnebornFranssonLundqvist_1997ApJ...477..848S,
       author = {{Sonneborn}, George and {Fransson}, Claes and {Lundqvist}, Peter and {Cassatella}, Angelo and {Gilmozzi}, Roberto and {Kirshner}, Robert P. and {Panagia}, Nino and {Wamsteker}, Willem},
        title = "{The Evolution of Ultraviolet Emission Lines From Circumstellar Material Surrounding SN 1987A}",
      journal = {\apj},
         year = 1997,
        month = mar,
       volume = {477},
       number = {2},
        pages = {848-864},
          doi = {10.1086/303720},
archivePrefix = {arXiv},
       eprint = {astro-ph/9610021},
 primaryClass = {astro-ph},
       adsurl = {https://ui.adsabs.harvard.edu/abs/1997ApJ...477..848S}
}

@ARTICLE{LundqvistFransson_1996ApJ...464..924L,
       author = {{Lundqvist}, Peter and {Fransson}, Claes},
        title = "{The Line Emission from the Circumstellar Gas around SN 1987A}",
      journal = {\apj},
         year = 1996,
        month = jun,
       volume = {464},
        pages = {924},
          doi = {10.1086/177380},
archivePrefix = {arXiv},
       eprint = {astro-ph/9512025},
 primaryClass = {astro-ph},
       adsurl = {https://ui.adsabs.harvard.edu/abs/1996ApJ...464..924L}
}

@ARTICLE{SaumonChabriervanHorn_1995ApJS...99..713S,
       author = {{Saumon}, D. and {Chabrier}, G. and {van Horn}, H.~M.},
        title = "{An Equation of State for Low-Mass Stars and Giant Planets}",
      journal = {\apjs},
         year = 1995,
        month = aug,
       volume = {99},
        pages = {713},
          doi = {10.1086/192204},
       adsurl = {https://ui.adsabs.harvard.edu/abs/1995ApJS...99..713S}
}

@ARTICLE{RasioShapiro_1995ApJ...438..887R,
       author = {{Rasio}, Frederic A. and {Shapiro}, Stuart L.},
        title = "{Hydrodynamics of Binary Coalescence. II. Polytropes with Gamma = 5/3}",
      journal = {\apj},
         year = 1995,
        month = jan,
       volume = {438},
        pages = {887},
          doi = {10.1086/175130},
archivePrefix = {arXiv},
       eprint = {astro-ph/9406032},
 primaryClass = {astro-ph},
       adsurl = {https://ui.adsabs.harvard.edu/abs/1995ApJ...438..887R}
}

@ARTICLE{Pols_1994A&A...290..119P,
       author = {{Pols}, O.~R.},
        title = "{Case A evolution of massive close binaries: formation of contact systems and possible reversal of the supernova order}",
      journal = {\aap},
         year = 1994,
        month = oct,
       volume = {290},
        pages = {119-128},
       adsurl = {https://ui.adsabs.harvard.edu/abs/1994A&A...290..119P}
}

@ARTICLE{LaiRasioShapiro_1993ApJ...412..593L,
       author = {{Lai}, Dong and {Rasio}, Frederic A. and {Shapiro}, Stuart L.},
        title = "{Collisions and Close Encounters between Massive Main-Sequence Stars}",
      journal = {\apj},
         year = 1993,
        month = aug,
       volume = {412},
        pages = {593},
          doi = {10.1086/172946},
       adsurl = {https://ui.adsabs.harvard.edu/abs/1993ApJ...412..593L}
}

@ARTICLE{SyerClarkeRees_1991MNRAS.250..505S,
       author = {{Syer}, D. and {Clarke}, C.~J. and {Rees}, M.~J.},
        title = "{Star-disc interactions near a massive black hole}",
      journal = {\mnras},
         year = 1991,
        month = jun,
       volume = {250},
        pages = {505-512},
          doi = {10.1093/mnras/250.3.505},
       adsurl = {https://ui.adsabs.harvard.edu/abs/1991MNRAS.250..505S}
}

@ARTICLE{PodsiadlowskiJossRappaport_1990A&A...227L...9P,
       author = {{Podsiadlowski}, P. and {Joss}, P.~C. and {Rappaport}, S.},
        title = "{A merger model for SN 1987A.}",
      journal = {\aap},
         year = 1990,
        month = jan,
       volume = {227},
        pages = {L9-L12},
       adsurl = {https://ui.adsabs.harvard.edu/abs/1990A&A...227L...9P}
}

@ARTICLE{BenzHills_1987ApJ...323..614B,
       author = {{Benz}, Willy and {Hills}, Jack G.},
        title = "{Three-dimensional Hydrodynamical Simulations of Stellar Collisions. I. Equal-Mass Main-Sequence Stars}",
      journal = {\apj},
         year = 1987,
        month = dec,
       volume = {323},
        pages = {614},
          doi = {10.1086/165857},
       adsurl = {https://ui.adsabs.harvard.edu/abs/1987ApJ...323..614B}
}

@ARTICLE{KolykhalovSyunyaev_1980SvAL....6..357K,
       author = {{Kolykhalov}, P.~I. and {Syunyaev}, R.~A.},
        title = "{The Outer Parts of the Accretion Disks around Supermassive Black Holes in Galaxy Nuclei and Quasars}",
      journal = {Soviet Astronomy Letters},
         year = 1980,
        month = jun,
       volume = {6},
        pages = {357-361},
       adsurl = {https://ui.adsabs.harvard.edu/abs/1980SvAL....6..357K}
}

@ARTICLE{Paczynski_1978AcA....28...91P,
       author = {{Paczynski}, B.},
        title = "{A model of selfgravitating accretion disk.}",
      journal = {\actaa},
         year = 1978,
        month = jan,
       volume = {28},
        pages = {91-109},
       adsurl = {https://ui.adsabs.harvard.edu/abs/1978AcA....28...91P}
}
\bibliographystyle{aasjournal}


\end{CJK*}

\end{document}